\documentclass[10pt]{article}
\usepackage{psfig,epsfig,amssymb} 
\usepackage{multicol}  
\input psfig.sty
\input cyracc.def
\input amssymb.sty
\usepackage{color}
\usepackage{bm}                       

\def\geq{\geqslant}
\def\le{\leqslant}
\def\leq{\leqslant}

\newfont{\yihao}{cmb10 at 18pt}
  
\newcommand{\xiaosi}{\fontsize{11.5pt}{13.5pt}\selectfont}

\DeclareSymbolFont{lettersA}{U}{txmia}{m}{it}
\DeclareMathSymbol{\piup}{\mathord}{lettersA}{25}
\DeclareMathSymbol{\muup}{\mathord}{lettersA}{22}      %
\renewcommand{\baselinestretch}{1.06} 
 \catcode`@=11

\let\@oddfoot\@empty  \let\@evenfoot\@empty
\def\@evenhead{\small\thepage\hfill {\small\it Frank M. RIEGER; Emma de ONA-WILHELMI, and Felix A. AHARONIAN, Front. Phys. }\hfill }
\def\@oddhead{\hfill{\small\it Frank M. RIEGER; Emma de ONA-WILHELMI, and Felix A. AHARONIAN, Front. Phys. }\hfill\small\thepage}

\definecolor{oranged}{cmyk}{0.0,0,0.0,0.5}
\newfont{\xbt}{cmb10 at 12pt}

\usepackage{graphicx}                 

\graphicspath{{figs/}}      
\newcommand\simlt{\lower.5ex\hbox{$\; \buildrel < \over \sim \;$}}
\newcommand\simgt{\lower.5ex\hbox{$\; \buildrel > \over \sim \;$}}

\begin{document}
\thispagestyle{empty}

\vspace*{-15mm} {\small Front. Phys.}\\
\vspace*{-3.5mm} {{\hspace*{-1mm}\psfig{figure=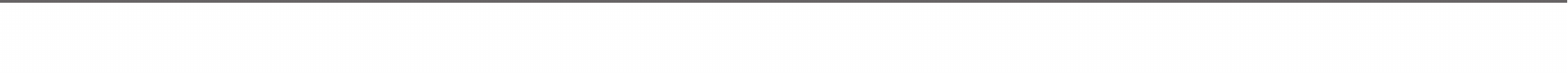}}}
\\[-1mm]
{\color{oranged}{\yihao R}{\bf\xiaosi \!E\!V\!I\!E\!W}{\bf\xiaosi
\!A\!R\!T\!I\!C\!L\!E}}
 \vspace*{8mm}

\begin{center}

{\usefont{T1}{fradmcn}{m}{n}\yihao TeV Astronomy}
\vspace*{5mm}

{\bf\small Frank M. RIEGER}$^{1,4}$, {\bf\small Emma de O\~NA-WILHELMI}$^{1,2}$, {\bf\small
Felix A. AHARONIAN}$^{1,3}$\vspace*{4mm}

{\footnotesize \it 
$^1$Max-Planck-Institut f\"ur Kernphysik, P.O. Box 103980, 69029 Heidelberg, Germany\\
$^2$Institut de Ciencies de L'Espai (IEEC-CSIC), Campus UAB, Torre C5, 08193 Bellaterra, Spain\\
$^3$Dublin Institute for Advanced Studies, 31 Fitzwilliam Place, Dublin 2, Ireland\\
$^4$European Associated Laboratory for Gamma-Ray Astronomy, jointly supported by CNRS and MPG
\\[1.5mm]
E-mail: $^\dag$frank.rieger@mpi-hd.mpg.de, wilhelmi@aliga.ieec.uab.es, felix.aharonian@mpi-hd.mpg.de\\[.5mm]
Received 2012; accepted 2013}
 \vspace{5mm}

\baselineskip 10pt
\renewcommand{\baselinestretch}{0.8}
\parbox[c]{152mm}
{\noindent{With the successful realization of the current-generation of ground-based detectors, TeV Astronomy 
has entered into a new era. We review recent advances in VHE astronomy, focusing on the potential of Imaging
Atmospheric Cherenkov Telescopes (IACTs), and highlight astrophysical implications of the results obtained 
within recent years.
\vspace{2mm}

 \vspace{2mm}

}}

\end{center}

\normalsize

\baselineskip 12pt \renewcommand{\baselinestretch}{1.06}
\parindent=10.8pt  \parskip=0mm \rm\vspace{5mm}

\begin{multicols}{2}
\setlength{\parindent}{1em}    


\centerline{Contents}\vspace{2.5mm}

\noindent 1\quad TeV Astronomy \hfill \vspace{0.0mm}

\hspace{1.5mm}1.1\quad Introduction\hfill 2\vspace{0.0mm}

\hspace{1.5mm}1.2\quad Ground-based Detection Technique \hfill 2\vspace{0.0mm}

\hspace{1.5mm}1.3\quad Future IACT Arrays \hfill 3\vspace{0.0mm}

\noindent 2\quad TeV Sources  \hfill \vspace{0.0mm} 

\hspace{1.5mm}2.1\quad Supernova Remnants\hfill 5\vspace{0.0mm}

\hspace{1.5mm}2.2\quad Pulsars\hfill 7\vspace{0.0mm}

\hspace{1.5mm}2.3\quad Pulsar Wind Nebulae\hfill 8\vspace{0.0mm}

\hspace{1.5mm}2.4\quad TeV Binary Systems\hfill 11\vspace{0.0mm}

\hspace{1.5mm}2.5\quad Galactic Centre\hfill 13\vspace{0.0mm}

\hspace{1.5mm}2.6\quad Blazars\hfill 16\vspace{0.0mm}

\hspace{1.5mm}2.7\quad Radio Galaxies\hfill 19\vspace{0.0mm}

\hspace{1.5mm}2.8\quad Starburst Galaxies\hfill 20\vspace{0.0mm}

\hspace{1.5mm}2.9\quad Candidates (GRBs, Clusters, Passive BHs)\hfill 20\vspace{0.0mm}

\noindent 3\quad Physics Impact of Recent Results\hfill \vspace{0.0mm}

\hspace{1.5mm}3.1\quad CR and Galactic Gamma-Ray Sources\hfill 21\vspace{0.0mm}

\hspace{1.5mm}3.2\quad Relativistic Outflows and AGNs \hfill 24\vspace{0.0mm}

\noindent 4\quad Conclusions and Perspectives\hfill 28\vspace{0.0mm}

\hspace{1.5mm}Acknowledgements\hfill \vspace{0.0mm}

\hspace{1.5mm}References\hfill 28\vspace{0.0mm}

\vspace*{6.0mm} \hrule\vspace{2mm} 
\noindent {\large \usefont{T1}{fradmcn}{m}{n}\xbt 1\quad TeV Astronomy}\vspace{2.5mm}
 
 \vspace*{5mm} \noindent {1.1\quad Introduction}\vspace{3.5mm}
 
The discovery of more than 100 extraterrestrial sources of  Very High Energy (VHE, $\geq100$ GeV) or TeV 
gamma-radiation belongs to the most remarkable achievements of the last decade in astrophysics. The strong 
impact of these discoveries on several topical areas of modern astrophysics and cosmology are recognised 
and highly appreciated by different astronomical communities. The implications of the results obtained with 
ground-based TeV gamma-ray detectors are vast; they extend from the origin of cosmic rays to the origin of 
Dark Matter, from processes of acceleration of particles by strong shock waves to the magnetohydrodynamics 
of relativistic  outflows, from distribution of atomic and molecular gas in the Interstellar Medium to the intergalactic 
radiation and magnetic fields.  

\noindent TeV gamma-rays are copiously produced in environments where effective acceleration of particles 
(electrons, protons, and nuclei) is accompanied by their intensive interactions with the surrounding gas and 
radiation fields. These interactions contribute significantly to the bolometric luminosities of young Supernova 
Remnants (SNRs), Star Forming Regions (SFRs), Giant Molecular Clouds (GMCs), Pulsar Wind Nebulae (PWNe), 
compact Binary Systems, Active Galactic Nuclei (AGNs) and Radio Galaxies (RGs), {\it etc.}.  

\noindent The fast emergence  of gamma-ray astronomy from an underdeveloped branch of cosmic-ray studies 
to a truly astronomical discipline is explained by the successful realization of the great potential of stereoscopic 
arrays of Imaging Atmospheric Cherenkov Telescopes (IACTs) which act as effective multifunctional tools for 
deep studies of cosmic gamma-radiation. 
 
\noindent Being recognised  as one of the most informative windows to the non-thermal Universe, the VHE 
domain of gamma-rays provides also means for probing fundamental physics beyond the reach of terrestrial 
accelerators. In particular, the indirect search for Dark Matter and tests of quantum gravity using these energetic 
gamma-rays are amongst the high priority objectives of the current and future projects with involvement of 
ground-based gamma-ray detectors. In this regard, the TeV gamma-ray astronomy is considered a key 
component of the new interdisciplinary research area called Astro-Particle Physics.

 \vspace*{5mm} \noindent {1.2\quad Ground-based Detection Technique}\vspace{3.5mm}

The atmosphere of the Earth is not transparent to gamma-rays, therefore their direct registration requires platforms
in space.
The currently operating {\it Fermi} Large Area Telescope ({\it Fermi}-LAT; formerly GLAST) is a powerful satellite-borne 
instrument designed for deep surveys with a very large field view of the order of 2 steradian. Presently, the study of 
the sky in MeV and GeV gamma-rays by {\it Fermi}-LAT is complemented by a somewhat smaller-scale telescope 
on the italian X-ray and gamma-ray satellite AGILE (Astro-rivelatore Gamma a Immagini LEggero). The angular 
resolution of these instruments below 1~GeV is quite modest (larger than $1^\circ$), but it becomes significantly 
better at higher energies, approaching $0.1^\circ$ above 10~GeV. {\it Fermi}-LAT covers a very broad energy 
region of primary gamma-rays extending from tens of MeV to hundreds of GeV (HE; up to 300 GeV). However, 
beyond 10 GeV the gamma-ray fluxes are generally very faint, so that the effective detection area of {\it Fermi}-LAT 
cannot provide adequate statistics for comprehensive spectral and temporal studies in the VHE domain.  

\noindent There is not much hope that space platforms could offer, in any time in the foreseeable future,  
instruments with detection areas significantly exceeding 1~m$^2$. This dramatically reduces the potential of 
studies of VHE gamma-rays from space. Fortunately, at these energies an alternative method can be exploited 
based on the registration of atmospheric showers, either directly or through their Cherenkov radiation. The faint 
and brief Cherenkov signal of relativistic electrons produced during the development of the electromagnetic 
cascades in the atmosphere, lasts only several nanoseconds, but it is sufficient for detection using 
large optical reflectors equipped with fast optical receivers. With a telescope consisting of a 10~m diameter 
reflector and a multichannel camera of pixel size of $\sim1/4$ degree and a field-of-view of $\sim3$ degree, 
primary gamma-rays of energy $E \geq 100$~GeV can be effectively collected across ground-level distances 
as large as 100~m providing a huge area for the detection of primary gamma-rays, $A_{\rm eff} \geq 10^4\ \rm m^2$. 
The total number of photons in the registered Cherenkov light image is proportional to the primary (absorbed 
in the atmosphere) energy, the orientation of the image correlates with the arrival direction of the gamma-ray 
photon, and the shape of the image contains information about the origin of the primary particle (a proton
or a photon?). The stereoscopic observations of air showers with two or more telescopes located at distances 
of about 100~m from each other, provide effective rejection of hadronic showers (by a factor of 100), as well 
as good angular resolution (better than $0.1^\circ$) and energy resolution (better than 15 per cent). At energies 
around 1 TeV, this results in a minimum detectable energy flux as low as $3 \times 10^{-13} \ \rm erg/cm^2 s$ 
(see e.g. [1]) This is much better than in any other gamma-ray domain, including the GeV energy band, where 
the sensitivity of {\it Fermi} LAT cannot compete with the performance already achieved by the H.E.S.S., MAGIC 
and VERITAS IACT arrays in the TeV energy band. Thanks to the very large collection area, the IACT technique 
provides large gamma-ray photon statistics even from relatively modest TeV gamma-ray emitters. In combination 
with good energy and angular resolutions, the gamma-ray photon statistics appears to be adequate for deep 
morphological, spectroscopic and temporal studies. This also makes the IACT arrays powerful multifunctional and 
multi-purpose tools for the exploration of a broad range of non-thermal objects and phenomena. The potential 
of the IACT arrays has been convincingly demonstrated by the H.E.S.S., MAGIC and VERITAS collaborations  
(see, e.g.~[2], and references therein).

\noindent The IACT arrays are capable to study not only point-like, but also extended sources with an angular 
size up to 1 degree or somewhat larger. Moreover, the high flux  sensitivity and relatively large ($\geq 4^\circ$)
field of view of  IACT arrays allow rather effective all-sky surveys as demonstrated by the H.E.S.S. collaboration. 
On the other hand, the potential of IACT arrays is rather limited for the search of very extended structures like
the galactic plane diffuse emission or the huge radio lobes of the nearby radio galaxy Centaurus A. IACT arrays have 
a limited capability for "hunting'' of solitary events like the possible VHE counterparts  of Gamma Ray Bursts. In this regard, 
the detection technique based on direct registration of particles (leptons, hadrons and photons) of extensive 
air showers (EAS) is a complementary approach to the IACT technique.

\noindent The traditional EAS technique, based on scintillators or water Cherenkov detectors spread over large 
areas, was originally designed for the detection of cosmic rays at PeV and EeV energies. In order to adopt this 
technique to gamma-ray astronomy, the energy threshold needs to be reduced by two or three orders of magnitude. 
This can be achieved using dense particle arrays located on very high altitudes.  The feasibility of both approaches 
recently have been successfully demonstrated by the ARGO and Milagro collaborations. In particular, several 
very extended sources have been reported by the Milagro group. These results, as well as the potential  for 
continuous monitoring of a significant part of the sky, which might lead to exciting discoveries of yet unknown 
VHE transient phenomena in the Universe, strongly support the proposals of constructing high altitude EAS 
detectors (see [1] for a review) like HAWC, a High Altitude Water Cherenkov Experiment, presently under 
construction on a site close to Sierra Negra, Mexico [3]. The 5~yr-survey sensitivity of HAWC above 1~TeV is 
expected to be comparable to the sensitivity of {\it Fermi}-LAT at 1 GeV. Thus  HAWC will be complementary 
to {\it Fermi} for continuous monitoring of more than 1 steradian fraction of the sky at TeV  energies. At higher 
energies, recently a new project called LHAASO (Large High Altitude Air Shower Observatory) has been 
suggested. The proposed huge detector facility at Yangbajing, China, will consist of several sub-arrays   
for the detection of the electromagnetic and muon components of air showers. They will cover a huge area, 
and can achieve an impressive sensitivity at energies of several tens of TeV (see Fig. 1). 
\vspace{3mm}
\centerline{\psfig{figure=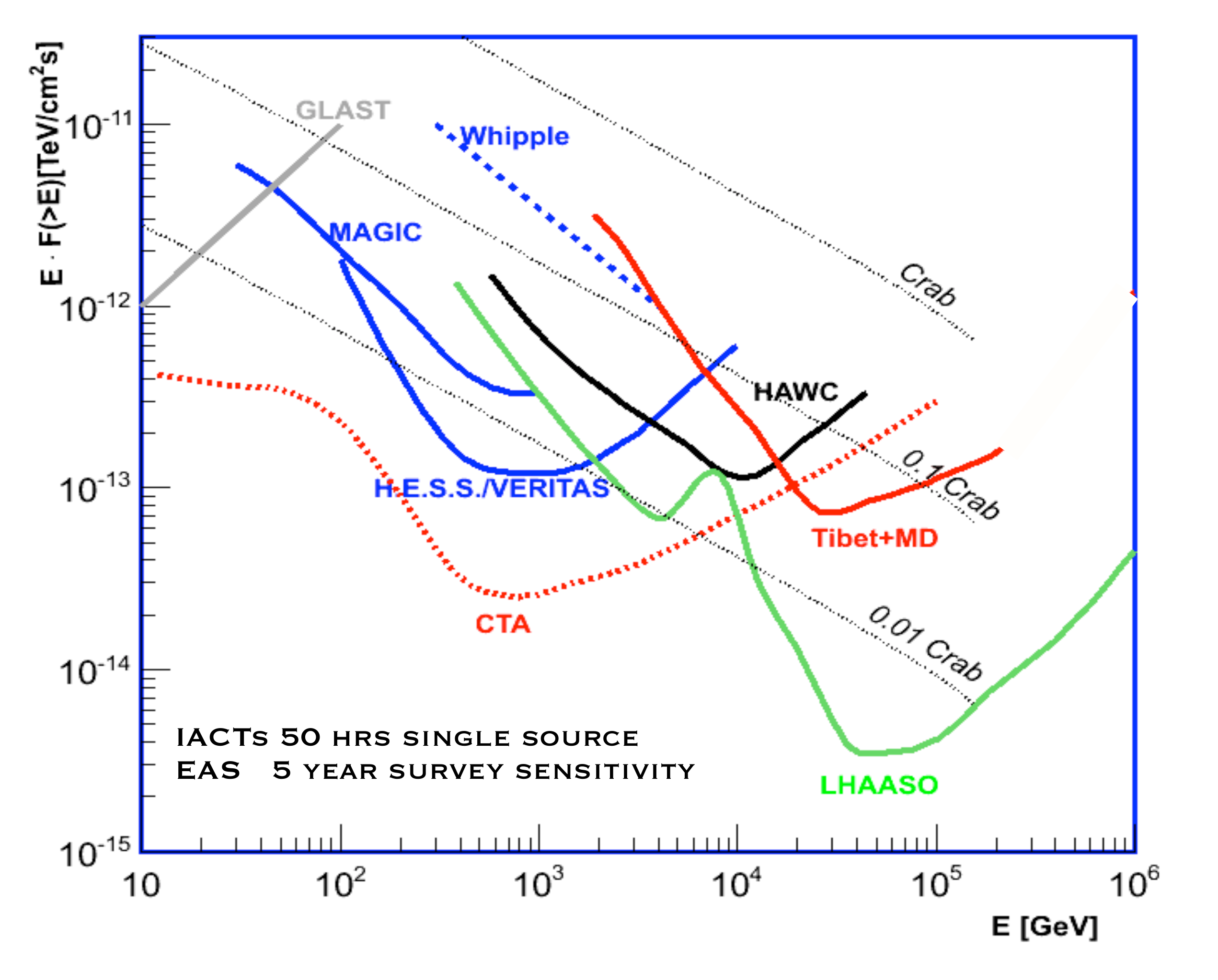, width=8cm}\vspace{0mm}}
{\baselineskip 10.5pt\renewcommand{\baselinestretch}{1.05}\footnotesize \noindent
{\bf Fig.~1}\quad Energy-flux sensitivities of current and future ground-based detectors - the IACT 
and EAS  arrays in the energy range $10^{10}$ to $10^{15}$ eV (courtesy of G.~Sinnis).}
\vspace{3mm}

\vspace*{5mm} \noindent {1.3\quad Future IACT Arrays}\vspace{3.5mm}

The future of observational gamma-ray astronomy, at least for the next 10-15 years, is connected with the 
next-generation IACT arrays, first of all with the observatory CTA (Cherenkov Telescope Array) [4], cf. also 
Fig.~2.\\  
The next generation of IACT arrays are aiming at (i) a significant (by an order of magnitude) improvement 
of the flux sensitivities in the standard 0.1-10 TeV energy interval ({\it TeV regime}), and (ii) an expansion 
of the energy domain of IACT arrays in both directions - down to 10 GeV ({\it multi-GeV regime}) and well 
beyond 10 TeV ({\it sub-PeV regime}):   

\vspace{3mm}
\centerline{\psfig{figure=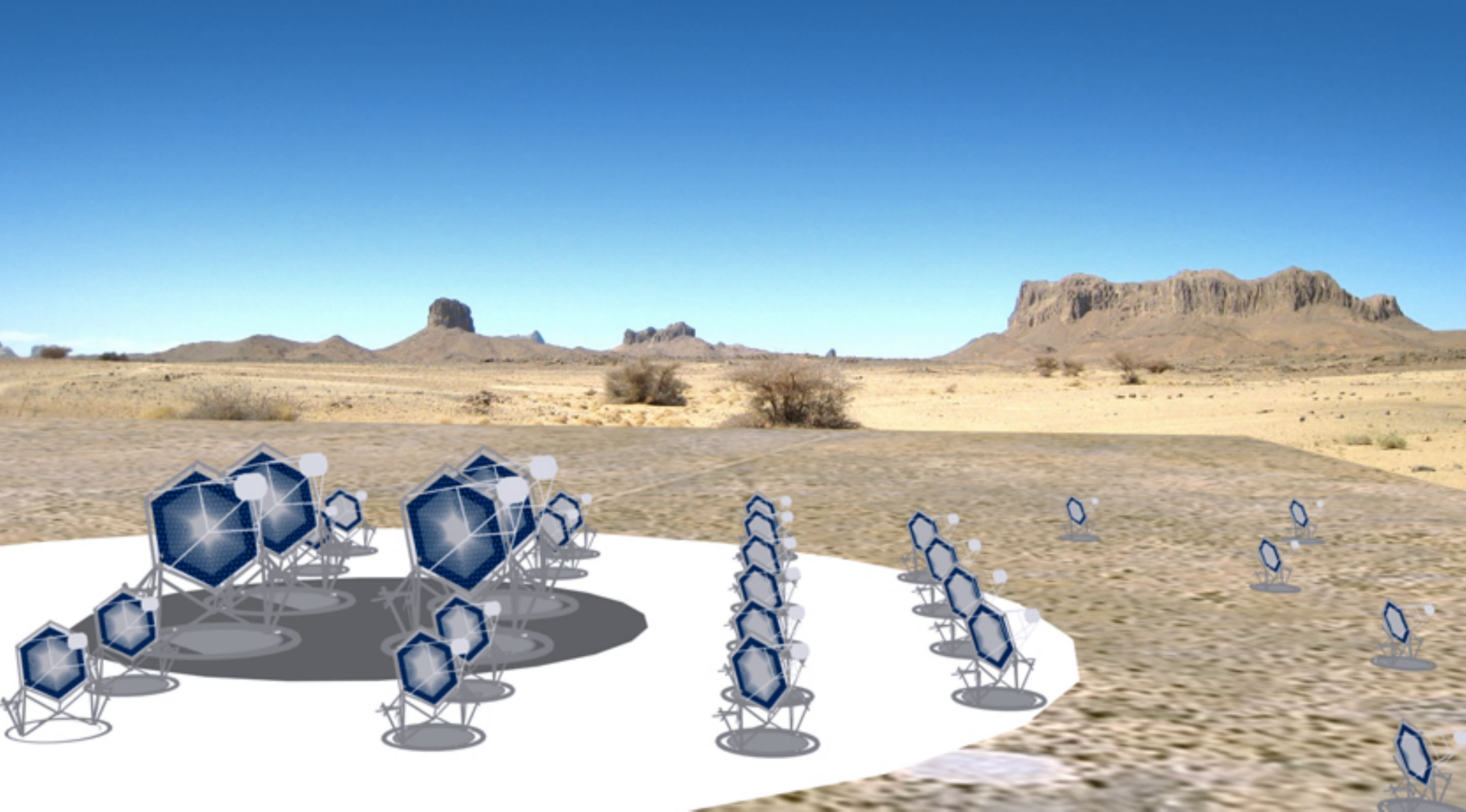, width=8cm}\vspace{1mm}}
{\baselineskip 10.5pt\renewcommand{\baselinestretch}{1.05}\footnotesize \noindent
{\bf Fig.~2}\quad Possible layout of the next-generation CTA instrument. From Ref. [4].}
\vspace{3mm}

\vspace{3mm}
\noindent {\bf $\bullet$ TeV regime:}
\vspace{2mm}\\
This is the "nominal" energy region where the IACT technique has achieved its best performance.  
The potential in this energy regime is still not saturated. With a stereoscopic array consisting of 
tens of 10~m-diameter (medium-size) class telescopes the minimum detectable energy flux could 
be reduced to the level of $10^{-14}$ erg/cm$^{2}$ s, and the angular resolution be improved to 
$\delta \theta  \leq$ 3 arc minutes. Generally, the optimum distance between the telescopes is 
considered to be around 100~m, the radius within which the Cherenkov light is distributed more or 
less homogeneously. However, if highest priority is given to the performance at energies around 
1 TeV and beyond, an increase of the distance between telescopes up to 300~m could be an attractive 
option. For a fixed number of telescopes this would increase the detection area by an order of magnitude, 
and, at the same time, improve the angular resolution to 1-2 arc minutes, although at the expense of 
a somewhat higher (by a factor of two or three) energy threshold. In any case, a reduction of the 
minimum detectable energy flux  around 1~TeV down to $10^{-14} \rm \ erg/cm^2 s$ seems to be a 
challenging but feasible "target". It will be a great achievement even by the standards of the most advanced 
branches of observational astronomy, allowing us to probe, in particular, potential TeV gamma-ray sources 
at luminosity levels of $10^{32}\ (d/10~\rm kpc)^2 \ \rm erg/s$ for galactic and $10^{40}\ (d/100~\rm Mpc)^2 \ 
\rm erg/s$ for extragalactic objects. Although for moderately extended sources, e.g. of angular size 
$\Psi \sim 1^\circ$, the minimum detectable energy flux will be by a factor of $\Psi / \delta \theta \sim
10-30$ higher, it would compete or be better than the energy flux sensitivities of the best current X-ray 
satellites, \textit{Chandra}, \textit{XMM}-Newton, \textit{INTEGRAL} and \textit{Suzaku}, and thus allow 
the deepest probes of non-thermal high energy phenomena in extended sources, in particular in 
shell-type SNRs, Giant Molecular Clouds, Pulsar-driven Nebulae (Plerions), Clusters of Galaxies, 
hypothetical Giant Pair Halos around AGN, \textit{etc.}  Such a system  of 10-12~m diameter class IACTs 
with a field of view (FoV) of 6-8 degrees, will most likely constitute the core of the Cherenkov Telescope 
Array (CTA) - an initiative towards a major ground-based gamma-ray detector (see Fig. 2).

\vspace{3mm}
\noindent
{\bf $\bullet$ Sub-PeV regime:}
\vspace{2mm}\\
External and intergalactic absorption of gamma-rays, the limited efficiency of particle acceleration, the  
escape of highest energy particles from the source \textit{etc.}, result in a suppression of fluxes at the 
highest energies. The general tendency of decreasing gamma-ray fluxes with energy becomes especially 
dramatic above 10 TeV. Therefore, any meaningful study of cosmic gamma-rays beyond 10 TeV typically 
requires detection areas  as large as   $1 \ \rm km^2$. An effective and straightforward approach would 
be the use of IACT arrays optimised for detection of gamma-rays in the region up to 100~TeV and beyond.  
This can be realised by modest, approximately 10-30~m$^2$-area reflectors separated from each other,   
depending on the scientific objectives and the configuration of the imagers, by 300 to 500~m.
The requirement on the pixel size of imagers is also rather modest, $0.25^\circ$ or so, however they should 
have large, up to 10 degree  FoV for simultaneous detection of showers from distances more than 300~m [5].
A sub-array consisting of several tens of such small-size telescopes is included in the concept of CTA with 
a primary goal to study the energy spectra of gamma-ray sources well beyond 10~TeV.  It will serve as a 
powerful tool for searches of galactic cosmic ray "PeVatrons", as well as nearby ($R \ll 10$~Mpc) radio 
and starburst galaxies.  

\vspace{3mm}
\noindent
{\bf $\bullet$ Sub-100 GeV regime:}
\vspace{2mm}\\
The energy threshold of detectors, $E_{\rm th}$, is generally defined as a characteristic energy at which
the gamma-ray detection rate for a primary power-law spectrum with a photon index  2-3 achieves its 
maximum. It is known from Monte Carlo simulations as well as from the experience of operation of  
previous generation of IACTs, that in practice the best sensitivity is achieved at energies exceeding several 
times $E_{\rm th}$. Thus, for optimisation of gamma-ray detection around 100 GeV, one should reduce 
the energy threshold of telescopes to $E_{\rm th} \leq 30 \ \rm GeV$. This can be done by using very large, 
20~m-diameter (large-size) class reflectors. On the other hand, the reduction of the threshold to 30~GeV is an 
important scientific issue in its own right; the intermediate interval between 30 and 300 GeV 
could be crucial for certain classes of galactic and extragalactic gamma-ray sources. A sub-array consisting 
of several large-size telescopes as foreseen in CTA (see Fig. 2) will indeed significantly broaden the topics 
and scientific objectives of CTA.  

\noindent Each of the IACT arrays discussed above covers at least two decades in energy with significant 
overlaps of the energy domains. Since these arrays contain the same basic elements, and generally have  
also common scientific motivations, an ideal arrangement would be if these sub-arrays are combined in 
a single facility which would have a sensitive and homogeneous coverage throughout the energy region 
from approximately 30 GeV to 300 TeV. The concept of CTA is based, to a large extent, on this argument [4].  
The high detection rates, coupled with good angular and energy resolutions over four energy decades will 
make CTA a powerful multi-purpose gamma-ray observatory with a great capability for spectrometric,
morphological and temporal studies  of a diverse range of persistent and transient high-energy phenomena 
in the Universe. 

\vspace{3mm}
\noindent
{\bf $\bullet$ Multi-GeV regime}: {\it  Gamma-Ray Timing Explorers}
\vspace{2mm}\\
Despite the recent great achievements of high energy (HE) gamma-ray astronomy, there are obvious 
shortcomings in the performance of the current so-called "pair-conversion"  tracking detection technique - 
the most effective approach used in satellite-borne instruments for detection of gamma-rays at energies 
above 100 MeV.    
One should note that the flux sensitivity of \textit{Fermi}-LAT at 1 GeV of about $10^{-12} \ \rm erg/cm^2 s$       
can be achieved only after one year all-sky survey. While for persistent gamma-ray sources this seems to be 
an adequate sensitivity (given that a huge number of sources  are simultaneously monitored within the large 
and homogeneous FoV), the small detection area significantly limits its potential, in particular for detailed 
studies of the temporal and spectral characteristics of highly variable sources like blazars or solitary events 
like gamma-ray bursts (GRBs). It will not be easy to improve the sensitivity achieved by {\it Fermi}-LAT at high 
energies by any future space-based mission, unless the Moon would be used in the (far) future as a possible 
platform for an installation of very large ($\gg 10 \rm m^2$) area pair-conversion tracking detectors. Apparently, 
the space-based resources of GeV gamma-ray astronomy have achieved a point where any further progress 
would appear extremely difficult and very expensive. In any case, for the next decades to come there is no 
space-based mission planned for the exploration of the high-energy gamma-ray sky. On the other hand, the 
principal possibility of an extension of the IACT technique towards 10 GeV promises a new breakthrough in 
gamma-ray astronomy [1]. The (relatively) large gamma-ray fluxes in this energy interval, together with the 
huge detection areas offered by the IACT technique, can provide the highest  gamma-ray photon statistics 
compared to any other energy band of cosmic gamma-radiation. Thus, in the case of a realization of 
10~GeV-threshold IACT arrays, the presently poorly explored interval between 10  and 100 GeV could become 
one of the most  advanced domains of gamma-ray astronomy with a great potential for the studies of highly 
variable phenomena.\\ 
The reduction of the energy threshold down to 10~GeV or even less is principally possible within the basic 
concept of the IACT  technique, but it requires an extreme approach of using  25~m diameter class telescopes 
with very high ($\geq  40\%$) quantum efficiency focal plane imagers to operate in a robotic regime at very high 
(5~km or) mountain altitudes [6].\\
The energy range from several GeV to 30 GeV has very specific astrophysical and cosmological objective: 
exploration of the highly variable non-thermal phenomena in the remote universe at redshifts of z = 5 (like large 
redshift quasars and GRBs), as well as the study of compact galactic sources such as pulsars and microquasars. 
A realization of such a gamma-ray timing explorer, hopefully during the lifetime of the {\it Fermi} observatory 
would be a great achievement for gamma-ray astronomy. 

\vspace*{6.0mm} \hrule\vspace{2mm} \noindent {\large
\usefont{T1}{fradmcn}{m}{n}\xbt 2\quad TeV Sources}\vspace{2.5mm}

\vspace*{5mm} \noindent {2.1\quad Supernova Remnants}\vspace{3.5mm}

\noindent Massive stars are believed to end their life undergoing a supernova explosion. This explosion blows 
off their other layers into a supernova remnant (SNR), which heats the surrounding medium and accelerates 
cosmic-rays (electrons and protons) to extremely high energies. The radiation from shell-like SNRs consists of 
thermal emission from shock-heated gas and non-thermal emission from shock-accelerated particles. The theory 
of diffusive shock acceleration (DSA) at shock fronts [7,8] predicts the production of a population of accelerated 
particles in SNRs that can interact with ambient magnetic fields, with ambient photon fields, or with matter. The 
amount of relativistic particles increases with time as the SNR passes through its free expansion phase, and 
reaches a maximum in the early stages of the Sedov phase. Correspondingly, the peak in gamma-ray luminosity 
typically appears some 10$^3$--10$^4$ years after the supernova explosion.\\ 
In the TeV domain, presently seven shell-type SNRs - Cas A [9-11], Tycho [12], SN 1006 [13], RX\,J1713.7--3946 
[14,15], RX\,J0852--4622 (Vela Junior) [16], RCW 86 [17], and G353.6--0.7 (HESS\,J1731--347)[18] have been 
firmly identified as VHE gamma-ray emitters (see Table~1).

\begin{table*}[ht]
\centering                                 
Table~1: Shell-like SNRs firmly detected at TeV energies\\[0.2cm]
\begin{tabular}{c c c c c c}      
\hline\hline                        
Name & Dist (kpc) &  Size (pc) & Age (yrs)  & L$_{\gamma}$ (10$^{33}$ erg/s) & $\Gamma$  \\
\hline    
RX\,J1713.7--3946 & 1 & 17.4 & 1.6 & 8 & 2.0  \\
RX\,J0852--4622 & 0.2(1) & 6.8(34) & 0.4(5)& 0.26(6.4) & 2.2  \\
RCW\,86 & 1(2.5) & 11(28) & 1.6(10) & 1(6) & 2.5 \\
SN\,1006 & 2.2 & 18.3 & 1 &  1.24 & 2.3 \\
Cas\,A & 3.4 & 2.5 & 350 & 7 & 2.4 \\
Tycho & 3.5 & 6 & 438 & 0.1 & 1.95 \\
SNR G353.6-0.7 & 3.2 & 27  & 2.5(14) &  10 &  2.3\\
\hline
\label{table:1}   
\end{tabular}
\end{table*}

\noindent Remarkably, while the first six sources are well established young SNRs, the object G353.6--0.7 is the first 
SNR discovered serendipitously in TeV gamma-rays, and only later confirmed by radio and X-ray observations [19,20]. 
Moreover, a possible new SNR candidate, HESS\,J1912+101, has been postulated recently [21] based solely on its 
shell-type morphology at TeV energies, although no counterpart at lower energies has been detected so far. The two 
latest examples demonstrate the potential of large field-of-view Cherenkov telescopes for serendipitously discovering 
extended SNRs (of typical size 0.2-1$^{\rm o}$ at a distance up to $\sim$3.5 kpc). Their relatively large sizes and $\gamma$-ray 
luminosities of about $(0.1-1)\times 10^{33}$ erg/s have enabled the detection of these objects up to distances of 
$\sim3.5$ kpc (e.g., Tycho) with current instrument sensitivities (cf. Fig.~1). 
If the VHE gamma-ray luminosities detected from these objects reflect the typical luminosity of the SNR population in 
the Galaxy, future instrument like CTA should be able to detect SNRs up to 15 kpc, thus sampling the whole Galaxy.
Taking the spatial distribution of SNRs in the Galaxy, their explosion rate, and the duration of the TeV emission (believed 
to last a few thousand years) into account, roughly $\sim$100 new SNRs could be discovered at TeV energies [22] (in 
a naive approximation, without considering energy cut-offs, hard/soft spectral indices, etc.). Such an enlarged population 
would allow the study of these objects at different evolutionary stages, sampling their spectral energy distribution from 
a few hundred of MeV (with {\it Fermi}-LAT and AGILE) up to 100 TeV, in the cut-off regime. 

\vspace{4mm}
\centerline{\psfig{figure=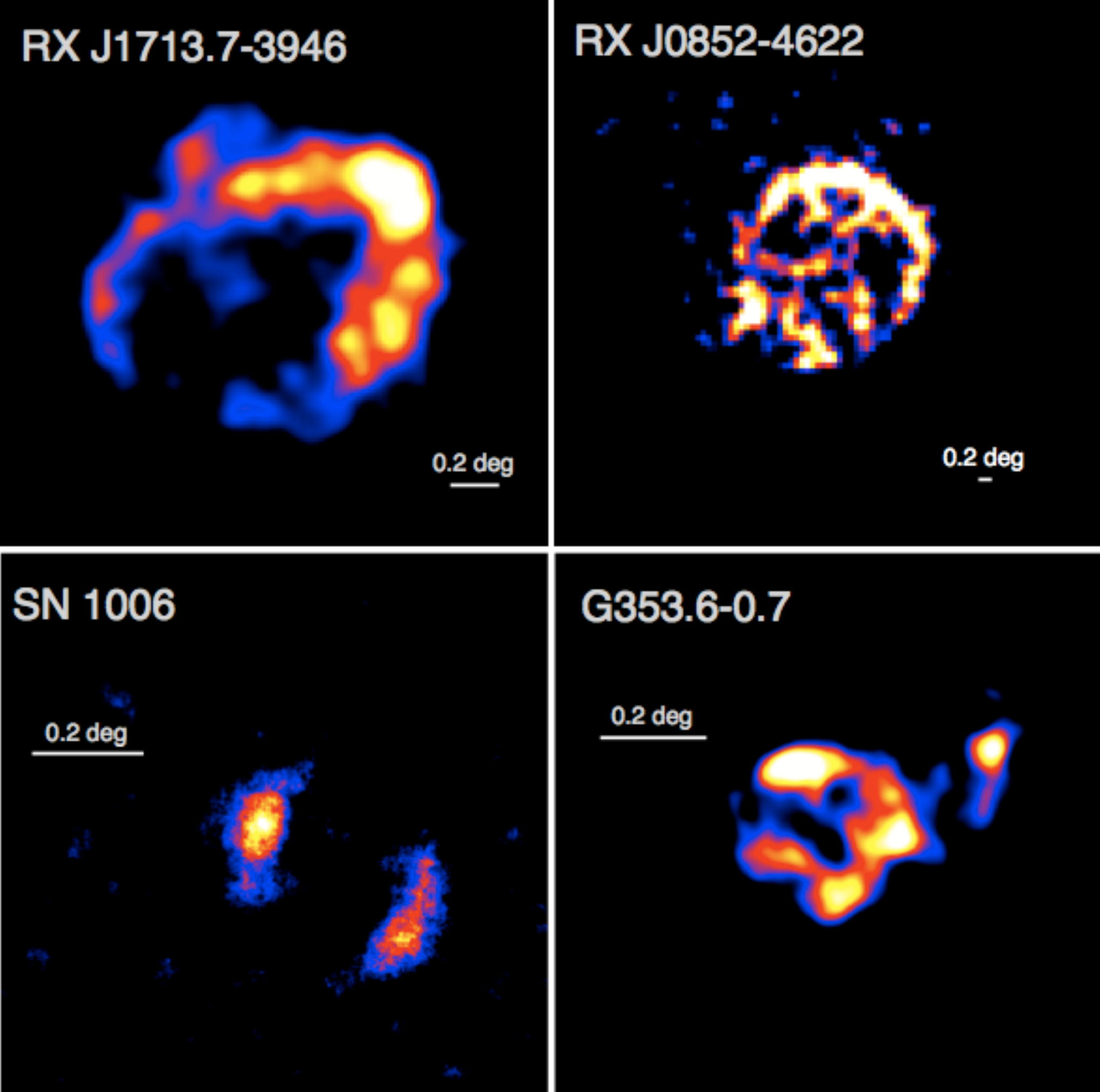, width=3in}\vspace{1mm}}
{\baselineskip 10.5pt\renewcommand{\baselinestretch}{1.05}\footnotesize \noindent
{\bf Fig.~3}\quad Example of four shell-type SNRs detected at TeV energies with the H.E.S.S. instrument.}
\vspace{3mm}

\noindent The sizes of several of these shell-like SNRs ($>0.1^{\rm o}$) has allowed to resolve them in VHE (see Fig. 3). 
The images of SNRs such as SN 1006, RX\,J1713.7--3946 or RX\,J0852--4622 have revealed a good correlation of the 
TeV emission sites with the non-thermal emission detected in X-rays, probing acceleration of relativistic particles up to 
multi-TeV energies. However, the relative contributions of accelerated protons and electrons to the gamma-ray production 
still remain unknown. The problem is that the ratio of gamma-rays produced by accelerated protons interacting with the 
surrounding gas, and by ultra-relativistic electrons up-scattering the 2.7K CMB radiation, is very sensitive to generally 
unknown parameters, in particular to the gas density and the magnetic field of the ambient medium (cf., e.g. [302]). The 
efficiency of inverse Compton (IC) scattering is especially high at TeV energies (up to E$_{\rm e}\approx$100 TeV, it 
proceeds in the Thomson regime, with a corresponding cooling time t$^{\rm IC}_{\rm cool}\propto$ 1/E$_{\rm e}\propto 
1/E_{\gamma}^{1/2}$). For example, the typical production time of a 1 TeV-photon by an electron and a proton of the 
same characteristic energy of about 20 TeV, are $\approx$5$\times$10$^4$yr and 5$\times$10$^7$(n/1 cm$^3$)$^{-1}$
yr, respectively (see, e.g. [23]). Correspondingly, at 1 TeV the ratio of the production rates of IC to $\pi^{\rm o}$-decay 
gamma-rays, is approximately 10$^3$ (W$_{\rm e}$/W$_{\rm p}$)(n/1 cm$^{-3}$)$^{-1}$, where W$_{\rm e}$ and 
W$_{\rm p}$ are the total energies in 20 TeV electrons and protons, respectively. Thus, even for a very small 
electron-to-proton ratio (at the stage of acceleration), e/p $=10^{-3}$, the contribution of the IC component will dominate 
over the $\pi^{\rm o}$-decay gamma-rays (in the shell with a typical gas density n$\le$1cm$^{-3}$), unless the magnetic 
field in the shell significantly exceeds 10$\mu$G. In this case, the accelerated electrons are cooled predominantly via 
synchrotron radiation, thus only a small fraction, $w_{\rm CMB}/w_{\rm B}\approx$  0.1(B/10$\mu$G)$^{-2}$, will be 
released in IC gamma-rays. Alternatively, the proton-to-electron acceleration ratio should exceed $e/p\sim10^{3}$ 
which, in principle, cannot be excluded given the uncertainty associated with one of the key aspects of DSA related to 
the so-called {\it injection problem} (see [24]).

\noindent In cases like RX\,J1713.7--3946, Tycho or Cas A, the magnetic field has been estimated from multi-wavelength 
observations to be $>$0.1~mG [25,26], restricting the contribution of the IC emission and in principle favouring an hadronic 
origin of the TeV emission. Nevertheless, if the IC and synchrotron components of the radiation are formed in different zones, 
these constraints are less robust. For instance, a difference of the magnetic field in the upstream and the downstream region 
could result in a positional shift of the production regions of synchrotron X-rays and IC gamma-rays, and more complex 
models implying multi-zone emission would need to be invoked [27,28]. In general, while the distribution of the X-ray 
radiation is dominated by the strength of the magnetic field, the TeV emission traces the particle distribution and does not 
depend on the magnetic field, allowing a more unbiased study of the particle acceleration in the shell. With the angular 
resolution of current instruments (of the order of $\approx$0.1$^{\rm o}$) those different sites are still indistinct, but the
future improvement of the angular resolution to a few arcmin should permit a detailed study of the TeV radial profile in
sources like RX\,J1713.7--3946 or SN\,1006 in comparison with the X-ray radiation profile. 

\noindent The spectral energy distribution (SED) of these young SNRs extends over almost five decades, from a few 
hundred MeV to a few tens of TeV. At low energies the SED part for some of these TeV shell-like SNRs has been 
detected with the {\it Fermi}-LAT telescope [29-32]. The coverage of the spectrum at low energies has improved our 
understanding of the origin of the gamma-ray emission, but also evidenced a more complicated scenario in which 
different regions can contribute to the total emission, such as the reverse shock [28] or dense clouds embedded in 
the shock [33,34]. The photon spectra of Tycho, RX\,J0852--4622 and Cas\,A continues to the MeV-GeV range with 
a rather hard spectral index of $\simeq$2.0 as predicted by the DSA theory [35-37]. This fact, together with the high 
magnetic field amplification derived from synchrotron X-ray filaments, preventing in principle a large IC contribution 
from leptons, favour an hadronic scenario in these SNRs. Moreover, high-energy radiation up to at least a few TeV 
has also been detected from these SNRs without an indication for a turnover in the spectrum. An extension of the 
high-energy emission by a factor two or three beyond 10 TeV could only be explained through hadronic interactions, 
given the fast Klein-Nishina-cooling suffered by 100 TeV-electrons emitting in this energy regime, and would robustly 
exclude an IC origin of the radiation. It would also provide a definitive probe of SNRs as origin of the cosmic-ray sea 
(see Section 3.1.).\\
RX\,J0852--4622 and RX\,J1713.7--3946, for which large magnetic fields have been estimated, face some difficulties 
when modeling their gamma-ray emission. These two SNRs have similar ages, sizes, and radio, X-ray and TeV 
gamma-ray spectra, although RX\,J1713.7--3946 shows a softer spectral index ($\simeq$1.5) in the 100 MeV to 1 
GeV-energy range, similar to the predicted indices in a leptonic scenario. In both cases, the apparent low gas density 
(n$\simeq$0.1~cm$^{-3}$) [38] poses troubles to standard hadronic scenarios [28,39,40]. Still, even in the case of a 
very low gas density of the shell, the contribution of hadronic gamma-rays could be significant, if accelerated protons 
interact with the dense cores of molecular clouds embedded in the shell [82]. In such a case, slow diffusion could 
prevent low-energy particles to penetrate into these dense cores, suppressing the low-energy gamma-ray emission 
and naturally explaining the hard gamma-ray spectrum measured in RX\,J1713.7--3946. At the highest energies, 
RX\,J1713.7--3946 shows a energy cut-off above few TeV, excluding PeV protons from this remnant. However, an
escape of high-energy protons that cannot be confined in the shell, can not be excluded and might be a plausible 
explanation. In fact, at GeV energies, a large number of mid-age SNRs has been discovered, while only a small 
fraction of them shines at TeV energies. The gamma-ray emission in these cases is likely related to interactions of 
cosmic-rays with dense gaseous complexes [34]. In cases like W51C, detected up to $\sim$5 TeV [41,42], an 
enhancement of the hadronic origin due to the large gas density in the region seems clearly favoured. On the other 
hand, the best example illustrating the escape of high-energy particles is the $10^4$ yr-old SNR W28 [43], where 
a clear correlation between the TeV emission and massive molecular clouds emitting in CO has been observed. 
Some of these clouds are also bright at GeV energies. Another example of this type of scenario is IC 443 [44-47], 
where the GeV and TeV emission appear shifted from each other. These images seem to support an escape 
scenario where, depending on the location of the massive clouds, the time of particle injection into the interstellar 
medium and the diffusion coefficient, a broad variety of energy distributions may be expected. 

\vspace*{5mm} \noindent {2.2\quad Pulsars}\vspace{3.5mm}

Pulsars -- rapidly rotating and highly magnetised neutron stars surrounded by a rotating magnetosphere and 
accompanied by relativistic outflows - emit radiation at all wavelengths. Charged particles (electrons and positrons) 
are thought to be efficiently accelerated in the electromagnetic fields of the pulsar, producing $\gamma$-radiation 
via e.g. curvature processes and supporting the formation of a cold relativistic outflow beyond the light cylinder. 
This pulsar wind carries almost the entire rotational energy of the pulsar in the form of Poynting flux and/or kinetic 
energy of the bulk motion, and creates a standing shock wave (the termination shock) when it interacts with the 
ambient medium. Particles accelerated at this shock are responsible for the steady and usually very extended 
non-thermal radiation observed (see Sec. 2.3). 

\noindent Although pulsars have been traditionally a subject of radio astronomy, with $\approx$1800 pulsars 
found beaming radio waves, most of their radiation is emitted at high-energies (a few percent of their spin-down 
power). Indeed, in the last three years, the number of gamma-ray pulsars has increased exponentially from half 
a dozen to more than 150 [48] thanks to the new sensitive instruments {\it Fermi}-LAT and AGILE. Despite the high 
Galactic background, the periodic gamma-ray emission stands out due to the high fluxes, hard spectral index 
and powerful timing analysis tools. The large statistics and good data quality has provided new insights into the
physics of pulsars. In general, it is believed that the pulsed, periodic gamma-ray radiation originates in regions of 
the magnetosphere, called {\it{gaps}}, where the electric field has a parallel component along the magnetic field 
lines. This electric field efficiently accelerates electrons and positrons to relativistic energies causing them to emit 
synchro-curvature radiation in the form of gamma-rays. There are currently a few models that differ, primarily, on 
the location of these gaps [49-51], which are capable to explain the light-curves and spectral energy distributions. 
Other mechanisms have also been suggested such as a magnetosphere with a force-free structure [52] or a striped 
wind topology [53]. The {\it Fermi}-LAT-measured light curves and energy spectra indicate that gamma-ray emission 
from the brightest pulsars is produced in the outer magnetosphere with fan-like beams scanning over a large portion 
of the celestial sphere. The energy spectra for most of the gamma-ray pulsars are best described by a power-law 
function with an exponential cutoff of the form E$^{-\Gamma}$$\exp{[-(E/E_0)^b]}$ with $b \leq 1$, and cut-off energy 
$E_0$ between 1 and 10 GeV [48]. The detection of gamma-rays beyond a few GeV without indication for a 
super-exponential attenuation (i.e., $b>1$) effectively excludes the so-called {\it polar cap} model and gives a 
preference to models of gamma-ray production in the outer magnetosphere (in order to avoid severe pair-production 
in the strong magnetic field in low-altitude zones). Most of the measured spectra can be well-fitted with a simple 
exponential attenuation ($b=1$) [48], which is in general well-explained by the mechanism of curvature radiation. 
However, an extension of the spectral measurements for the brightest gamma-ray pulsars towards both, higher 
and lower energies, has revealed that the spectra beyond the cut-off could be smoother (b$\simeq$0.5). For 
example, the phase-averaged spectrum of the Crab pulsar is better fitted with the combination of parameters 
$b=0.43$, $\Gamma=1.59$ and $E_0=0.50$~GeV [54], rather than $b=1$, $\Gamma=1.97$ and $E_0=5.8$~GeV 
as reported earlier by the {\it Fermi}-LAT collaboration based on smaller gamma-ray statistics [55].  In any case, if 
the above noted fit of the energy spectra is extrapolated to higher energies, a dramatic decrease of gamma-ray 
fluxes well beyond 10 GeV is expected, preventing the detection of pulsed emission with the current instrument at 
$\sim$100\,GeV. The MAGIC telescope, using a novel trigger system detected sub-100\,GeV pulsed emission 
from the Crab pulsar [56], favouring models with exponential or sub-exponential cut-offs (slot gap and outer gap 
models). 

\vspace{4mm}
\centerline{\psfig{figure=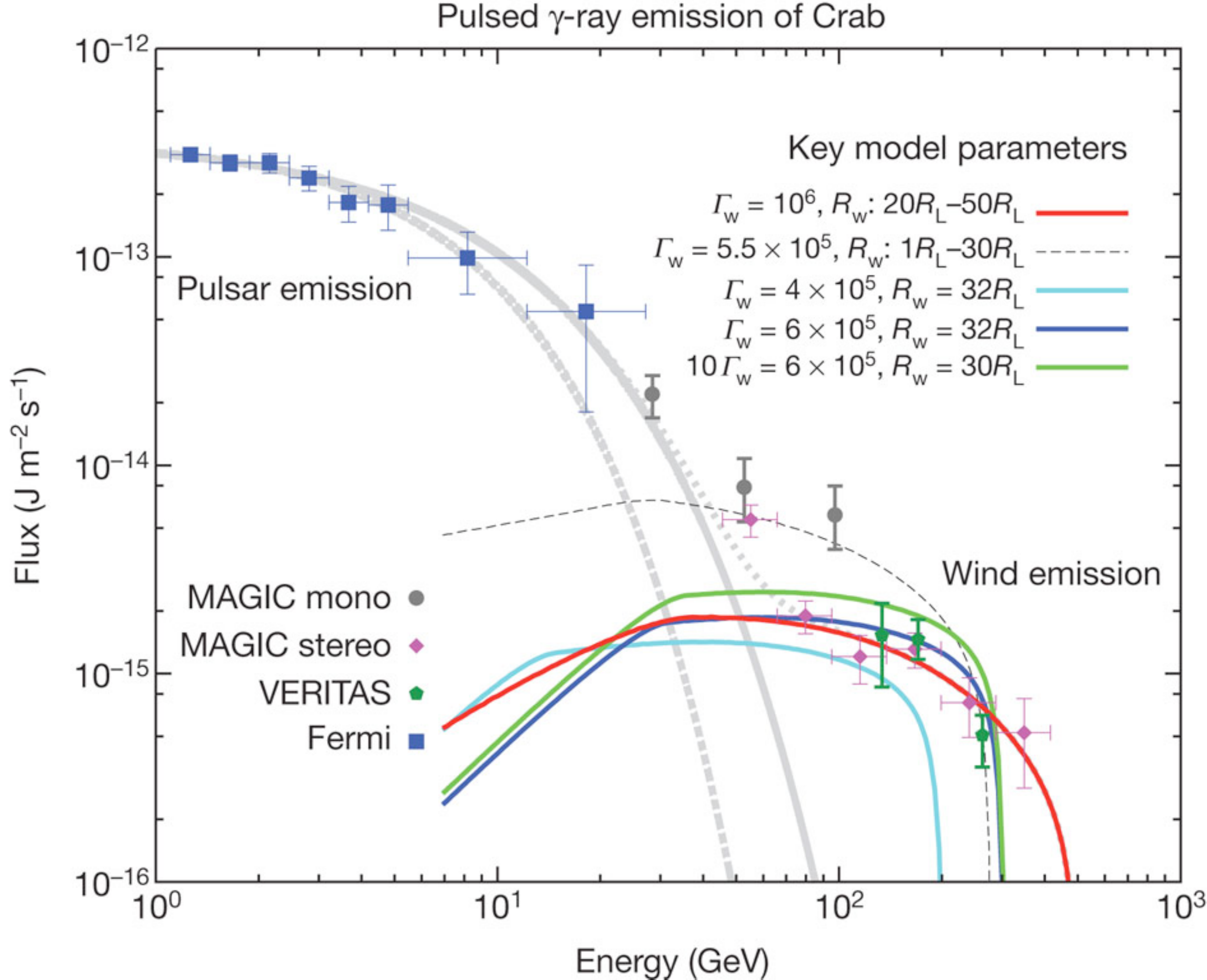, width=3.3in}\vspace{1mm}}
{\baselineskip 5.5pt\renewcommand{\baselinestretch}{0.05}\footnotesize \noindent
{\bf Fig.~4}\quad Spectral energy distribution (SED) of the pulsed gamma-ray emission from the direction of the 
Crab pulsar and nebula. {\it Fermi}-LAT points are shown (blue squares) together with MAGIC (grey and pink) 
and VERITAS (green) points. A {\it Fermi}-LAT-points best-fit, using two different hypotheses ($b=1$, $E_0=5.8$
GeV and $\Gamma=1.97$ and $b=0.85$, $E_0=7$~GeV and $\Gamma=1.97$), is displayed in grey. The pulsed 
VHE radiation can be successfully accounted for (light blue, blue, green and red curves) by inverse Compton 
up-scattering of the pulsed magnetospheric X-ray emission by a cold ultra-relativistic pulsar wind (see Sec. 3.2). 
From Ref.~[60].}
\vspace{3mm}

\noindent Yet unexpectedly, pulsed $\gamma$-ray emission above 100 GeV and up to 400 GeV of unknown origin 
was recently detected from the Crab with the VERITAS and MAGIC telescopes [57,58], cf. Fig.~4, challenging 
models for the origin of the periodic emission in neutron stars. Different explanations could be pursued to 
accommodate these new experimental findings within current models, such as secondary emission of electrons 
in the outer magnetosphere [59] or IC emission from energetic electrons in the ultra-relativistic pulsar wind [60] (cf. 
also Sec. 3.2). Approaches like these predict different spectral shapes and light-curve behaviour at GeV and TeV
energies. The detected phase-averaged, pulsed emission (Fig. 4) could in principle be fitted by extrapolating the
reported Fermi fluxes to the VHE domain as a power law with photon index of $3.8\pm0.5$ and a flux of 1\% of the 
flux of the Nebula at 150\,GeV, but the nature of such an extrapolation seems rather difficult to justify on physical 
(magnetospheric) grounds [60]. The VHE light curve shows a double peak structure well-aligned with the light 
curve at lower energies, although narrower by a factor of two or three than those measured by {\it Fermi}-LAT. The 
spectrum of the narrow peaks, extending no more that 10$\%$ of the rotational period, does not show a significant 
deviation in its shape from the global spectral fit. Assuming a common (magnetospheric?) origin, a smooth 
connection of the VHE points with the HE points can be achieved by fitting the data with a broken power-law 
function, but to the exclusion of an exponential cut-off. An alternative explanation consists in considering the entire 
gamma-ray region as a superposition of two separate components, a nominal (magnetospheric) GeV one and an 
additional VHE component produced by IC up-scattering of the magnetospheric emission by the fast pulsar wind 
[60]. Measuring the spectral shape with high precision in the near future will provide constrains on these models 
and allow to investigate the connection with the low-energy points around 50 GeV and the spectral extension 
above 400 GeV. Up to now, the observed $\gamma$-ray features make the Crab a unique source of this kind at 
VHE. An increase of the sample by observing the brightest {\it Fermi}-LAT pulsars, such Vela or Geminga will be 
pursued by H.E.S.S. II, MAGIC II and VERITAS (and CTA in the future), providing more input to understand the 
origin of this pulsed VHE radiation [61]. 

\vspace*{5mm} \noindent {2.3\quad Pulsar Wind Nebulae}\vspace{3.5mm}

Relativistic winds from energetic pulsars carry most of the rotational power into the surrounding medium, accelerating 
particles to high energies, either during their expansion or at the shocks produced in collisions of the winds with the 
sub-sonic environment. Accelerated leptons can interact with magnetic fields and low-energy radiations fields of 
synchrotron, thermal or microwave-background origins. As a result, non-thermal radiation is produced from the lowest 
possible energies up to $\simeq$100 TeV. For magnetic fields of few $\mu$G, freshly injected electrons (and positrons) 
create a synchrotron nebula around the pulsar, ranging from the radio to the X-ray and, in some cases, to the MeV band. 
At high energies a second component appears as a result of Comptonization of these soft photon fields by the relativistic 
leptons, creating an extended IC-nebula around the pulsar [56]. 

\vspace{1mm}
\centerline{\psfig{figure=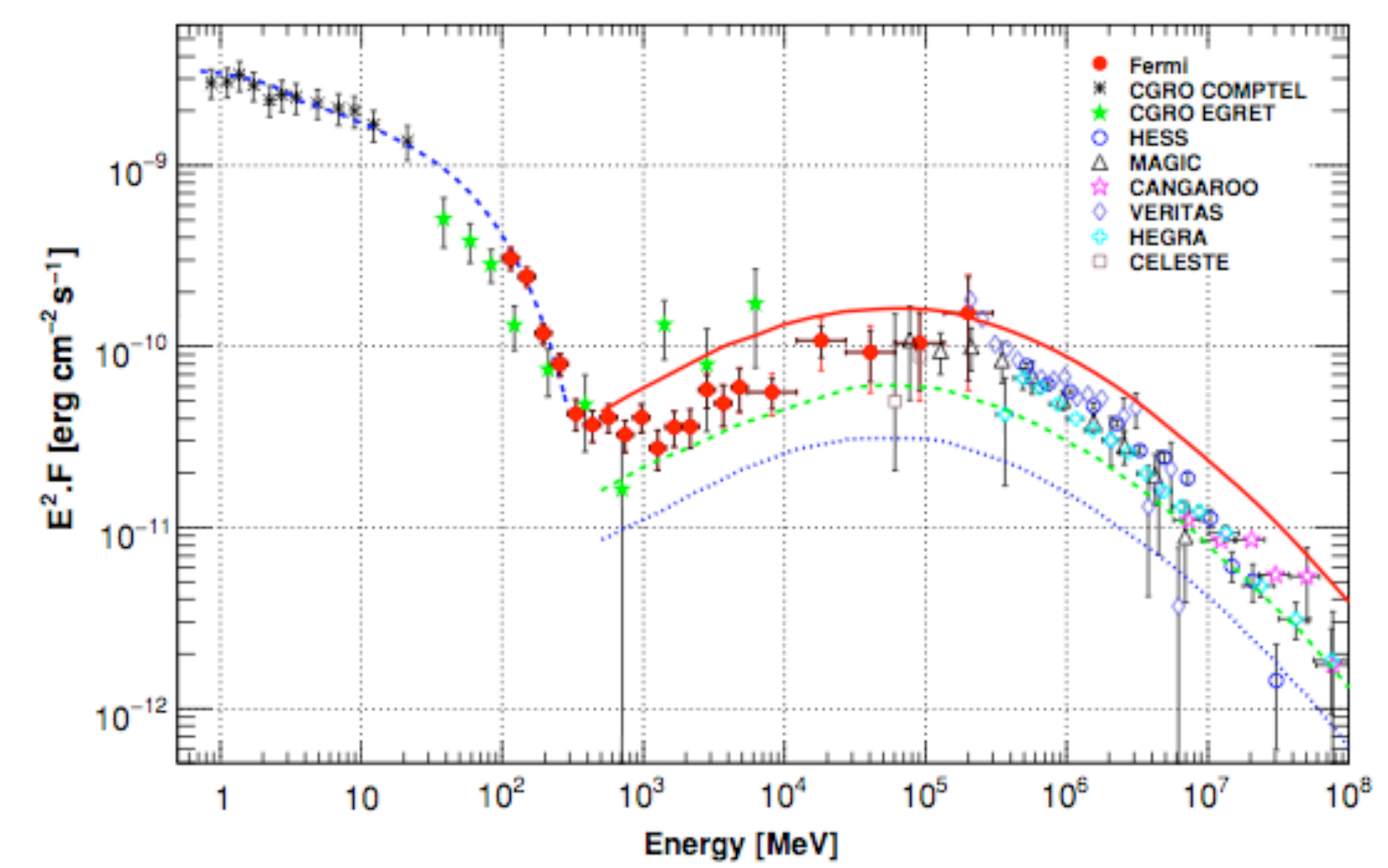, width=3.5in}\vspace{1mm}}
{\baselineskip 5.5pt\renewcommand{\baselinestretch}{0.05}\footnotesize \noindent
{\bf Fig.~5} \quad Spectral energy distribution (SED) of the Crab Nebula in the high- and very high energy gamma-ray
domain. The spectral points from low to VHE gamma-rays are shown together with a fit of the synchrotron component 
(blue dashed line) and predictions for IC gamma-rays calculated for three different values of the mean magnetic field: 
$B=100 \ \mu$G (solid red line), $B=200 \ \mu$G, and the equipartition field of the nebula of $300 \ \mu$G. From 
Ref.~[55], reproduced by permission of the AAS.}
\vspace{3mm}

\noindent VHE observations of these pulsar wind nebulae (PWNe) have revealed PWNe to be the most effective Galactic
objects for the production of VHE gamma-rays, allowing the detection of such systems even outside our own Galaxy (in 
the LMC [63]). As recently as of 2004, only the Crab PWN was detected with a steady gamma-ray flux above 1\,TeV of 
(2.1$\pm$0.1$_{\rm stat}$)$\times$10$^{\rm -11}$cm$^{\rm -2}$s$^{-1}$ [64,65]. The development of the new sensitive 
IACTs in the last years has raised the number of likely PWNe detected to at least 27 sources, whereas many of the 
unidentified gamma-ray sources are widely believed to be PWNe (or old relic PWNe) [23].

\noindent For many years, the Crab nebula was considered as a standard candle for the cross-calibration of VHE 
detectors, as the brightest persistent point-like TeV gamma-ray source seen effectively from both hemispheres. The 
main features of its non-thermal emission, extending over 21 decades of frequencies, has been satisfactorily described 
by the formation of a PWNe based, to a large extent, on a simple MHD model for the interaction of a cold ultra-relativistic 
electron-positron wind with the interstellar medium [66]. Recent detailed two-dimensional MHD simulations [67,68] have
confirmed such a concept, at least for the Crab Nebula. The IC emission detected at TeV provides crucial information 
about the conditions in the nebula even when it only constitutes a small fraction of the synchrotron luminosity of the 
nebula. In particular, a comparison of the X-ray and TeV gamma-ray fluxes observed from the Crab Nebula has lead to 
a robust estimate of the average nebular magnetic field of less than $100 \ \mu$G, in good agreement 
with predictions for the termination of the wind in MHD theory [66]. Figure 5 shows the high-energy coverage of the Crab 
Nebula spectrum. While the COMPTEL and EGRET data carry information about the synchrotron radiation in the cut-off 
region, the {\it Fermi}-LAT data reveal the sharp transition from the synchrotron to the IC component at around 1~GeV. 
At an energy E$\simeq$100\,GeV, a clear indication of the IC maximum is supported by both satellite ({\it Fermi}-LAT [55], 
and ground-based (MAGIC [69]  and VERITAS [58]) measurements, which show remarkable agreement with each other. 
The measurements with ground-based IACTs have almost approached 100~TeV [64,70,71], where the IC component 
should still extend to the energy region set by the maximum energy of the accelerated electrons,  i.e., 1 PeV. Although 
the production of gamma-rays at such energies takes place in the Klein-Nishina regime, and is therefore strongly 
suppressed, future instrument such CTA should be able to detect this emission.\\ 
Yet, despite the large coverage and deep observations, many aspects of this unique source are still unresolved. For 
instance, rapid high-energy flares with rise time as short as 6 hours from the Crab PWN have been reported by the 
{\it Fermi}-LAT and the AGILE collaboration [54,72]. This amazing discovery has opened new questions such as how 
these flares connect with the pulsar energy release or as to their origin (are they related to the inner pulsar wind or 
to the magnetosphere?, see e.g., [73-77]). 
The exceptionally high fluxes during the active state in April 2011 allow detailed spectroscopy for different flux levels 
[54]. In order to study the spectral evolution of the flaring component, a steady-state (constant) background has been 
assumed with a steep power-law spectrum described by a photon index $\Gamma_{\rm b}=3.9$. The spectrum of the 
flaring component has been assumed in the form of power-law with exponential cutoff, $\nu F_\nu=f_0 
E^{2-\Gamma_{\rm f}}\exp[{-(E/E_0)^\kappa}]$. The results show that the spectra during all selected windows can be 
well described by the same photon index $\Gamma_{\rm f}=1.27 \pm 0.12$ and exponential cutoff index  $\kappa=1$, 
but with variable total flux $f_0$ and the cut-off energy $E_0$. A variation by a factor of two allows a good fitting of the 
data, but the total flux has to be changed more than an order of magnitude in this approach. While different theories 
(including synchrotron radiation and reconnection) have been put forward to explain these flares, many key issues 
are still unresolved.

\noindent Even as one of the strongest sources in the TeV sky, the Crab nebula is very inefficient in producing 
gamma-rays through IC scattering, and only its extremely high spin-down power compensates for this.The energy 
density of the magnetic field (of the order of $\sim100\,\mu$G) exceeds by more than two orders of magnitude 
the radiation energy density. Thus, less than one per cent of the energy of the accelerated electrons is released in IC 
gamma-rays, the rest being emitted through synchrotron radiation. In other systems, the pulsar wind is not as powerful 
as the one in Crab, resulting in weaker magnetic fields in the nebula of the order of a few $\mu$G. This low magnetic 
field translates into a more efficient emission via IC at VHE due to the sharing of the electron energy losses between 
synchrotron and IC mechanism. For instance, in the case of the cosmic microwave radiation (CMB), the two radiation 
components are related through $L_\gamma/L_{\rm X}=w_{\rm CMB}/w_{\rm B} \simeq 1 \, (B/3 \mu \rm G)^{-2}$. This 
implies that in a PWN with a nebular magnetic field of about $10 \ \mu$G or less, the IC gamma-ray production efficiency 
could be as large as 10\%. Given that the rotational energy of pulsars is eventually released in relativistic electrons 
accelerated at the termination shock, PWNe associated with young pulsars with spin-down luminosities $L_0 \geq 
10^{34} (d/1 {\rm kpc})^2 \ \rm erg/s$ were expected to be detected [78]. 
These expectations have been confirmed by the results obtained with MAGIC and VERITAS, but overall by the survey 
performed with H.E.S.S. The Galactic plane survey (GPS) as seen by H.E.S.S. in Fall 2012 is shown in Fig. 6. The 
survey, covering a range between [-85$^{\rm o}$, 60$^{\rm o}$] in longitude and [-2.5$^{\rm o}$, 2.5$^{\rm o}$] in 
latitude, has revealed more than fifty new VHE $\gamma$-ray sources, out of which more than half are believed to be 
gamma-ray PWNe, located in the close vicinity of young and energetic pulsars.

\vspace{3mm}
\centerline{\psfig{figure=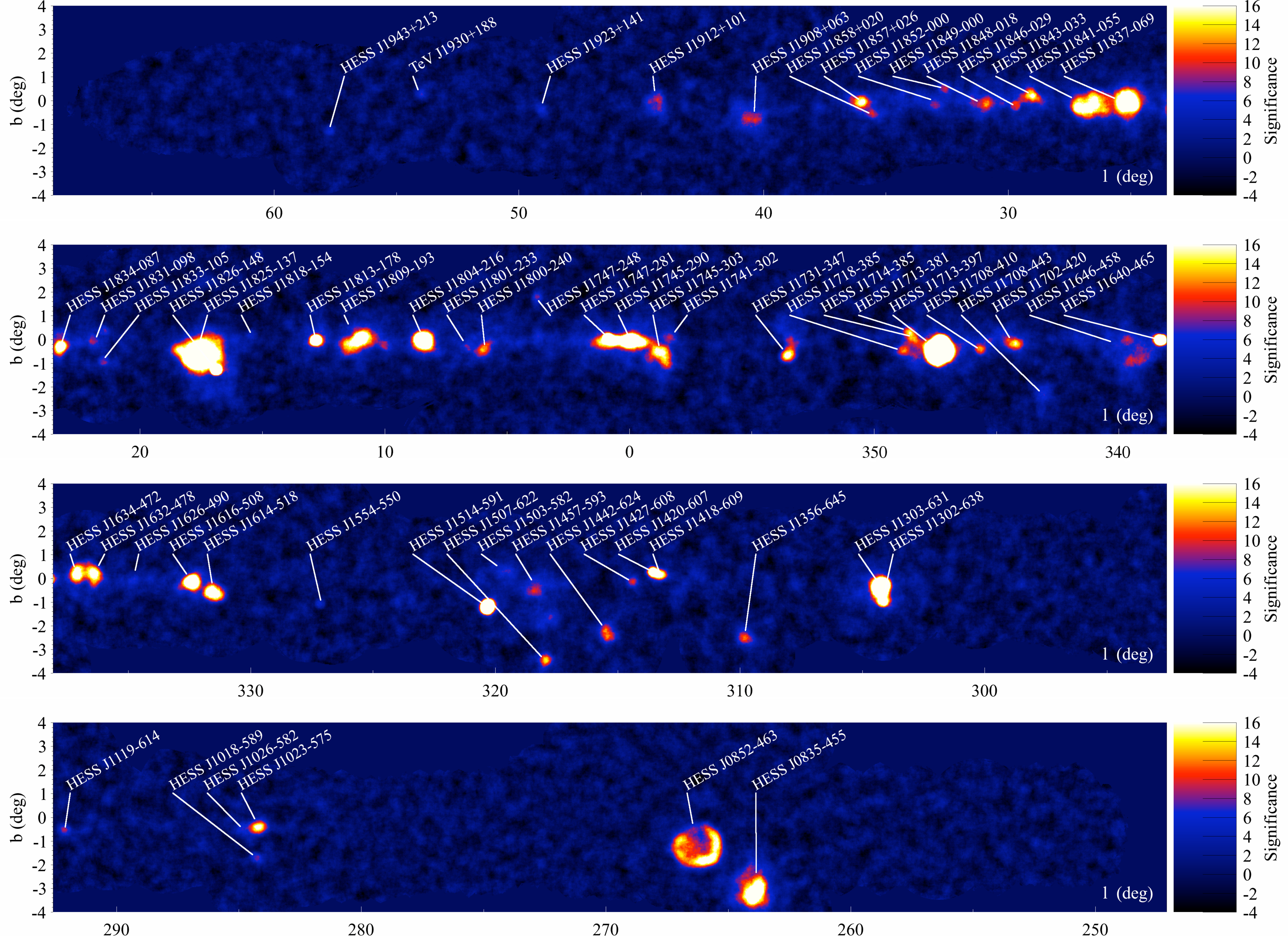, width=3.5in}\vspace{1mm}}
{\baselineskip 5.5pt\renewcommand{\baselinestretch}{0.05}\footnotesize \noindent
{\bf Fig.~6}\quad Significance (pre-trial) map of the Galactic plane survey by H.E.S.S. From Ref. [296].}
\vspace{3mm}

\noindent Presently PWNe constitute the largest galactic TeV source population. Many previously dubbed  "dark"  TeV 
gamma-ray sources, including the first unidentified TeV gamma-ray source discovered by the HEGRA collaboration, 
TeV\,J2032+4130 [79], have later been identified as PWNe. Most of these identifications with PWNe are quite convincing, 
yet still tentative, except for several ones which are firmly identified, either by excellent radio/X-ray morphological 
correlations, such as the Kookaburra complex, MSH 15-52 and Vela\,X [80,81], or by observations of an energy-dependent 
morphology, tracing the cooling mechanisms in the leptonic population injected by the pulsar (as observed in
 HESS\,J1825--137 or HESS\,J1303--631 [82,83], cf. Fig. 7). 

\vspace{4mm}
\centerline{\psfig{figure=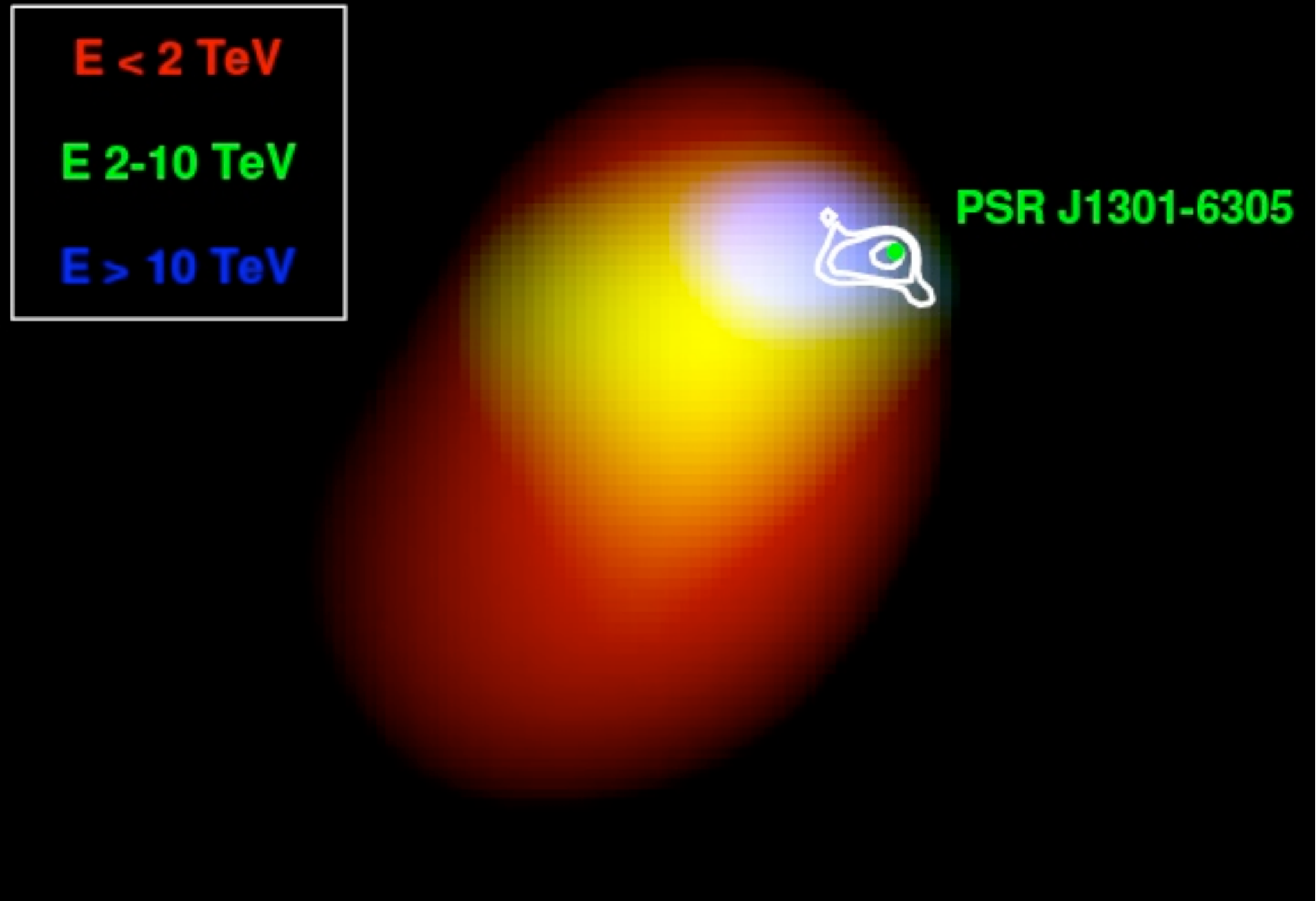, width=3.1in}\vspace{1mm}}
{\baselineskip 5.5pt\renewcommand{\baselinestretch}{0.05}\footnotesize \noindent
{\bf Fig.~7}\quad VHE image of the TeV pulsar wind nebula candidate HESS 1303-631 at different energy ranges. 
The highest-energy photons originate near to the pulsar. X-ray (XMM) contours are shown in white. See Ref.~[297]}
\vspace{3mm}

Out of the PWNe detected at VHE two different populations of PWNe seem to be emerging: PWNe associated to young, 
compact X-ray PWNe, often still embedded in their associated supernova remnant; and evolved (extended and resolved) 
sources, in which the TeV emission seems to be due to a "relic" population of electrons, whereas the associated shell 
has already faded away. In the latter group, the centre of gravity of the extended TeV images is often offset with respect to
the position of the powering pulsar. Asymmetric, one-sided images of these PWNe have also been found in X-rays, but on 
significantly smaller scales. Although the mechanism which causes PWN offsets from the pulsar positions is not yet firmly 
established, this effect could be linked to the propagation of a reverse shock created at the termination of the pulsar wind 
in a highly inhomogeneous medium [62]. The significantly larger extension of the TeV emission region can be understood 
as a result of several factors: 
(i) Generally, for PWNe with magnetic field of order of $10 \ \mu$G or less, as apparently the case for most TeV PWNe, 
the electrons responsible for the X-ray emission are more energetic than the electrons emitting TeV gamma-rays. 
Therefore, synchrotron-burning of the highest-energy electrons results in a smaller size of the X-ray source.  
(ii) When electrons diffuse beyond the PWN boundary, they emit less synchrotron radiation (due to the reduced magnetic 
field), but they can still effectively radiate gamma-rays via inverse Compton scattering of the universal CMB. 
(iii) Finally, because of the high X-ray background, the sensitivities of X-ray detectors like Chandra and XMM-Newton are 
dramatically reduced beyond several angular minutes. This significantly limits the potential of these instruments for weak, 
extended X-ray sources. In contrast, the sensitivity of IACT arrays remains almost unchanged approximately within a 
1$^{\rm o}$ radius of  field-of-view. This flat response makes IACT technique the most powerful tool for studying the 
non-thermal population of electrons in PWNe. 

\vspace{2mm}
\centerline{\psfig{figure=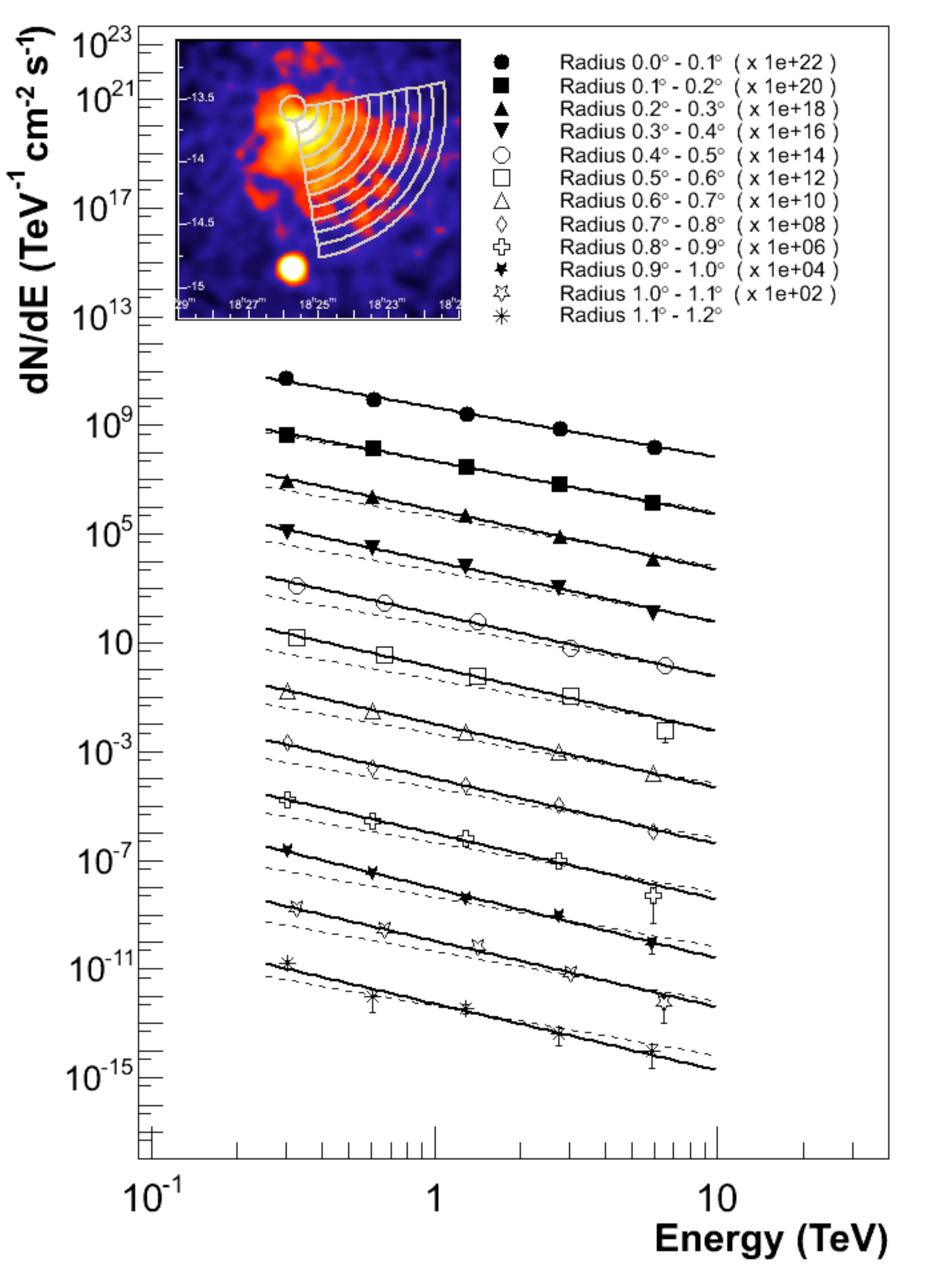, width=3.3in}\vspace{1mm}}
{\baselineskip 5.5pt\renewcommand{\baselinestretch}{0.05}\footnotesize \noindent
{\bf Fig.~8}\quad Energy-dependent VHE morphology of pulsar wind nebula HESS J1825-137, 
showing a softening of the spectra with increasing distance from the pulsar. The plot shows the 
energy spectra in radial bins as indicated in the inset (with the dashed line from the innermost 
region for comparison). From Ref. [82]}
\vspace{3mm}

The asymmetry observed in those PWNe has been explained as a consequence of the propagation of the precursor 
supernova explosion in the inhomogeneous interstellar medium [84], resulting in a faster evolution of the associated 
PWN in the opposite direction of the denser environment or/and a high kick-off velocity of the pulsar, displacing it from 
the centre of the supernova explosion.

\noindent The accumulation of particles with time, the continuous injection and the ubiquitous presence of a soft photon 
target (CMB) make these objects extremely efficient in the production of VHE emission. The high flux and extension of 
these TeV PWNe have permitted the investigation of the spectral behaviour with good statistics in different regions of the 
nebula, unveiling a softening of the gamma-ray spectral index as a function of the distance from the pulsar (see Fig. 8). 
This effect is due to the radiation of uncooled electrons which quickly leave the compact region near the pulsar, suffering
significant radiative losses as they propagate away. This seems also to be the case for Vela~X, a nearby PWN related 
to the powerful pulsar PSR~J0835-4510 ($\tau \approx$11,000~yr, $L_0=7 \times 10^{36} \ \rm erg/s$). Vela~X has been 
established [81] as one of the strongest TeV gamma-ray sources in the Galaxy. The energy spectrum of this source is 
quite different from other galactic sources; it is very hard at low energies, with photon index $\Gamma \approx 1.5$, 
and contains a high-energy exponential cut-off resulting in a distinct maximum in the SED at 10~TeV. Because of the 
nearby location of the source ($d  \approx 300$~pc) we see, despite the large angular size of the gamma-ray image 
of order of 1 degree, only the central region with a linear size less than several pc. In this regard, Vela~X is a perfect  
object for the exploration of processes in the inner parts of the nebula close to the termination shock.  
The significantly improved sensitivity of the future CTA instrument and its superior angular resolution (one to two arc 
minutes at 10 TeV) should allow a unique probe of the relativistic electrons inside the region of the termination shock, 
i.e., at the very heart of the accelerator. 

\noindent Along with these evolved nebula, a large number of compact objects have also been identified recently (see, 
e.g. [85,86]), in which the PWN is still expanding within the shell. A text-book example is the composite SNR\,G327.1--1.1 
(HESS\,J1554--550) [87], in which the detected TeV emission is spatially coincident with the X-ray and radio PWN, well 
inside the remnant. A similar case is the newly detected source HESS\,J1818--154 [88], embedded in the SNR G15.4+0.1. 
The latter was discovered after a long exposure of 145 h with a flux of 1.5\% of the Crab Nebula flux, and no 
X-ray or radio PWN has been detected yet, allowing SNR G15.4+0.1 to be identified as a composite SNR by means of 
VHE observations only. Those objects display a very low magnetic field in comparison to the Crab Nebula of the order 
of a few $\mu$G, compensating so the lower spin-down power luminosity with a particle-dominated wind, which allows 
an enhancement of the inverse-Compton emission at very high energy.

\vspace*{5mm} \noindent {2.4\quad TeV Binary Systems}\vspace{3.5mm}

The number of TeV binary systems - sources emitting variable, modulated VHE emission composed of a massive star 
and a compact object - has increased steadily in the last years, thanks to the large time coverage and the deep and 
uniform exposure of the Galactic plane by MAGIC, VERITAS and H.E.S.S. The TeV emission is believed to arise from 
the interactions between the two objects, either in an accretion-powered jet ({\it{microquasar}} scenario), or in the
shock between a pulsar wind and a stellar wind ({\it{wind-wind}} scenario) (see e.g. [89-93], cf. also Sec. 3.2). In 
the {\it{microquasar}} scenario, particle acceleration takes place in a jet which originates from an accretion disk. 
This scaled-down version of an active galactic nucleus opens the possibility to obtain significant insights into the 
mechanism of jet production. In the {\it{wind-wind}} scenario, on the other hand, particle acceleration occurs in the 
interaction region between a ultra-relativistic pulsar wind and the dense radiation field provided by the companion star. 
Likewise, X-rays and high-energy components are expected due to radiative (synchrotron and inverse-Compton) cooling 
of relativistic electrons accelerated at the termination shock [94,95].

\noindent Four periodic binary systems have been firmly identified at VHE (PSR\,B1259--63 [96], HESS\,J0632+057 [97,98], 
LS\,5039 [99] and LSI\,+61\,303 [100-102]), whereas two more sources (HESS\,J1018-589 [103] and Cyg\,X-1 [104]) are 
less certain and still pending confirmation. The observed variability implies a compact emission region which translates 
into a point-like source morphology at a distance of 1 to 5 kpc. Indeed, the majority of point-like sources detected in the 
H.E.S.S. Galactic Survey have been identify as TeV binary systems. This univocal identification is based on the observed
VHE variability/periodicity and correlations with flux variation at other wavelengths. They exhibit a maximum flux of $\sim 
5-15$\% of the Crab Nebula flux and apparent similar spectral indices (2.0 to 2.7), but the enlargement of the TeV (and 
GeV) binary sample has indicated a very diverse behaviour from one system to the other, demanding a detailed 
source-to-source investigation.\\ 
The first TeV binary established was the pulsar-B2Ve star system associated to PSR\,B1259--63 (or LS\,2833) in 2004, 
which was anticipated before its detection [95]. In this system, a 48 ms pulsar is moving around a massive Be star, 
crossing its disk every 3.4 years, on a highly eccentric (e=0.87) orbit. The observations show a complex light curve, and 
the VHE emission can be satisfactorily explained in a pulsar-wind stellar-wind scenario, although the different year-to-year 
observations still challenge current models. Moreover, the source exhibited a large post-periastron orphan flare at GeV 
energy that was not observed in the TeV range [105,106,290], which lasted approximately two weeks with an enhanced 
flux above 100 MeV at the level of $3\times10^{-10}$ erg cm$^{-2}$s$^{-1}$. Several scenarios have been proposed to 
account for this phenomenon, involving energy-dependent absorption processes and/or Comptonization of the photon 
field provided by the star by the cold ultra-relativistic pulsar wind [107].\\ 
The second, very-long-period ($\sim$320 days)-system was discovery serendipitously in the H.E.S.S. survey, being one 
of the very few point-like ($<$2$^{\prime}$) unidentified sources. HESS\,J0632+057 was finally identified in a joint 
H.E.S.S. and VERITAS campaign, followed up by MAGIC observations, through long-term X-ray observations with SWIFT, 
that succeeded in confirming its nature. Strong evidence for periodic X-ray variability with a very long period of 321$\pm$5 
days has been reported [108], implying the discovery of a binary for the first time on the basis of TeV observations (see 
also Fig. 9).

\vspace{0mm}
\centerline{\psfig{figure=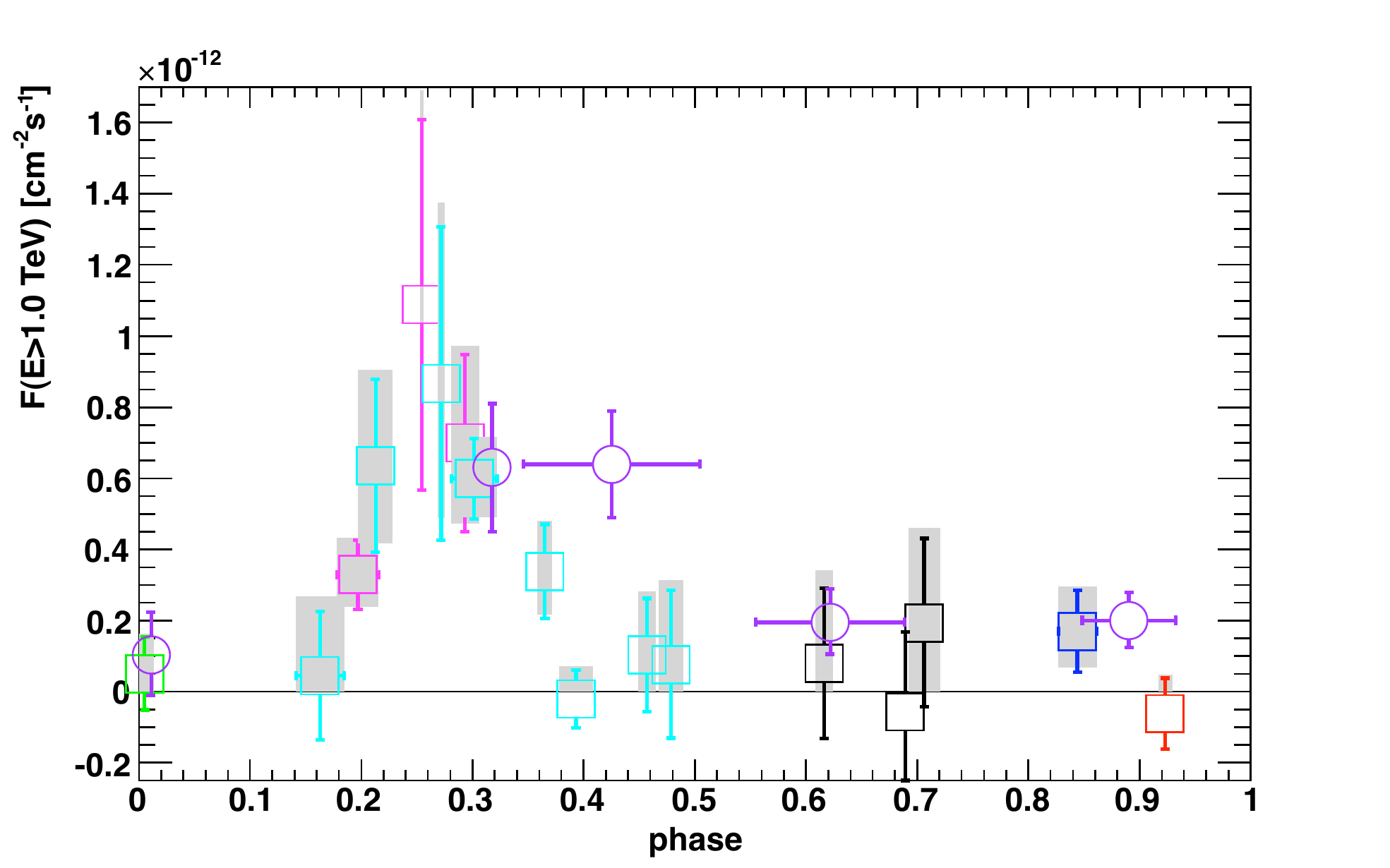, width=3.5in}\vspace{1mm}}
{\baselineskip 10.5pt\renewcommand{\baselinestretch}{1.05}\footnotesize \noindent
{\bf Fig.~9} \quad VHE observations of the binary system HESS J0632+057 folded with a period of 321 days. 
The H.E.S.S. (circular markers) and VERITAS (open squares) measurements are shown in different colours for 
different observational periods. From Ref. [97].}
\vspace{2mm}

The last two mentioned VHE binary systems, LSI +61\,303 and LS\,5039, show short-periodic 
orbital variability, of the order of days, allowing a larger integration of VHE data and deeper investigation of their 
light curve. However, they behave quite differently from each other. While LS\,5039 (P$\sim$3.9 days) exhibits 
are very regular light curve, LSI+61\,303, with a period of $\sim$26.5 days, shows a quite erratic behaviour, likely 
related with a 1667 super-orbital variability [109]. The nature of the compact object for both system is unknown: 
It could be anything from a 1.4 M$_{\circ}$ neutron star to a (3.7)4 M$_{\circ}$ black hole. No pulsation has been 
found in radio or X-ray searches. It seems likely, however, that any pulsed radiation would be absorbed in the 
optical-thick dense ambient due to Compton scattering [22]. 

\begin{table*}
\vspace{1mm}
\centerline{\psfig{figure=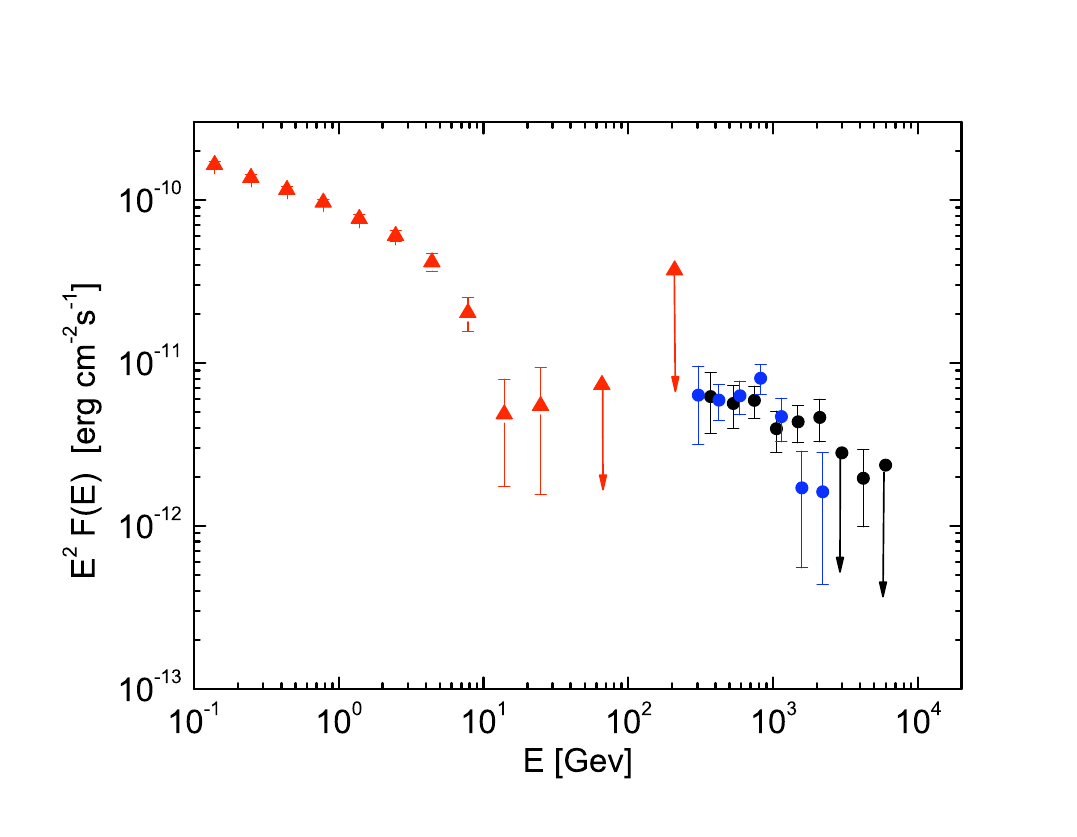, width=3.3in}\vspace{1mm}
\psfig{figure=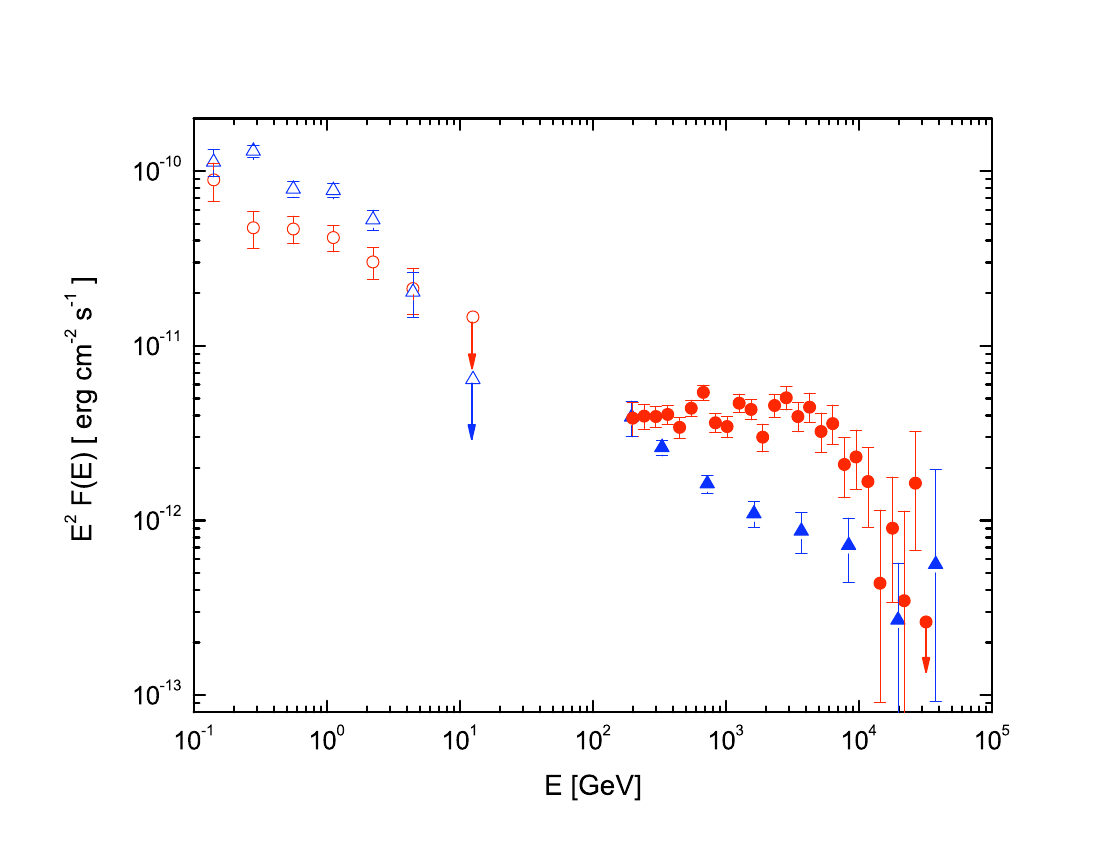, width=3.3in}\vspace{1mm}}
{\baselineskip 10.5pt\renewcommand{\baselinestretch}{1.05}\footnotesize \noindent
{\bf Fig.~10}\quad High-energy ({\it Fermi}-LAT) and VHE (MAGIC, VERITAS and H.E.S.S.) observations of 
LSI +61\,303 (left) and LS\,5039 (right), cf. Refs.~[298,299]. The right figure shows the spectral data at inferior 
conjunction in red circles whereas the observations in the superior conjunction are shown in blue triangles.}
\vspace{2mm}
\end{table*}

These two binary systems have also been detected with the Fermi-LAT telescope above 100 MeV. The spectrum 
of LS\,5039 shows a clear hardening in the 0.3 to 20 TeV region (see Fig. 10), while the GeV component shows 
a softening in inferior conjunction. On the other hand, at superior conjunction an opposite behaviour is observed. 
LSI +61\,303 on the contrary, does not show variation of the spectral index, but its emission vanished after October 
2008, reappearing again in 2010, accompanied by a change in the high-energy flux with decrease of the orbital 
modulation in 2009 [111-113]. From the multi-wavelength data it is clear that more sophisticated scenarios are 
needed to understand the acceleration and emission processes involved in these two sources.

Finally, two more VHE regions have been associated with binary systems: MAGIC has reported a 4$\sigma$ 
evidence for VHE emission from the direction of the Microquasar Cyg X-1 [104], correlated with an increase in 
soft and hard X-rays, but this was not confirmed during later, similarly high X-ray flux flaring events; and the 
GeV 16.5 days binary system 1FGL\,J1018.6--5856 [114], coincident with the H.E.S.S. source HESS\,J1018--589. 
For the latter, no VHE variability has been discovered yet, making the association somewhat unclear. Deep 
observations and uniform exposure in time with H.E.S.S. will help to clarify the origin of its VHE emission.

\vspace*{5mm} \noindent {2.5\quad Galactic Centre}\vspace{3.5mm}

The Galactic Centre (GC) harbours many remarkable objects, including a few potential sites for particle acceleration 
and gamma-ray production, in particular the compact radio source Sgr~A*, a suspected super-massive black hole 
located at the dynamical centre of the Galaxy.\\  
The GC contains a strong gamma-ray source (cf. Figs. 11 and 12) with a broad-band spectrum that spans from 
100 MeV [115] to 30 TeV [116]. Assuming that gamma-rays from the entire interval are linked to the same source, 
the spectrum has an interesting form with several distinct features: Hard at low energies, with a photon index 
$\Gamma \approx 2.2$, it becomes significantly steeper by $\Delta \Gamma \approx 0.5$ above 2~GeV [115], 
then hardens again at TeV energies with a photon index $\Gamma \simeq 2.1$ and an apparent break or cut-off 
above 10 TeV (see Fig. 12). 

\vspace{3mm}
\centerline{\psfig{figure=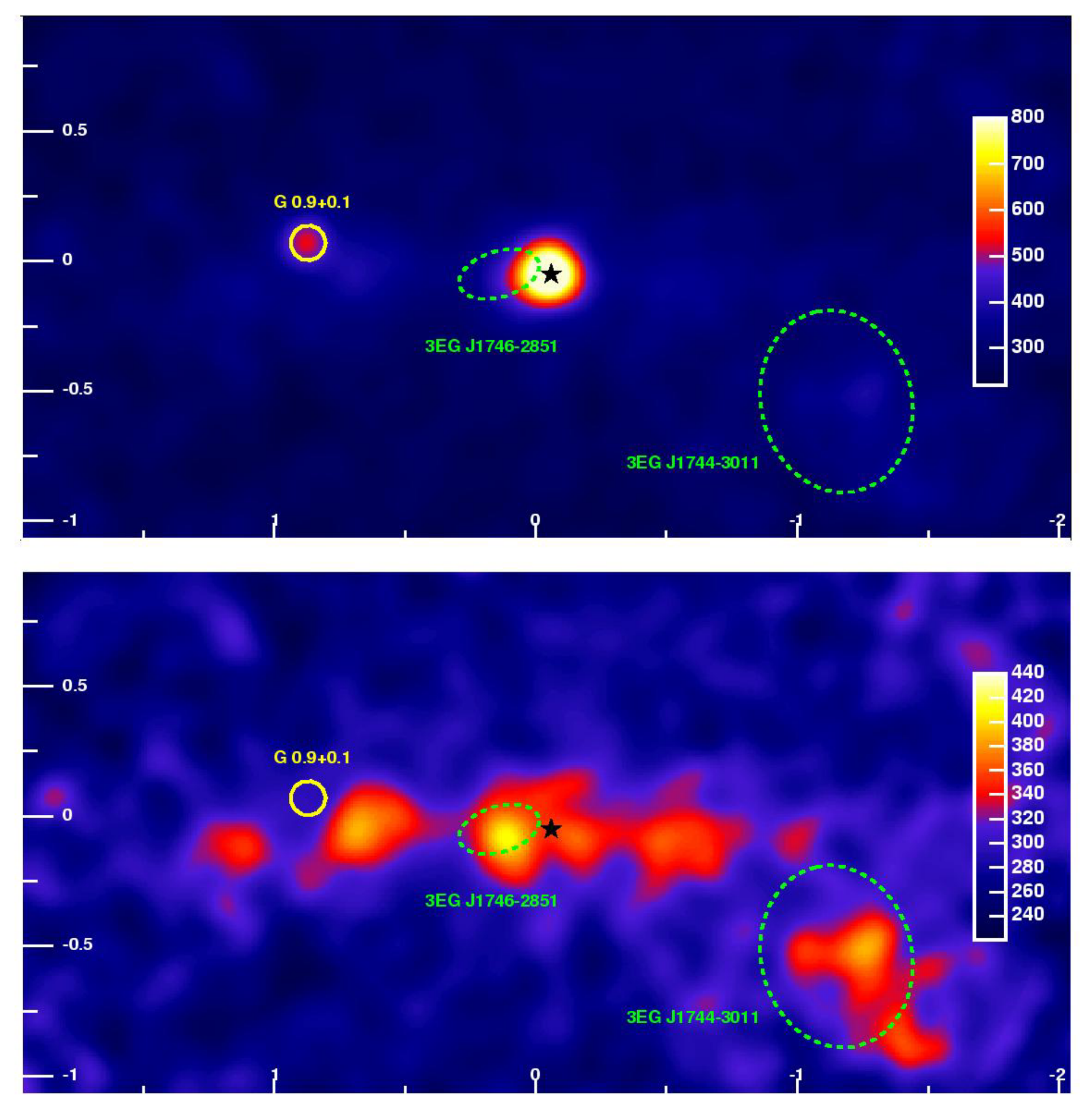, width=8cm}\vspace{1mm}}
{\baselineskip 10.5pt\renewcommand{\baselinestretch}{1.05}\footnotesize \noindent
{\bf Fig.~11}\quad The image of the several-hundred parsec region of the Galactic Centre in TeV gamma-rays 
(top: $\gamma$-ray count map; bottom: same map after subtraction of the two point sources). It contains a point 
like source (angular radius less than a few arc-minutes), the gravity centre of which coincides with an accuracy 
of 13 arc-seconds with the compact  radio source Sgr A* (marked with black star) - a supermassive black hole at 
the dynamical centre of the Milky Way [120,121]. The second point-like source located about one degree away 
positionally coincides with the composite supernova remnants G09+0.1 [85]. A prominent feature of this region 
is the ridge of diffuse emission tracing several well-identified giant molecular clouds (lower panel; cf. Ref.~[122] 
for more details). This complex region contains some other, not yet firmly identified, "hot spots''.}
\vspace{2mm}

\noindent Although the gamma-ray source spatially coincides with the position of Sgr~A*  (see Fig. 11), the upper limit 
on the angular size of the TeV source of a few arc minutes is still too large to exclude the link to other potential sources 
located within the central $\leq 10$~pc region. The detection of variability of the gamma-ray flux would greatly 
contribute to the localisation of the gamma-ray production region in Sgr~A*. However, unlike the observations at radio 
and X-ray wavelengths, no variability has been observed both at GeV and TeV energies. This disfavours, but still cannot 
discard Sgr~A* as a possible gamma-ray source, especially given that several radiation mechanism, associated with 
the accretion flow, are capable of explaining the reported gamma-ray fluxes [117].\\
Perhaps a more plausible site of gamma-ray production could be the central, dense extended region of radius of 
10~pc. However, even in this scenario Sgr A* remains a potential source indirectly responsible for the gamma-ray signal 
through interactions of runaway particles accelerated in Sgr A*, but later injected into the surrounding dense gas 
environment [118,119]. The analysis of the combined {\it Fermi}-LAT and H.E.S.S. data show that  the complex shape 
of the GeV-TeV radiation can be indeed naturally explained by the propagation effects of protons interacting with the 
dense gas within the central 10 pc region [115,119]. A good agreement between the data and calculations is shown 
in Fig.~12, where the radial profile of the gas density has been carefully taken into account. The flat spectra in the 
segments of the proton spectrum around  1 GeV,  and at TeV energies (below 10 TeV)  have different explanations. 
While at GeV energies the protons are diffusively trapped, so that they lose a large fraction of their energy before they 
leave the dense 3~pc region, at TeV energies they propagate rectilinearly. At intermediate energies the protons start 
to effectively leave the inner 3~pc-region, and the steepening of the energy spectrum can be naturally referred to the 
energy-dependent diffusion coefficient. What concerns the proton injection spectrum, it should be a hard power-law, 
close to $E^{-2}$, with an intrinsic cut-off around 100~TeV. The required total energy of protons currently trapped in 
the gamma-ray production region, $W_{\rm p} \simeq L_\gamma t_{\rm pp \rightarrow \gamma} \simeq 10^{49} 
(n/10^{-3} \rm cm^{3})^{-1} \ \rm erg$ is quite modest, given that the density in the circum-nuclear ring could be as 
large as $10^{5} \ \rm cm^{-3}$ [119].

\vspace{3mm}
\centerline{\psfig{figure=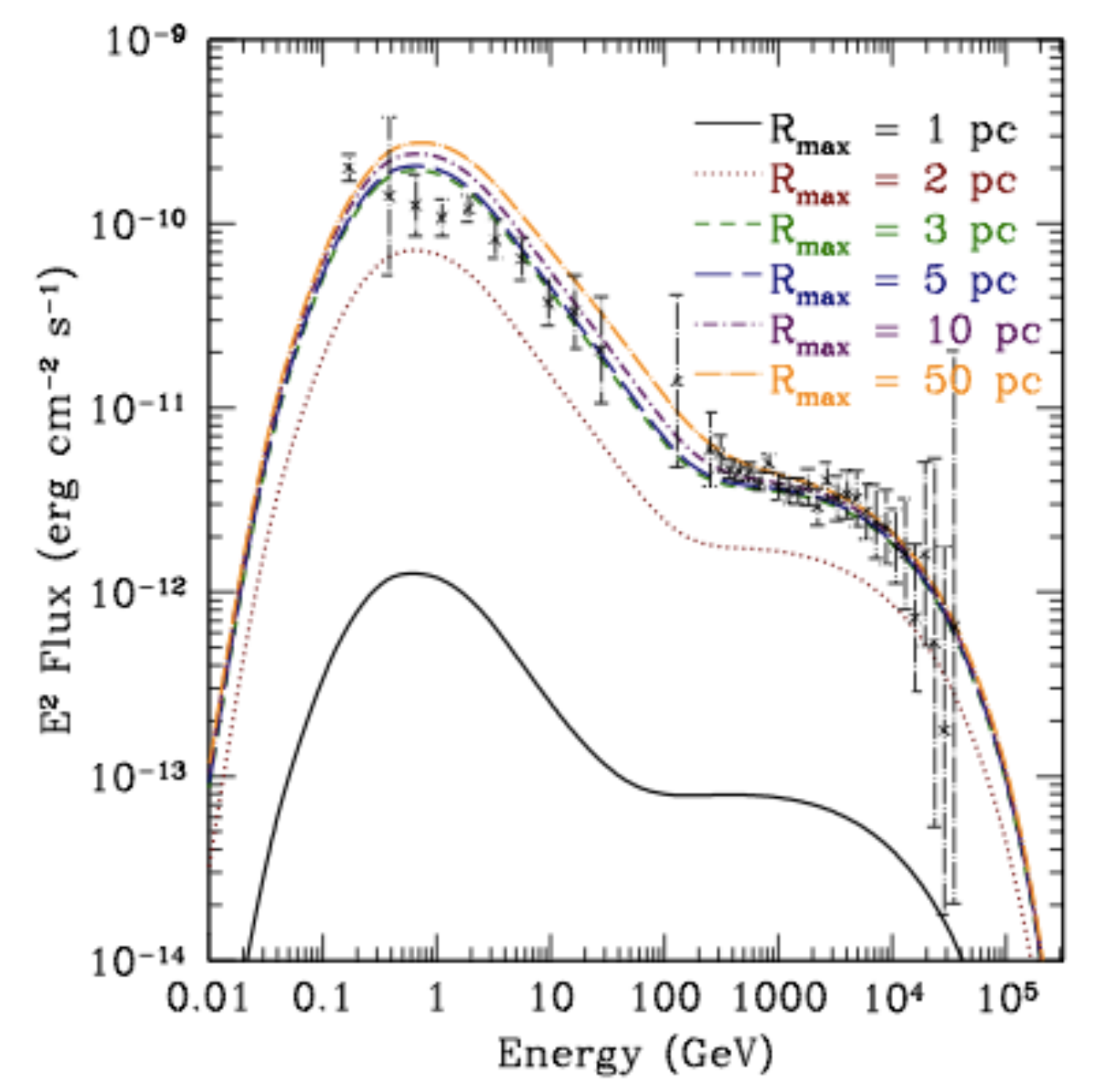, width=7cm}\vspace{1mm}}
{\baselineskip 10.5pt\renewcommand{\baselinestretch}{1.05}\footnotesize \noindent
{\bf Fig.~12}\quad Energy spectra of  gamma-ray emission from GC. The {\it Fermi}-LAT [115] and H.E.S.S. 
data [116] are shown together with calculations of $\gamma$-rays from {\it pp}-interactions within radial cones 
of various size up to  50~pc. The flux falls off rapidly after 3 pc because the main contribution comes from the 
1.2-3 pc circum-nuclear ring. From Ref. [119], reproduced by permission of the AAS.}\label{lindenFig2}
\vspace{2mm}

\noindent The interpretation of the spatially unresolved gamma-ray emission towards Sgr A* by interactions of 
runaway protons with the dense gas in the central (several pc) ring, predicts a smooth transition to another radiation 
component formed in more extended regions of the GC. The energy and spatial distributions of this radiation 
depend on the injection history of protons and the character of their diffusion. The H.E.S.S. observations of the 
so-called Central Molecular Zone (CMZ) of radius $\approx 200$pc indeed revealed an extended TeV gamma-ray 
emission [122] with a clear correlation with the most prominent giant molecular clouds located in CMZ  (see Fig. 11).  
Using the maps of TeV gamma-ray emission, and maps of the CS (J=1-0) emission which contain information 
about the column density in dense cores of molecular clouds, the cosmic-ray density in these clouds has been
derived. It appears to be significantly enhanced (by an order of magnitude at multi-TeV energies) relative to the 
local cosmic-ray flux in the solar neighbourhood. This indicates to a strong non-thermal activity accompanied with 
proton acceleration which in the past was perhaps higher than at the present epoch. An additional support for 
this hypothesis comes from the spatial distribution of gamma-rays. The H.E.S.S. observations show that the ratio of
gamma-ray flux to the molecular gas column density varies with galactic longitude, with a noticeable "deficit'' at 
$l\approx1.3^{\circ}$. This interesting feature can be interpreted as a non-uniform spatial distribution of cosmic 
rays, i.e.  the relativistic protons accelerated in Sgr A* have not yet had time to diffuse out to the periphery of the 
200~pc region. The epoch of the high activity of the accelerator depends on the proton diffusion coefficient. 
Assuming, for example, that the propagation of multi-TeV protons in the GC proceeds with a speed similar to the 
one in the Galactic Disk, the epoch of high activity of the accelerator and the total energy release in relativistic 
particles during the outburst are estimated to be $10^4$ yr  and $10^{50}$ erg, respectively [122]. 

\noindent High-energy processes that take place in the GC apparently play a key role in the formation of  two 
enormous gamma-ray structures recently discovered in the {\it Fermi}-LAT  data set  - the {\it Fermi bubbles} [123]. 
Centred on the core of the Galaxy, these structures symmetrically extend to approximately 10 kpc above the 
Galactic plane. The parent relativistic particles (e.g., protons) could be accelerated in the nucleus of GC, and 
then injected into Fermi Bubbles. Alternatively, protons and electrons could be produced {\it in situ} through 
first- and/or second-order Fermi acceleration mechanisms supported by hydrodynamical shocks or plasma 
waves in a highly turbulent medium. The processes that create and support these structures could originate 
either from an AGN-type activity related to the central black hole (Sgr~A*) or from ongoing star formation in 
the galactic nucleus.\\  
The luminosity of gamma-rays with hard, $E^{-2}$-type, spectrum in the energy interval 1-100 GeV (see Fig. 13)
is  $L_\gamma \approx 4\times10^{37} \ \rm erg/s$. Given the overall limited energy budget of the GC, particle 
acceleration and gamma-ray emission in the {\it Fermi} bubbles should proceed with very high efficiency. Despite 
the significant differences of the models proposed for the origin of the {\it Fermi} bubbles, only two radiation 
mechanism can be responsible for gamma-rays  - IC emission by relativistic electrons or decays of neutral 
pions produced in {\it pp}-interactions. Because of severe radiative energy losses, however, the mean free path 
of $\geq 100$~GeV electrons is significantly shorter than the size of the {\it Fermi} bubbles. Therefore, one has 
to postulate {\it in situ} electron acceleration throughout the entire volume of the bubbles [123,124]. Such a 
scenario could be realised through stochastic (second-order Fermi) acceleration [125] or due to series of 
shocks propagating through the bubbles and accelerating relativistic electrons [126]. Importantly, the 
suggested acceleration mechanism seem unable to boost the electron energy beyond 1~TeV, thus in order 
to explain the extension of the observed gamma-ray spectrum up to 100~GeV by IC,  one has to invoke FIR 
and optical/UV background emission supplied by the galactic disk (see Fig.~13). This model provides robust 
predictions. In particular, since the FIR and optical/UV contributions to the target field for IC scattering decrease 
quickly with distance from the disk, the spectrum of gamma-rays from high latitudes should contain a cut-off 
above tens of GeV. The limb brightening at highest energies is another characteristic feature predicted by this 
model. These spectral and spatial features can be explored in the near future, after the gamma-ray photon 
statistics in the {\it Fermi}-LAT data set has achieved an adequate level.\\ 
An hadronic origin for the observed gamma-rays is an alternative interpretation suggested for the {\it Fermi} 
bubbles [124,127].
Despite the low plasma density in the {\it Fermi} bubbles, $n \leq 10^{-2} \ \rm cm^{-3}$, the efficiency of proton 
interactions can be very high. Indeed, if protons would have been continuously injected and trapped in the bubbles 
over timescales of approximately $10^{10} \ \rm $yr, the main power in accelerated protons would be lost in 
$pp$-collisions given that the characteristic time of the latter, $t_{\rm pp} =1/(k_p n \sigma_{\rm pp} c) \approx 5 
\times 10^9 (n/10^{-2}\rm cm^{-3})^{-1}  \ \rm yr$, is shorter than the confinement time. This implies that one deals 
with a so-called "thick target'' scenario, when the system is in saturation. The hadronic gamma-ray luminosity is 
equal to $L_\gamma \approx W_{\rm  p}/t_{\rm pp \rightarrow \pi^0}$, where $W_{\rm p}$ is the total energy of 
protons in the bubbles, and $t_{\rm pp \rightarrow \pi^0}$ is the timescale for neutral pion production in 
$pp$-interactions. In the saturation regime, $W_{\rm p}=\dot{Q_{\rm p}} t_{\rm pp}$ (with $\dot{Q}_{\rm p}$ the 
injection rate of protons), assuming that the energy dissipation through $pp$-collisions is the dominant loss process. 
Since $t_{\rm pp}=1/3~t_{\rm pp \rightarrow \pi^0}$, we have $L_\gamma=\dot{Q_{\rm p}}/3$, thus about a third of 
the power injected into relativistic CRs emerges in gamma-rays (of all energies) independent of the local density, 
interaction volume and the injection time. Note that since the timescale of $pp$-interactions is comparable to the 
supposed age of the bubbles of $10^{10}$yr, the efficiency would be somewhat less. Also, one should take into 
account that at low energies, ionisation and adiabatic losses of protons play a non-negligible role, thus the overall 
efficiency for a broad energy spectrum of protons would be reduced to several percent. The fluxes of hadronic 
gamma-rays shown in Fig.~13 confirm these simple estimates. Note that independent of the history of injection 
of relativistic protons, the current total energy in protons should be as high as $W_{\rm p}= L_\gamma 
t_{\rm pp \rightarrow \pi^0} \simeq 10^{55}$ erg which is comparable to the magnetic field energy in the bubbles
(cf. Ref.~[301]). 

\vspace{3mm}
\centerline{\psfig{figure=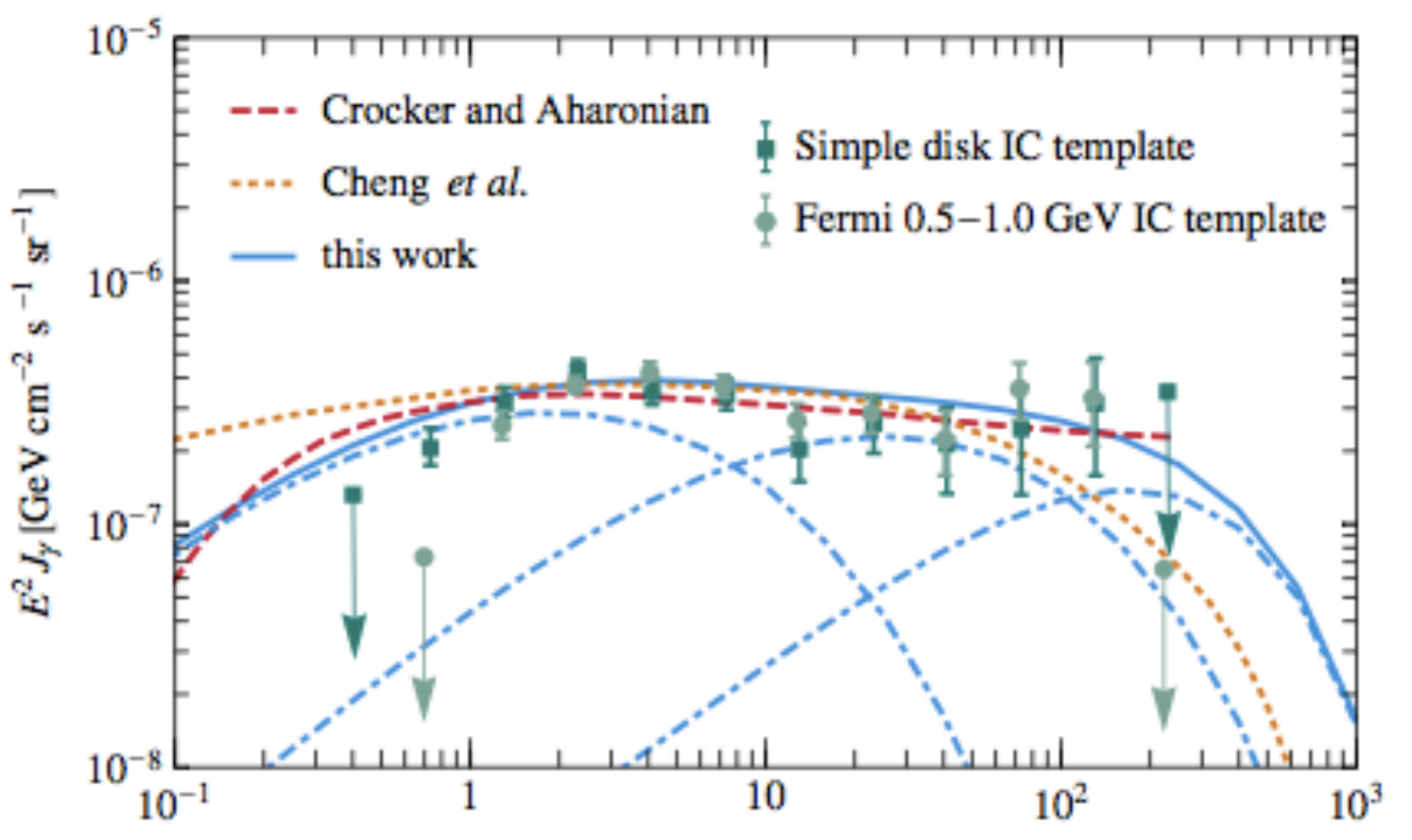, width=8cm}\vspace{1mm}}
{\baselineskip 10.5pt\renewcommand{\baselinestretch}{1.05}\footnotesize \noindent
{\bf Fig.~13}\quad The spectral energy distribution of gamma-rays from the {\it Fermi} bubbles compared to 
theoretical predictions. ({\it i}) IC model of Ref.~[125] (solid line) assuming stochastic acceleration of electrons 
in the bubbles  (the contributions from the scattering on the CMB, FIR, and optical/UV backgrounds are shown 
separately); ({\it ii}) IC model of Ref.~[126] (dotted line) assuming diffusive shock acceleration of electrons; 
({\it iii}) hadronic model of Ref.~[124] (dashed line). The figure is from Ref.~[125].}\label{MertschFig2}
\vspace{2mm}

The above noted hadronic model of gamma-ray emission of the {\it Fermi} bubbles does not exclude other 
"hadronic" scenarios with faster energy release related to the activity of the central black hole Sgr A*.  A fast 
energy release can be provided, for example, by the capture of stars by Sgr A* over the last 10 Myr with an 
average capture rate of $3 \times 10^{-5} \ \rm yr^{-1}$ and energy release of $3 \times 10^{52} \ \rm erg$ per 
capture [128]. It has been argued in Ref.~[140]  that quasi-periodic injection of hot plasma could produce a 
series of strong shocks in the {\it Fermi} bubbles which can (re)accelerate protons beyond the "knee", up to 
energies of about $10^{18}$ eV. If confirmed by independent detailed hydrodynamical simulations, this could 
appear a viable solution for the origin of one of the most "problematic'' (poorly understood) energy intervals  
of cosmic rays.   

\vspace*{5mm} \noindent {2.6\quad Blazars}\vspace{3.5mm}

\noindent Most of the detected extragalactic gamma-ray sources belong to the {\it blazar} 
class, which comprises BL Lac objects and Flat Spectrum Radio Quasars ({\it FSRQs}). 
The central engine in these active galaxies (AGNs), a supermassive black hole (BH) of 
mass $\simgt 10^7 M_{\odot}$ surrounded by an accretion disk, is commonly believed to 
eject a relativistic jet pointing almost directly towards the observer. Doppler boosting effects 
 results in strong flux amplification, thus naturally favouring the detection of blazars 
on the extragalactic sky.\\ 
{\it Fermi}-LAT, for example, has detected over 1000 extragalactic high-energy (HE) sources 
in two years of survey (2LAC), most ($>90\%$) of which are blazars~[129]. In comparison, 
non-blazar sources like starburst galaxies (SBs) or radio galaxies ({\it RGs}) only make 
out a minor fraction (in numbers).\\ 
At the time of writing more than 50 extragalactic VHE sources, populating the whole
sky, are listed in the online TeV Catalog (TeVCat).$^1$\footnote{$^1$http://tevcat.uchicago.edu/} 
The majority of them ($\sim90\%$) are again of the blazar type, with the so-called 
high-frequency-peaked BL Lac objects ({\it HBLs}, with low-energy component peaking 
at $\nu_p > 10^{15}$ Hz, in contrast to {\it LBLs}=low-frequency-peaked BL Lacs, 
peaking at $\nu_p<10^{14}$ Hz) constituting the dominant ($\geq 70\%$) sub-class, yet 
also including three FSRQs (3C279 at $z=0.536$; PKS 1510-089 at $z=0.361$ and 
PKS 1222+21 at $z=0.432$). FSRQs are typically distinguished from BL Lac objects by
the presence of strong and broad (rest-frame equivalent width $\geq5${\AA}) optical 
emission lines. Almost all Fermi-detected FSRQs for which $\nu_p$ can be estimated 
are of the low-frequency-peaked ($\nu_p <10^{14}$ Hz) type. Note that AGNs, which
have been detected at TeV are typically characterised by a harder GeV photon index 
than the majority of 2LAC sources.\\ 
At present, blazar sources out to redshift $z\sim0.6$ (i.e., 3C279 at $z=0.536$ [163] and 
BL Lac KUV 00311-1938 at $z>0.51$, tentative $z=0.61$ [164]) have been detected at 
VHE energies, cf. Fig.~14 for their redshift distribution. Blazar population studies at lower 
(radio-X-ray) frequencies indicate a redshift distribution for BL Lacs objects that seems 
to peak at $z\sim 0.3$, with only few sources beyond $z\sim 0.8$ (under the proviso of 
some bias as for a substantial fraction of BL Lacs the redshift is not known), while the 
FSRQ population is characterised by a rather broad maximum between $z\sim(0.6-1.5)$ 
[160].

\vspace{1mm}
\centerline{\psfig{figure=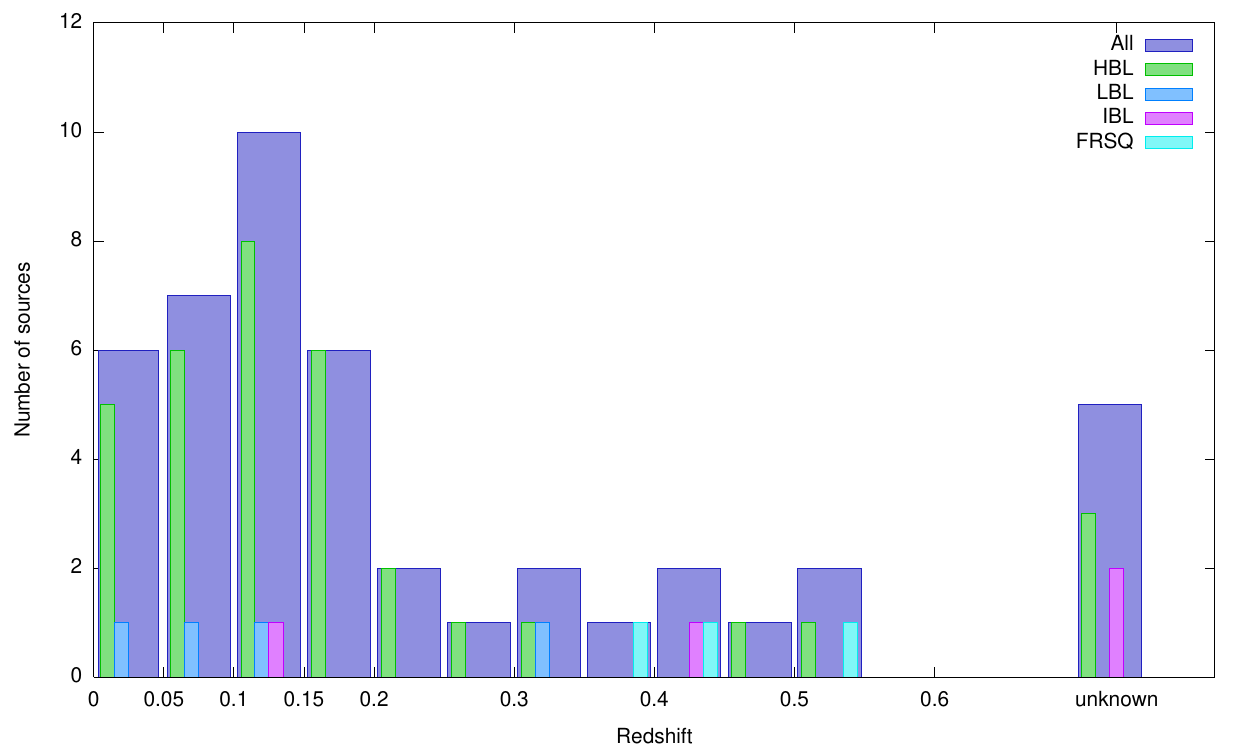, width=3.5in}\vspace{1mm}}
{\baselineskip 10.5pt\renewcommand{\baselinestretch}{1.05}\footnotesize \noindent
{\bf Fig.~14}\quad Distribution of redshift for the VHE-detected blazars. Redshift data are 
taken from TeVCat. Most objects are within $z\simlt 0.2$.}
\vspace{4mm}

\noindent Observed VHE flux levels typically range from $\sim 1\%$ (for the average/steady 
states) up to $\sim15$ times the Crab Nebula steady flux in high activity stages (e.g., for the 
2006 flaring state of PKS~2155-304 [137]). Gamma-ray emission beyond 10 TeV has been 
established, with measured photon energies reaching, e.g. $\sim20$ TeV in Mkn 501 
(z=0.034; 1997 VHE flare [130]) and Mkn 421 (z=0.031; 2010 VHE flare [131]). The observed 
VHE spectra usually vary between hard and soft power laws; for the HBL sources, for 
example, inferred photon indices (assuming a single power law $dN/dE \propto 
E^{-\Gamma}$) range from $\sim 2.3$ to $\sim 4.5$, with some indications for spectral 
hardening with increasing activity. Non-HBL sources (i.e., {\it IBL}=intermediate-peaked 
BL Lacs and LBLs) are usually detected during high states only, with low states expected 
to fall below current sensitivities.\\
In a $\nu F_{\nu}$ representation (SED), blazars often show a double-humped structure, 
cf. Figs. 15 and 16. The first hump (sometimes reaching peak frequencies up to $\sim10^{19}$ 
Hz, as in the HBL object RGB J0710+591 [132]) is commonly attributed to non-thermal electron 
synchrotron emission; the second one is widely believed to be due to leptonic inverse 
Compton processes (on the synchrotron photons in SSC, or on ambient photons in 
External Inverse Compton [=EC] models), although hadronic scenarios often remain possible. 
Different blazar population studies seem to suggest that there is a continuous spectral trend 
(see Fig. 15) from FSRQ $\rightarrow$LBL$\rightarrow$IBL$\rightarrow$HBL, often called 
the "blazar sequence", characterised by a decreasing source luminosity, increasing synchrotron 
peak frequency and a decreasing ratio of high- to low-energy component [133,134] (but cf. 
also [135] for caveats due to selection effects and unknown redshift). 

\vspace{0mm}
\centerline{\psfig{figure=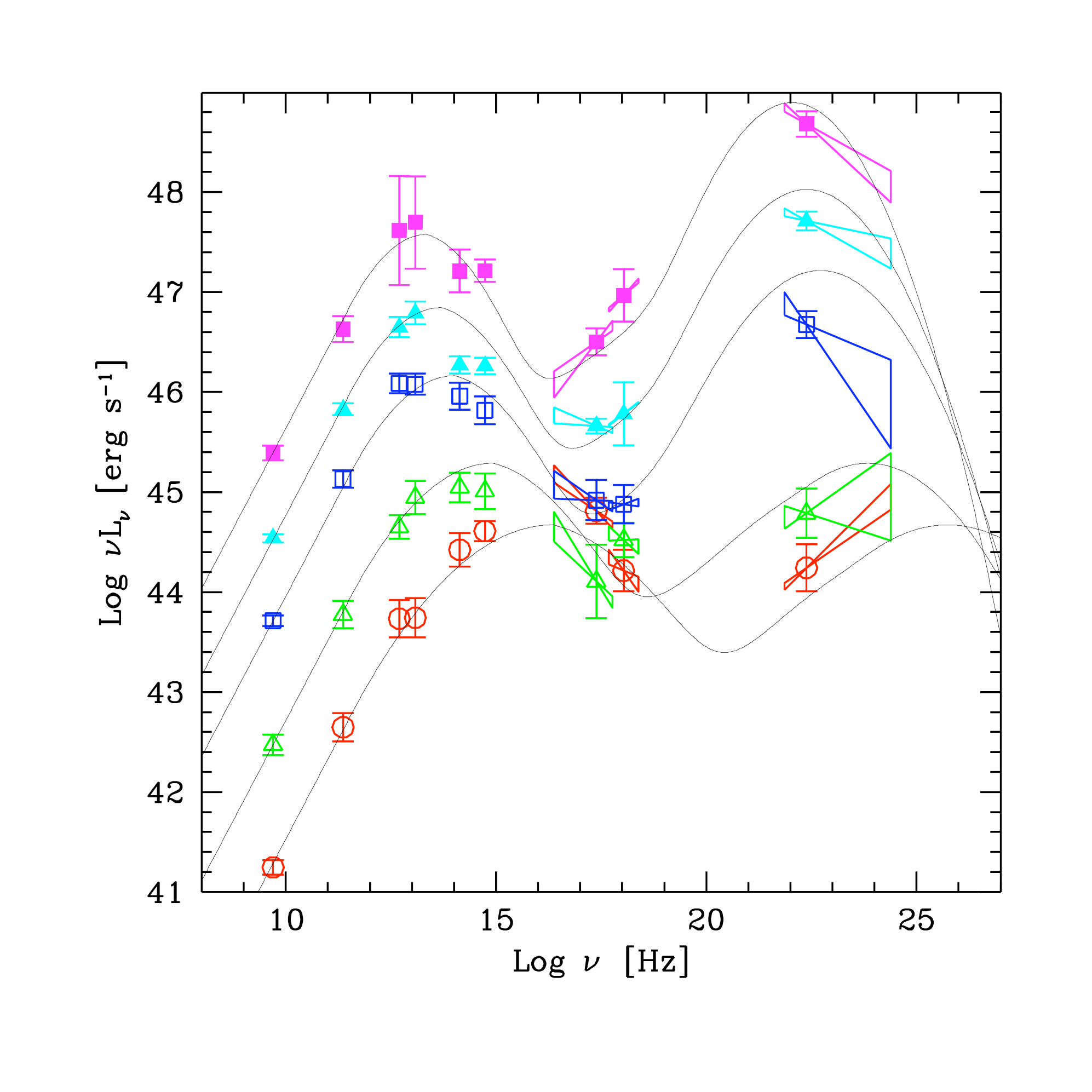, width=3.3in}\vspace*{-3mm}}
{\baselineskip 10.5pt\renewcommand{\baselinestretch}{1.05}\footnotesize \noindent
{\bf Fig.~15}\quad Sequence of characteristic blazar SEDs as a function of source luminosity
from FSRQ (top curve) to HBL objects (bottom curve). From Ref. [134], reproduced with 
permission \copyright ESO.} 
\vspace{4mm}

Blazar SEDs can span almost 20 orders of magnitude in energy, making simultaneous multi-wavelength 
observations a particular important diagnostic tool to disentangle the underlying non-thermal processes.
A variety of leptonic and hadronic emission models have been discussed in the literature (see, e.g.,
[136] and reference therein). A significant correlation between TeV and X-ray flux variations for example, 
which is often found, could favour a leptonic synchrotron-Compton interpretation, but counterexamples 
("orphan TeV flares") do exist [141]. Short-term variability is usually more difficult to account for in hadronic 
models because of longer cooling timescales, but strong magnetic fields (for proton synchrotron, e.g.
[142]) or high target matter densities (pp-interactions triggered by jet-star interactions, e.g. [143]) may 
partly compensate. While for HBL objects, homogeneous (one-zone) leptonic SSC modeling often seems 
to provide a reasonable SED characterization (but see, e.g., [144] for a possible exemption), this does not 
apply in a similar way to LBL objects. Among the four LBLs detected, for example, AP Lib ($z=0049$) 
represents an intriguing example where the 2nd bump seems extremely broad (stretching from keV to 
TeV), defying a simple homogeneous SSC interpretation [145].\\

\vspace{1mm}
\centerline{\psfig{figure=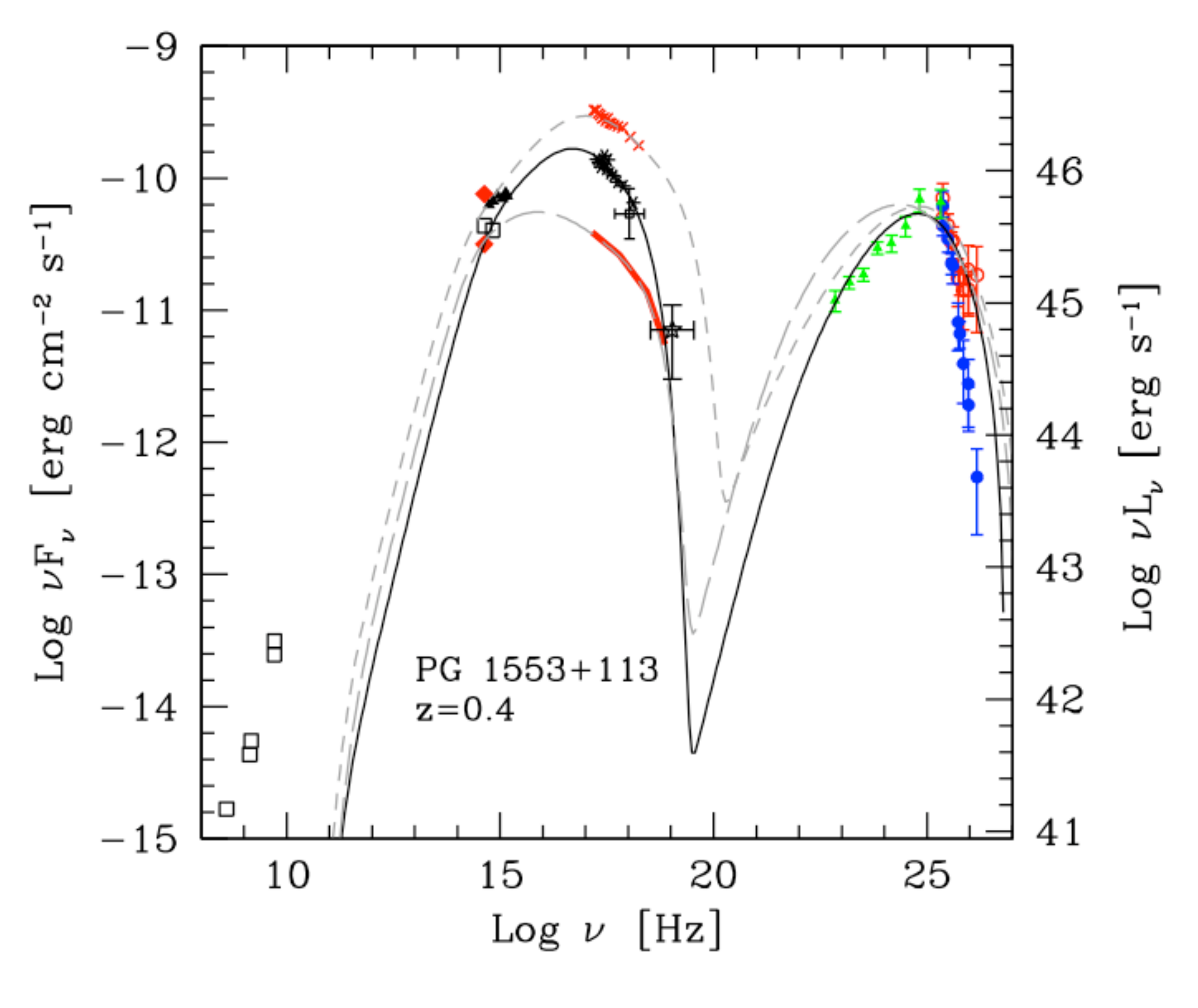, width=3.3in}
\vspace{1mm}}
{\baselineskip 10.5pt\renewcommand{\baselinestretch}{1.05}\footnotesize \noindent
{\bf Fig.~16}\quad A recent, double-hump-structured SED example: The high-frequency-peaked 
(HBL) BL Lac object PG 1553+113 as based on VHE (MAGIC, 2005-2009) observations and 
archival data. Pronounced variability (on yearly time scale) is seen in the X-ray band. The 
average SED has been modelled with a one-zone SSC model (continuous black line). 
From Ref.~[159], reproduced by permission of the AAS.}
\vspace{4mm}

\begin{center}
\begin{tabular}{c|c|c|c}
AGN                          &   type    & redshift  &    $\Delta t_{\rm VHE}$ \\ \hline 
PKS 2155-304        &   HBL    &  0.116    &   $\sim3$ min    \\
Mkn 501                   &   HBL    &  0.034    &   $\sim3$ min    \\   
PKS 1222+21         &  FSRQ  &  0.432    &   $\sim10$ min \\    
Mkn 421                   &   HBL    &  0.031    &   $\sim10$ min \\  
BL Lac                      &   LBL   &  0.069     &   $\sim15$ min \\  
W Comae                 &   IBL     &  0.102    &   $\sim1$  day   \\  
M87                           &   RG     &  0.004    &   $\sim1$  day   \\
\end{tabular}
\end{center}
{\baselineskip 10.5pt\renewcommand{\baselinestretch}{1.05}\footnotesize \noindent
{\bf Tab.~2}\quad VHE variability in AGN: Characteristic minimum VHE variability timescale 
$\Delta t_{\rm VHE}$ as observed with current instruments for an exemplary set of AGN.} 
\vspace{4mm}

\noindent Despite the limited temporal coverage of the current IACTs more than half of 
the AGN detected in the TeV domain shown variability, albeit often weak. For the majority 
of them, variability timescales above one month have been found. In about a quarter of 
them there is clear evidence for short-term VHE variability on observed timescales of 
less than one day, cf. Table~2. The HBL class currently reveals the most rapid and dramatic 
VHE gamma-ray flux variability with observed variability timescales $< 5$ min, as found 
by the H.E.S.S. and MAGIC experiments for PKS 2155-304 ($z=0.116$)~[137]  and Mkn 501
[138], respectively, cf. Fig.~17. Given the limited angular resolution ($\sim 0.1^{\circ}$) 
of IACTs, this implies that one of the most constraining requirements on the jet kinematics 
and the high-energy emitting region comes from VHE variability studies. Fast VHE variability 
from distant blazars can also be used to derive constraints on an energy-dependent 
violation of Lorentz invariance (energy-dependent speed-of-light) as predicted in various 
models of Quantum Gravity [146,158]. 

\vspace{2mm}
\centerline{\psfig{figure=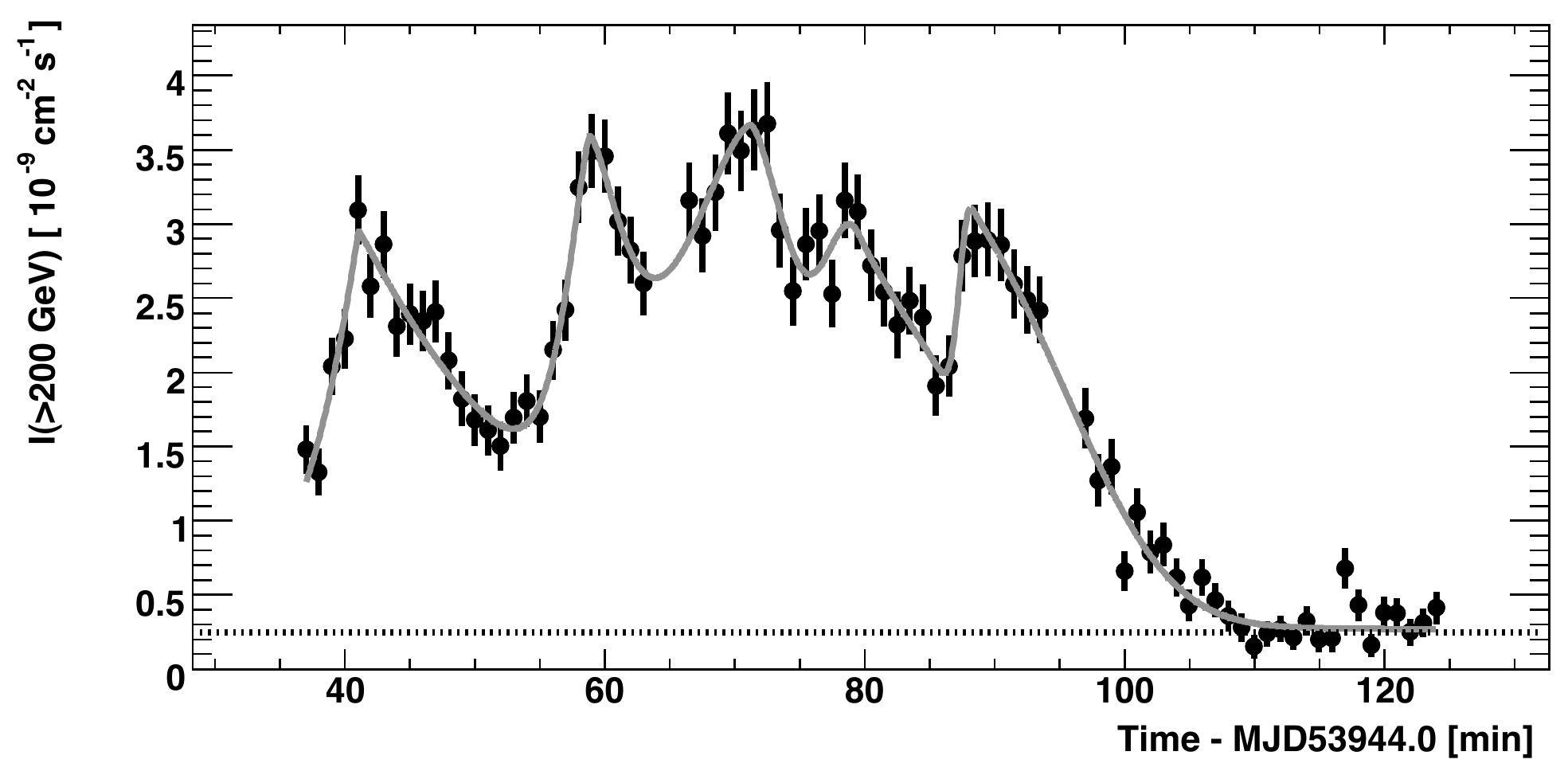, width=3.45in}\vspace{1mm}}
{\baselineskip 10.5pt\renewcommand{\baselinestretch}{1.05}\footnotesize \noindent
{\bf Fig.~17}\quad Light curve: Integrated flux $I(>200$ GeV) versus time as observed by 
H.E.S.S. for PKS 2155-304 on July 28, 2006. The data are binned in 1-minute intervals. 
The horizontal line gives the steady flux from the Crab Nebula for comparison. From 
Ref.~[137].}
\vspace{4mm}

\noindent The detection of a large number of gamma-ray emitting blazars has opened a new research 
area - "observational gamma-ray cosmology". The underlying idea is based on the energy-dependent 
absorption of $\gamma$-rays from distant extragalactic objects caused by interactions 
($\gamma_{\rm VHE}~\gamma_{\rm EBL}\rightarrow e^+~e^-$) with the Extragalactic Background Light 
({\it EBL}) that extends from UV to far IR wavelengths. The identification of absorption features in the 
spectra of $\gamma$-rays above 10 GeV, as well as detection of characteristic angular and time distributions 
of gamma-rays produced during the cascade development in the intergalactic medium on large ($\geq 
100$~Mpc) scales, should allow us to derive unique cosmological information about the EBL and the 
intergalactic magnetic fields ({\it IMFs}). The realization of these exciting possibilities requires not only precise 
spectroscopic measurements from a large number of extragalactic objects located at different redshifts, 
but, more importantly, a good understanding of the intrinsic gamma-ray spectra. So, far the most significant 
contribution in this area comes from the measurements of gamma-rays from blazars with redshifts between 
0.1-0.2. In particular, based on such observations, the H.E.S.S. collaboration has first reported a quite 
meaningful upper limit on the EBL at near and mid-infrared wavelengths [147]. Remarkably, the inferred 
upper limit appeared to be very close to the lower limit given by the measured integrated light of resolved 
galaxies (galaxy counts), cf. also [148,150,151] for related inferences. Very recently, a similar result has 
been reported by the {\it Fermi}-LAT collaboration [152] for the EBL at optical and UV bands. One should 
mention, however, that the inferred upper limits are not model-independent. The H.E.S.S. result, for example, 
is based on the assumption that the differential intrinsic spectrum is not harder than $E^{-1.5}$. The {\it Fermi}-LAT 
result is based on the detection of cutoffs in the averaged spectra of three samples of BL Lac objects combined 
in three different intervals of redshift, assuming that these cut-offs are caused by intergalactic absorption. 
Although both assumptions sound quite reasonable, and the derived upper limits agree with most of the 
theoretical/phenomenological predictions for the EBL, one should keep in mind that they are not free of model 
assumptions. It is believed that future measurements by next-generation detectors, in particular by CTA, based 
on a much larger sample of AGN should significantly increase the source statistics and improve the quality of 
data, and consequently reveal details in the EBL. This optimistic view may, however, underestimate the 
difficulties related to the uncertainties of the intrinsic source spectra. On the other hand, the limits in which 
presently the EBL fluxes are robustly constrained, are quite tight, so one can "recover" the intrinsic gamma-ray 
spectra with a reasonable accuracy. Interestingly, in the case of some blazars, the gamma-ray spectra after 
correction for intergalactic absorption, appear extremely hard with photon indices $\leq 1.5$ or even close 
to 1, see e.g. [147,150,153]. This challenges conventional radiation models, but still cannot be considered 
as a failure of the standard blazar paradigm and as a need for new physics. Such spectra can still be explained, 
assuming, for example, the prevalence of certain conditions for the formation of the parent electron spectra 
(e.g., stochastic acceleration) or specific internal gamma-ray absorption, see e.g. [154-156]. Nevertheless, 
the growing number of VHE blazars with redshift exceeding $z\sim0.5$ tells us that one should perhaps be 
prepared for even more dramatic assumptions, including violation of Lorentz invariance or "exotic" interactions 
involving hypothetical axion-like particles. An alternative interpretation of gamma-rays from very distant blazars 
(in case of their detection) exists in the framework of standard physics: TeV gamma-rays can in principle be 
observed even from a source at $z\geq1$, if the observed gamma-rays are secondary photons produced in 
hadronic interactions (with CMB or EBL background photons) of energetic cosmic-ray protons, originating in 
the blazar jet and propagating over cosmological distances almost rectilinearly. In the case of a detection of 
TeV gamma-rays from a blazar with $z\geq 1$, this model could in principle provide a viable interpretation 
consistent with conventional physics, but with an extreme assumption on the strength of the IMF in the range 
of $10^{-17}-10^{-15}$~G (see, e.g. [157]). On the other hand, if VHE $\gamma$-rays from distant blazar 
attenuate through pair-production with EBL photons, constraints on the strength of the IMF can be derived by 
modeling the anticipated GeV emission from the electromagnetic cascades, taking the possible deflections 
of pairs in the IMF into account. According to a recent study, this method suggests a lower bound on the IMF 
of $B \simgt10^{-17}$ G [149].

\vspace*{5mm} \noindent {2.7\quad Radio Galaxies}\vspace{3.5mm}

\noindent Misaligned (non-blazar) AGNs, characterised by jets substantially inclined 
with respect to the observer, represent a particularly interesting class of VHE emitters. 
Nearby radio galaxies (RGs) are especially attractive as their proximity may allow us to 
resolve the radio jets down to sub-parsec scales and to study possible multi-wavelength 
correlations. The absence of strong Doppler boosting could make a VHE detection 
challenging, yet also allow to get unique insights into emission regions otherwise 
hidden.

\noindent Out of $\sim1000$ high-energy (HE) sources (886 in the "Clean Sample"), 
{\it Fermi}-LAT has reported the detection of only about ten misaligned RGs at GeV 
energies, with a predominance of the Fanaroff-Riley-type~I (FR~I) [129,161,162]. At 
TeV energies, only four RGs have been identified by current IACTs: The nearest AGN 
Cen~A ($d\simeq 3.8$ Mpc), the giant RG M87 ($\simeq 16.7$ Mpc), and the Perseus 
Cluster ($d\sim77$ Mpc, $z\sim 0.018$) RGs NGC 1275 and IC 310. A detection of the 
RG 3C66B was initially reported by MAGIC (2007 observations [139]), but the VHE 
emission seems not sufficiently disentangled from the nearby (separation $\theta\sim 
0.12^{\circ}$) IBL blazar 3C66A to include it here.

\noindent {\it Cen~A} was detected at VHE in a deep ($>$120h) exposure by H.E.S.S. 
with a integral flux above 250 GeV of $\sim0.8\%$ of the steady flux of the Crab 
Nebula (corresponding to an apparent isotropic luminosity of $L(>250$ GeV) 
$\simeq 2 \times 10^{39}$ erg/s) [165]. The measured VHE spectrum extends up to 
$\sim5$ TeV and is consistent with a power-law of photon index $2.7\pm0.5$. 
No significant variability has been found. {\it Fermi}-LAT has also detected HE 
emission up to 10 GeV from the core of Cen~A, with the HE light curve (15 d 
bins) being consistent with no variability and the HE spectrum described by a 
comparable photon index [166]. A simple extrapolation of the Fermi HE power-law
to the VHE domain, however, tends to under-predict the observed TeV flux. This 
could be indicative of an additional contribution to the VHE domain beyond the 
common synchrotron-Compton emission, emerging at the highest energies [167,303]. 
While the giant radio lobes are also detected at GeV energies (with evidence for a 
spatial extension beyond the radio image [168]), they are clearly excluded (given 
the angular resolution of H.E.S.S.) as source of the detected TeV emission.

\noindent The giant radio galaxy {\it M87} was the first RG detected at TeV energies [169]. 
Commonly considered as a FR I-type RG, M87 is known to host a highly massive black hole 
of $M_{\rm BH} \simeq (2-6) \times 10^9\,M_{\odot}$ and to exhibit a relativistic jet misaligned 
by an angle $\theta\simeq(15-25)^{\circ}$, consistent with modest Doppler boosting $D=
1/[\Gamma_j (1-\beta\cos\theta)] \simlt 3$ (for a recent review, see e.g. [170]). M87 has 
shown particularly interesting VHE features over the past years, including rapid variability 
(day-scale flux doubling time scales $\Delta t_{\rm obs} \sim 1$) above 350 GeV during 
high source states (with levels sometimes exceeding $10\%$ of the Crab Nebula), and a 
hard spectrum (compatible with a power law of photons index $\Gamma \simeq 2.2$) 
extending beyond 10 TeV [171-174]. Both the hard VHE spectrum and the 
observed rapid VHE variability are remarkable features for a misaligned AGN. 
M87 is the only RG where clear evidence for day-scale TeV variability (see Fig. 18) 
has been found, although there are now hints for day-scale variability in IC~310 as well. 
The VHE variations in M87 are the fastest observed in any other waveband so far, 
and already imply a compact size of the $\gamma$-ray emitting region ($R < Dc
\Delta t_{\rm obs}/[1+z]$), comparable to the Schwarzschild radius ($r_s=(0.6-1.8)
\times10^{15}$ cm) of the black hole in M87. Consistent with radio evidence for 
the ejection of a new component from the core during a flare in 2008 [175], this 
could point to the black hole vicinity as the most likely origin of the variable VHE 
radiation.

\vspace{2mm}
\centerline{\psfig{figure=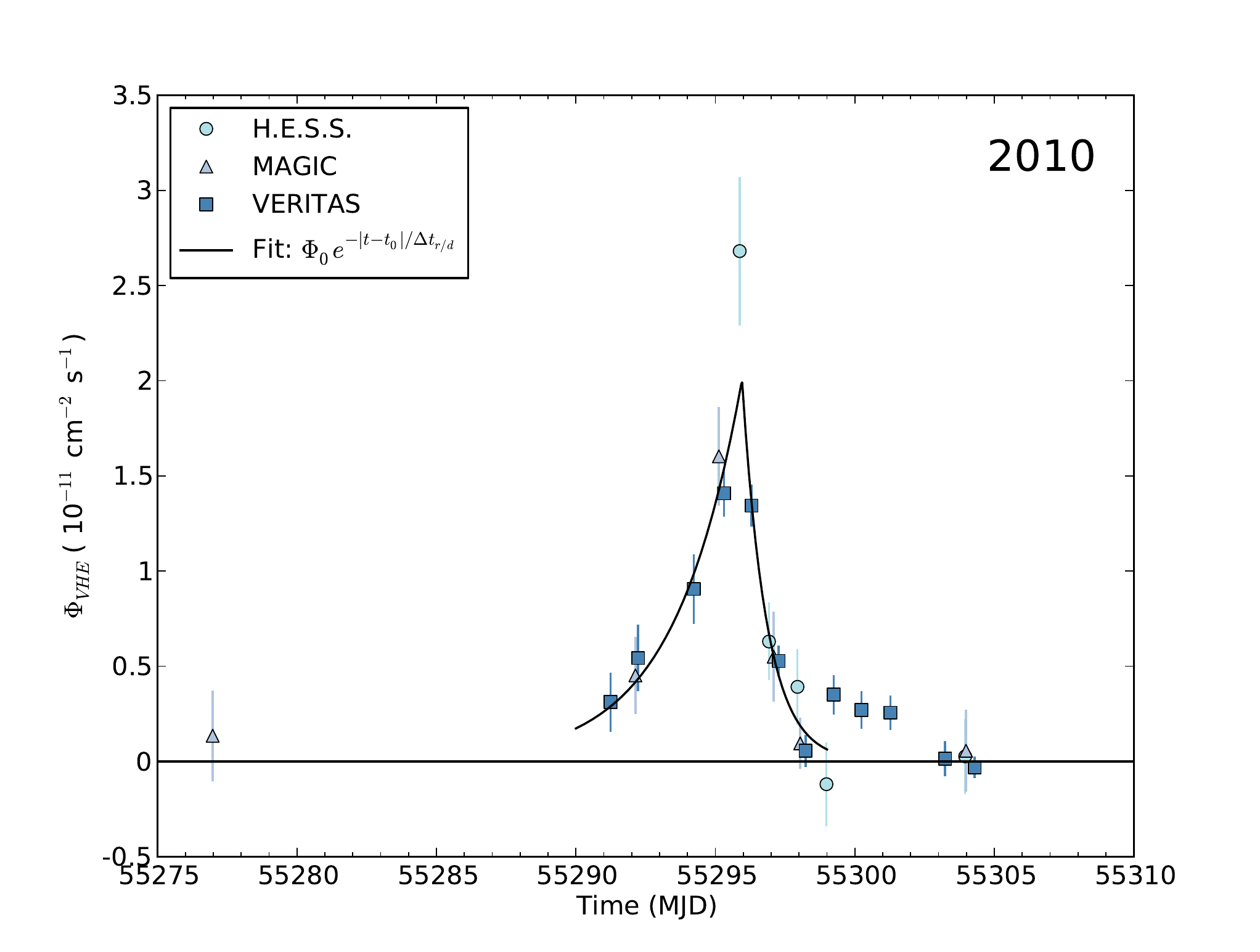, width=3.3in}\vspace{1mm}}
{\baselineskip 10.5pt\renewcommand{\baselinestretch}{1.05}\footnotesize \noindent
{\bf Fig.~18}\quad VHE flare of the radio galaxy M87 as observed in April 2010 by 
different IACTs. Significant day-scale activity is evident. The curve shows a fit of 
an exponential function to the data. From Ref.~[174].}
\vspace{4mm}

\noindent IC~310 and NGC~1275 are relatively new additions, being discovered 
at VHE as part of a Perseus galaxy cluster campaign by MAGIC [176,178] . {\it IC~310}, 
has been originally classified as a head-tail RG. However, according to a recent 
high-resolution VLBA study, there are little indications for jet bending [180], rather 
supporting (along with the jet-to-counter-jet flux ratio) the case that IC~310 may 
instead rather be a weakly beamed blazar [177]. The source has been detected 
above 300 GeV in about 21h of data (taken in 2009/2010) at an average level of 
$\sim3$\% of the Crab Nebula [176]. The VHE spectrum between 150 GeV and 7 TeV 
is very hard (even harder than in M87) and compatible with a single power law of photon 
index $\Gamma \simeq 2.0$. There is clear evidence for VHE variability on yearly 
and monthly time scales, with indications for day-scale activity found in a new 
analysis, features that are all reminiscent of the VHE activity seen in M87. \\ 
On the other hand, the central dominant (FR I) cluster galaxy {\it NGC~1275} (having 
radio jets misaligned by $\simgt30^{\circ}$), has been recently detected above 
$\sim100$ GeV during enhanced high energy ({\it Fermi}-LAT) activity in 46h of data 
(taken between 08/2010-02/2011). While the {\it Fermi}-LAT data reveal evidence for 
flaring activity above 0.8 GeV down to time scales of days [179], the situation at VHE
energies is less evident. No evidence of variability has been found in the 08/2010
to 02/2011 VHE light curve. A recent, improved analysis of an earlier (10/2009
-02/2010) data set, however, seems to provide hints for a possible month-type 
VHE variability. NGC~1275 shows a steep VHE spectrum ($\Gamma \simeq 4.1$) 
extending up to $\sim500$ GeV [178] and a hard HE ({\it Fermi}-LAT) spectrum (photon 
index $\Gamma\simeq 2.1$), indicative of a break or cut-off in the SED around 
some tens of GeV.

\vspace*{5mm} \noindent {2.8\quad Starburst Galaxies}\vspace{3.5mm}

\noindent Starburst Galaxies (SGs) are galaxies showing a very high rate of star 
formation ("starburst") in a localised region, the burst sometimes being triggered 
by a close encounter with another galaxy. The resultant highly increased supernova 
(SN) explosion rate and the expectation that the remnants (SNR) of those are 
efficient cosmic-ray (CR) proton accelerators (possibly up to $\sim10^{16}$ eV [184]), 
suggest that starburst regions may possess a high cosmic-ray density. Because 
of the very high ambient gas densities ($n>100$ cm$^{-3}$), hadronic interactions 
(inelastic proton-proton collisions and subsequent  $\pi^0$-decay) could then 
lead to efficient $\gamma$-ray production, making SGs promising targets for HE 
and VHE astronomy.\\
The spiral galaxy NGC~253 is the closest ($d\sim 2.6-3.9$ Mpc) SG in the southern 
sky, harbouring a compact ($\sim 100$ pc in extension) starburst region with an 
estimated SN rate of $\sim 0.03$ year$^{-1}$. The mean density of interstellar gas 
in this region ($n\sim 600$ protons/cm$^3$) is almost three order of magnitude 
higher than the average gas density in the Milky Way. NGC 253 was detected 
above 220 GeV in a deep exposure ($>119$h) in 2005-2008 by H.E.S.S. at a flux 
level of $\sim0.3\%$ of the Crab Nebulae [181]. A recent spectral analysis of an 
extended (-2009) data set (Fig. 19) shows that the VHE spectrum can be described 
by a power law with photon index $\Gamma \simeq 2.2$, being compatible with a 
power-law extension of the HE ({\it Fermi}-LAT) $\gamma$-ray spectrum and little
evidence for a spectral break or turnover between 200 MeV and $\sim5$ TeV [182].

\vspace{2mm}
\centerline{\psfig{figure=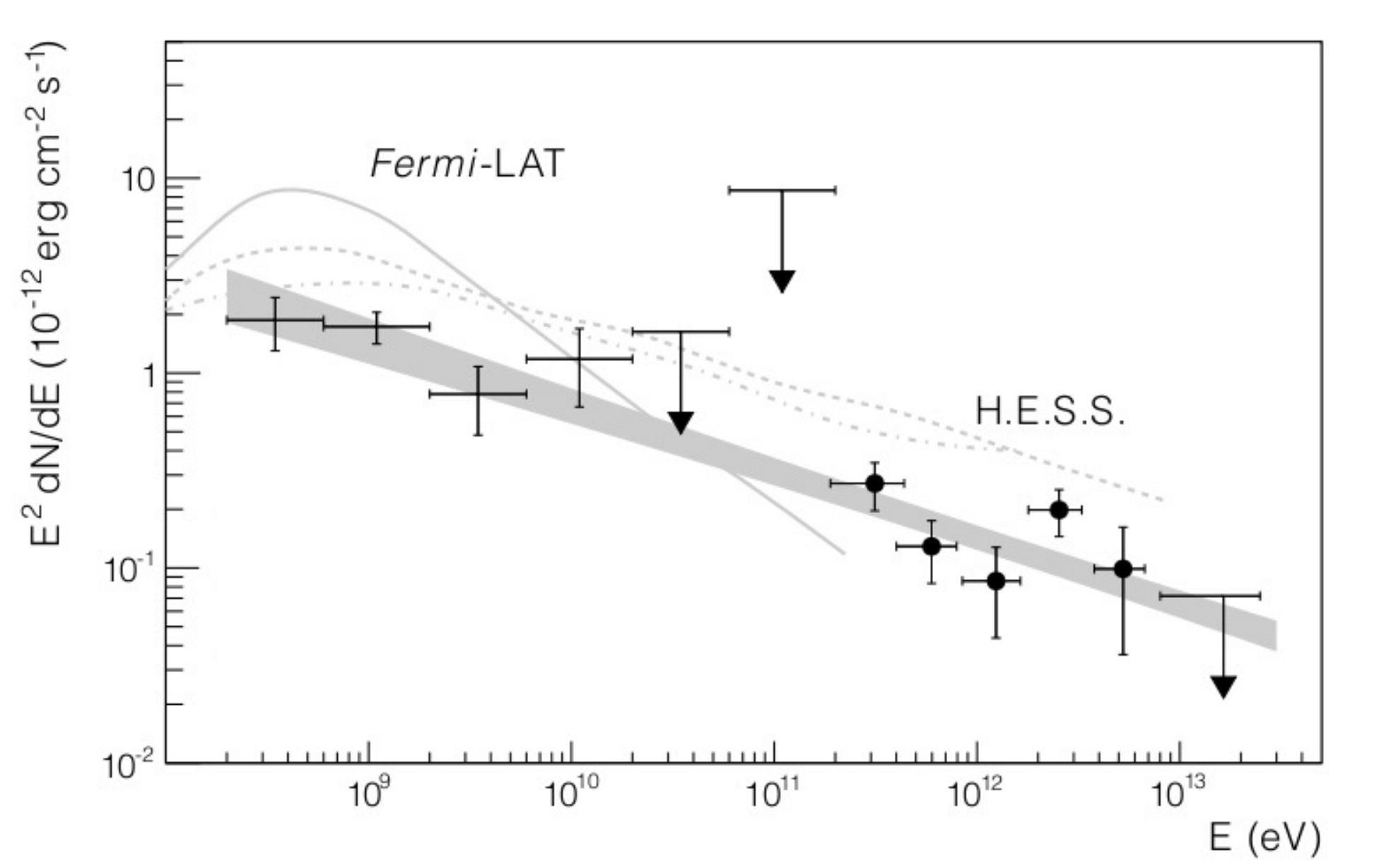, width=3.3in}\vspace{1mm}}
{\baselineskip 10.5pt\renewcommand{\baselinestretch}{1.05}\footnotesize \noindent
{\bf Fig.~19}\quad Differential energy spectrum of the starburst galaxy NGC 253 from
high-energy (Fermi-LAT, crosses) to the VHE (H.E.S.S., circles). The shaded area 
shows the 1$\sigma$ confidence band from a simultaneous fit to the high-energy and 
VHE data. From Ref.~[182].}
\vspace{4mm}

\noindent M82, on the other hand, is a well-known SG in the Northern Sky at a distance 
of  $d\sim 3.9$ Mpc, harbouring an active starburst region with a high SN rate of $0.1-
 0.3$ year$^{-1}$. This source was detected above a 700-GeV analysis threshold 
 in a deep exposure ($>137$ h) in 2008-2009 by VERITAS at a level of $0.9\%$ of 
 the Crab Nebula (corresponding to $L_{\gamma}(>700~\mathrm{GeV})\simeq 2
 \times 10^{39}$ erg/s) [183].\\
The noted SG findings are roughly compatible with hadronic CR ($pp$-interaction) 
models and energy-independent (advective) particle transport from the starburst 
region (but see also [304,305] for an alternative approach related to the TeV emission
from PWNe). If CRs would lose most of their energy due to inelastic collisions before they 
escape from the starburst region ("quasi-calorimeter" scenario), the energy required 
in TeV protons would be of the order $W_p\simeq L_{\gamma} t_{pp}\sim 10^{52}\
(200\,\mathrm{cm}^{-3}/n)$ erg, with $L_{\gamma}\sim 10^{39}$ erg/s the characteristic 
VHE $\gamma$-ray luminosity, $t_{pp} \sim (c\sigma_{pp}k_pn)^{-1}$ the typical 
pp-collision time scale, and $n$ the ambient gas number density. This would imply a 
typical CR energy density $W_p/V \sim 10^{-10}$ erg/cm$^{3}$, i.e., roughly two orders 
of magnitude higher than the average value ($\sim 1$eV/cm$^3$) for the Milky Way. 
Note however, that in starburst region is likely to be mildly calorimetric, with only 
$\sim 20-30\%$ of the CR energy channeled into pion-production [182].

\vspace*{5mm} \noindent {2.9.\quad Candidates: GRBs, Galaxy Clusters
and Passive Black Holes}\vspace{3.5mm}

Several additional source classes are thought to be possible VHE emitters, but have
not yet been detected. This includes GRBs, Galaxy Clusters and Passive Black Holes:  

\noindent {\it Gamma-Ray Bursts} (GRBs) have been observed to emit intense flashes 
of $\gamma$-rays up to a least some tens of GeV energies, and their intrinsic photon 
spectra are likely to extend to TeV energies due to synchrotron-inverse Compton 
processes (by electrons accelerated at internal or external shocks) in leptonic models 
or electromagnetic cascades (triggered by $p\gamma$- and $pp$-interactions from 
CR-accelerated protons) in hadronic models [185]. Extending the energy range for 
GRBs to the VHE domain could thus be particularly important to disentangle the origin 
of their high-energy emission. Significant $\gamma$-ray emission is expected to occur 
on time scales from second to at most hours, so that VHE instruments need to be 
re-posited very quickly to catch the dominant part of the outburst. The VHE prospects 
are, yet, generally constrained by (i) $\gamma\gamma$-pair production within the source 
(e.g., see the {\it Fermi}-LAT evidence for a spectral break around one GeV in GRB 090926A 
[189]), and (ii) in particular, interactions with EBL photons, limiting the  detectability of VHE 
GRBs from cosmological distances to relatively nearby sources ($z\simlt0.6$).
Despite several attempts, no significant VHE excess from GRBs has been found yet
and only upper limits have been reported (e.g., [186-188]). It will be highly interesting 
to see whether a VHE detection becomes possible in the near future with the new 
HESS~II instrument, having a drive system designed as to reduce the response time 
to GRB alerts and a lower energy threshold as to reduce EBL absorption.\\
 
\noindent Massive {\it Cluster of Galaxies}, harbouring powerful AGNs and being surrounded 
by strong accretion shocks represent another potential VHE source class. Gamma-ray
emission could be produced by several processes [190], e.g., IC up-scattering of CMB 
and other soft photon fields by electrons injected into the inter-cluster medium and further 
accelerated by turbulent processes [191,192]. Alternatively, hadronic (pp) interactions of 
relativistic CR protons with the ambient cluster gas could result in gamma-rays via 
$\pi^0$-decay. Given that galaxy clusters are thought to be capable of accelerating protons 
to energies $E>10^{18}$ eV [193], $\gamma$-ray emission could also be related to IC 
emission by secondary electrons generated via Bethe Heitler pair production with CMB 
photons ($p+ \gamma\rightarrow e^+e^-+p$) [194,195].
Finally, self-annihilation of a dark matter particle (e.g., WIMP) could possibly also lead to 
gamma-ray production [196]. However, despite these expectations, no extended 
gamma-ray signal has been found so far in VHE observations of local clusters. The 
Perseus Cluster, for example, one of the closest ($d\sim77$ Mpc) and largest clusters 
has been regarded as a particularly promising candidate. This cluster is characterised 
by intense X-ray emission and hosts a luminous radio halo (size $\sim 200$ kpc), 
indicative of the presence of a high density of CR protons. Large intra-cluster fields 
($\sim 1-10\mu$G) suggest it to be an excellent place for particle acceleration. Yet, 
no (extended) gamma-ray excess (but two RG, see above) has been found in recent 
VHE observations (by MAGIC for a total of 99h in 10/2009-02/2011) [197]. Comparing 
this with simple CR model prediction suggests that the average ratio of CR-to-thermal 
pressure has to be below several percent. Similar (yet somewhat less stringent) CR 
constraints have also been inferred based on the VHE non-detection of the Coma 
Cluster ($d\sim 100$ Mpc) by VERITAS and H.E.S.S. [198,199] (for constraints on CR
mixing conditions in the Hydra Cluster, see [200]), while the VHE non-detection of the 
Fornax Cluster ($d\sim 19$ Mpc) has been used to put constraints on the dark matter 
(velocity-weighted $<\sigma v>$) annihilation cross-section [201] (cf. also Ref.~[300] 
for recent $<\sigma v>$ upper limits based on central Galactic Halo data).\\ 

\noindent Nearby quiescent or {\it "Passive" supermassive Black Holes} (PBHs), i.e., 
very massive black holes hosted in galactic nuclei without bright signatures of broad-band 
emission and with very low luminosity at lower frequency, could potentially also produce 
ultra-high energy cosmic rays (UHECR) and VHE gamma-rays by a number of magnetospheric 
processes (see [202] for review). A recent analysis [203] based on the non-detection of 
one of the prime candidate sources, NGC~1399 (distance$\sim 20$ Mpc; black hole mass 
$\sim 5\times10^8 M_{\odot}$), with current HE \& VHE instruments allows to put constraints 
on, e.g., gap-type particle acceleration scenarios, suggesting that this source is unlikely 
to be capable of accelerating protons to energies beyond $4\times10^{18}$ eV.

\vspace*{6.0mm} \hrule\vspace{2mm} 
\noindent {\large \usefont{T1}{fradmcn}{m}{n}\xbt 3\quad Physics Impact of Recent Results}
\vspace{2.5mm}

\vspace*{5mm} \noindent {3.1\quad CR and Galactic Gamma-Ray Sources}\vspace{3.5mm}

Cosmic-rays (CRs) are highly energetic protons, nuclei and electrons which are measured from 
$\simeq10^{9}$ eV up to $\simeq$10$^{21}$ eV with an almost featureless continuous spectrum. 
The most prominent deviation from a pure power-law (with index of about $-2.8$) happens in the 
"knee", where the index changes to $\sim-3.1$ at about 4~PeV [204], and the "ankle", a second 
steepening at about 5$\times$10$^{17}$ eV. It is widely believed that CRs below the knee are 
accelerated inside our Galaxy. Those CRs carry on average as much energy per unit volume as 
the energy density of star light, the interstellar magnetic fields, or the kinetic energy density of the
interstellar gas. One hundred years after their discovery [205], the question of their origin is still 
open. The acceleration, accumulation and effective mixture of non-thermal particles, through their 
diffusion and convection in galactic magnetic fields, produce the so-called "{\it sea}" of Galactic 
CRs. SNRs are thought to be the main contributor to this CR sea. The main (phenomenological) 
argument in favour of this hypothesis is the CR production rate in the Galaxy, $\dot W_{\rm CR} 
\approx (0.3-1) \times 10^{41} \ \rm erg/s$. This rate can be easily explained if one assumes that 
$\simeq$10\% of the kinetic energy released in supernova explosions (which can reach up to 
$2\times10^{39}$ erg s$^{-1}$ kpc$^{-2}$) is devoted to the acceleration of CRs (see, e.g., 
Refs.~[206-208]).  
Moreover, this highly efficient conversion of kinetic energy of bulk motion to relativistic particles can 
be achieved through diffusive shock acceleration (DSA) [7,8,184], which takes place in the shocks 
created in SNRs. A straightforward test of the acceleration of CRs in SNRs would be the detection of 
hadronic gamma-rays (TeV $\gamma$-rays resulting from hadronic interactions through the production 
and decay of $\pi^0$-mesons) directly from young remnants [209] and/or from dense clouds overtaken 
by the expanding shells [210]. The DSA theory predicts observational features that can be tested with 
Cherenkov telescopes: The characteristic feature of shock acceleration in the non-linear regime is a 
concave shape of the particle energy distribution. At low (GeV) energies, the particle spectrum is 
relatively steep with a differential spectral index larger than 2, whereas it becomes very hard at the 
highest energies. In the presence of a strongly modified shock, the proton spectrum, just before the 
high-energy cut-off, can be as hard as $E^{-1.5}$ (see, e.g. [24] ). These features should then be reflected 
in the spectrum of secondary gamma-rays [28,40,211,212]. However, energy-dependent propagation 
effects of the particles could introduce significant modifications in the proton spectra, in particular in 
dense regions where most of the gamma-rays are produced. This applies to massive molecular clouds 
located outside mid-age SNRs [213,214], as well as possible, dense compact condensations inside 
young SNR shells [28]. Such propagation effects can lead to a substantial deviation of the observed 
gamma-ray spectra from the parent protons spectrum.

\noindent The new sensitive, wide-field-of-view TeV instruments, such as H.E.S.S., have been able to 
image SNRs in TeV gamma-rays, probing the SNR-shell acceleration of either electrons or hadrons up 
to at least 100 TeV (in case of leptonic emission) or a few hundred TeV (for hadronic acceleration). 
The acceleration sites are spatially coincident with the sites of non-thermal X-ray emission, strengthening 
the hypothesis that primary Galactic CRs up to the "knee" of the energy distribution are accelerated in 
SNRs. However, even if radio and X-ray data suggest that SNRs are indeed the sources of CR electrons, 
no compelling evidence for the acceleration of protons in SNRs has been found up to now, and it is not 
clear whether proton and electron accelerators are of different nature. In fact, only a relatively small 
number of SNRs has been detected at VHE (see Sec. 2.1). A possible explanation for this is related to 
the evolution of SNRs, which may only be able to accelerate CRs to PeV energies in the first $\sim1000$ 
yr, while later the high-energy hadrons escape from the system. For instance, a spectral cut-off is observed 
in the bright SNR RX~J1713.7$-$3946 at $\sim$4 TeV which implies an exponential cut-off in the proton 
spectrum at $\sim25$ TeV. 

\vspace{2mm}
\centerline{\psfig{figure=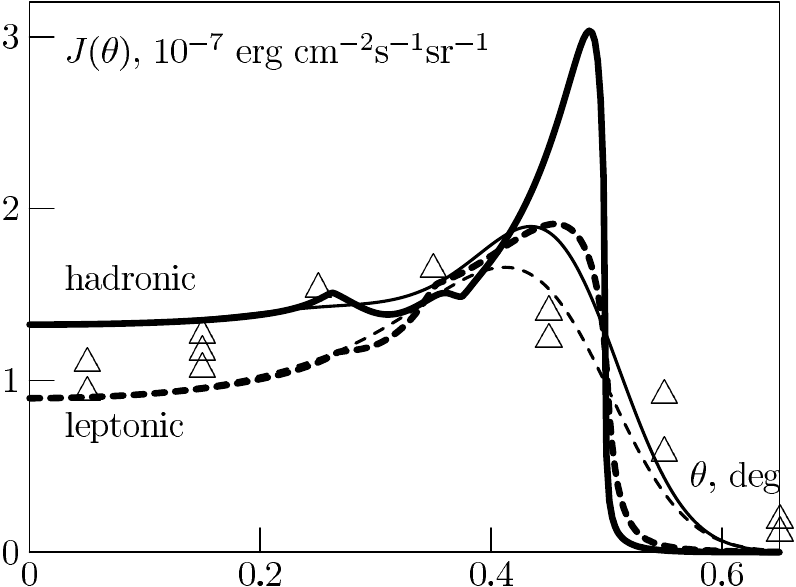, width=2.8in}\vspace{1mm}}
{\baselineskip 5.5pt\renewcommand{\baselinestretch}{0.05}\footnotesize \noindent
{\bf Fig.~20}\quad Radial profiles of 1 TeV gamma-rays calculated for the hadronic and electronic 
scenarios of Ref.~[28] in a uniform medium (solid) and for a leptonic scenario with unmodified forward 
shock (dashed). The profiles, smoothed with a Gaussian point spread function with $\sigma =0.05^{\circ }$, 
are also shown (thin lines). The triangles correspond to the azimuthally-averaged TeV gamma-ray 
radial profile as observed by H.E.S.S. From Ref.~[28].}
\vspace{4mm}

\noindent Still, the production of gamma-rays via $pp$-interactions in dense gas condensations embedded 
in low-density shells represents an interesting scenario which maintains a hadronic origin as a viable option 
with several attractive features. The increase in photon statistic in future observations with {\it Fermi}-LAT and 
H.E.S.S. should help, but may hardly be sufficient to distinguish unambiguously the contributions of leptonic and 
hadronic interactions to the different bands of the gamma-ray spectrum. In this regard, CTA has an important 
role to play. This applies, first of all, to a precise measurements of the energy spectrum below 1~TeV (down 
to tens of GeV), and above 10~TeV (up to 100~TeV). Morphological studies will provide independent and 
complementary information about the radiation mechanism. The low magnetic field, which is a key element 
in any IC model, allows multi-TeV electrons to propagate to large distances, and thus to fill a large volume. 
Because of the homogeneous distribution of target photon fields, the spatial distribution of the resulting IC 
gamma-rays will appear quite broad. Hadronic models, on the other hand, predict narrower and sharper 
spatial distributions, mainly due to the enhanced emission in the compressed region of the shock, as can be
seen in Fig. 20. Using currents instruments with a typical point spread function of $\simeq 3$~arc minutes, 
this effect is smeared out. An angular resolution of 1-2\,arc minutes, however, as reachable with CTA, would 
be sufficient to investigate the radial profile with enough precision to test these scenarios. 

\noindent A second univocal proof of the origin of CRs in SNRs would be the detection of $\gamma$-ray emission 
at extremely high energies, the so-called {\it PeVatron} (1 PeV=$10^{15}$ eV). In SNR shocks with relatively 
low acceleration rates, synchrotron losses typically prevent the acceleration of electrons to energies beyond 
100~TeV. In addition, at such high energies IC emission is suppressed due to Klein-Nishina effects. Therefore, 
a contribution of IC gamma-rays to the radiation above 10~TeV is expected to gradually fade out. Thus, 
detecting gamma-rays up to 100~TeV would unambiguously establish a hadronic origin of the radiation. 
 
\noindent Figure 21 shows an example for the expected X-ray and gamma-ray emission of a $T=1000$~yr-old 
proton PeVatron calculated for three different distributions of the accelerated protons. Both radiation components 
are initiated by the interactions of accelerated protons with ambient gas (of density $n=1 \ \rm  cm^{-3}$) and 
magnetic fields $B=300 \ \mu$G. The gamma-ray emission arises directly from the decays of $\pi^0$-mesons, 
while the X-rays is due to the synchrotron radiation from secondary electrons (the products of $\pi^\pm$-decays). 
The lifetime of electrons producing X-rays, $t_{\rm syn} \simeq 1.5\ B_{\rm mG}^{-3/2}(E_{\rm X}/1\ \rm  keV)^{-1/2}$ 
yr, is very short ($\leq  50$~yr) compared to the age of the source. Roughly the same fraction of the energy of the 
parent protons is shared between the secondary electrons and the gamma-rays from $\pi^0$. However, since the 
energy of the sub-TeV electrons is not radiated away effectively, the direct ($\pi^0$-decay) gamma-ray luminosity 
exceeds the synchrotron luminosity. The ratio $L_{\rm X}/L_\gamma$ depends on the proton spectrum and the 
injection history, and typically does not exceed 0.2-0.3. The noted calculations are based on a constant injection
rate $L_{\rm p}=10^{39}\rm erg/s$, giving a total energy in protons of $W_{\rm p}=L_{\rm p} \cdot T \simeq 3 \times 
10^{49} \rm  erg$.\\ 
In order to estimate the X-ray and gamma-ray energy fluxes (in units of $\rm erg/cm^2 s$) from an arbitrary PeVatron, 
the luminosities shown in Fig. 21 need to be multiplied with a factor $\kappa \approx 10^{-44} \rm (nW_{\rm p}/3 
\times 10^{49} \rm erg/cm^3)(d/1 \ \rm kpc)^{-2}$. This indicates that with the future CTA instrument, with an 
anticipated sensitivity at 10 TeV of about $10^{-13} \ \rm erg/cm^2 s$ (cf. Fig. 1), all Galactic sources capable 
of accelerating particles to PeV energies with $\rm nW_{\rm p}  \geq 10^{49}$ erg/cm$^3$ should become 
detectable up to distances of 10~kpc.

 \vspace{2mm}
\centerline{\psfig{figure=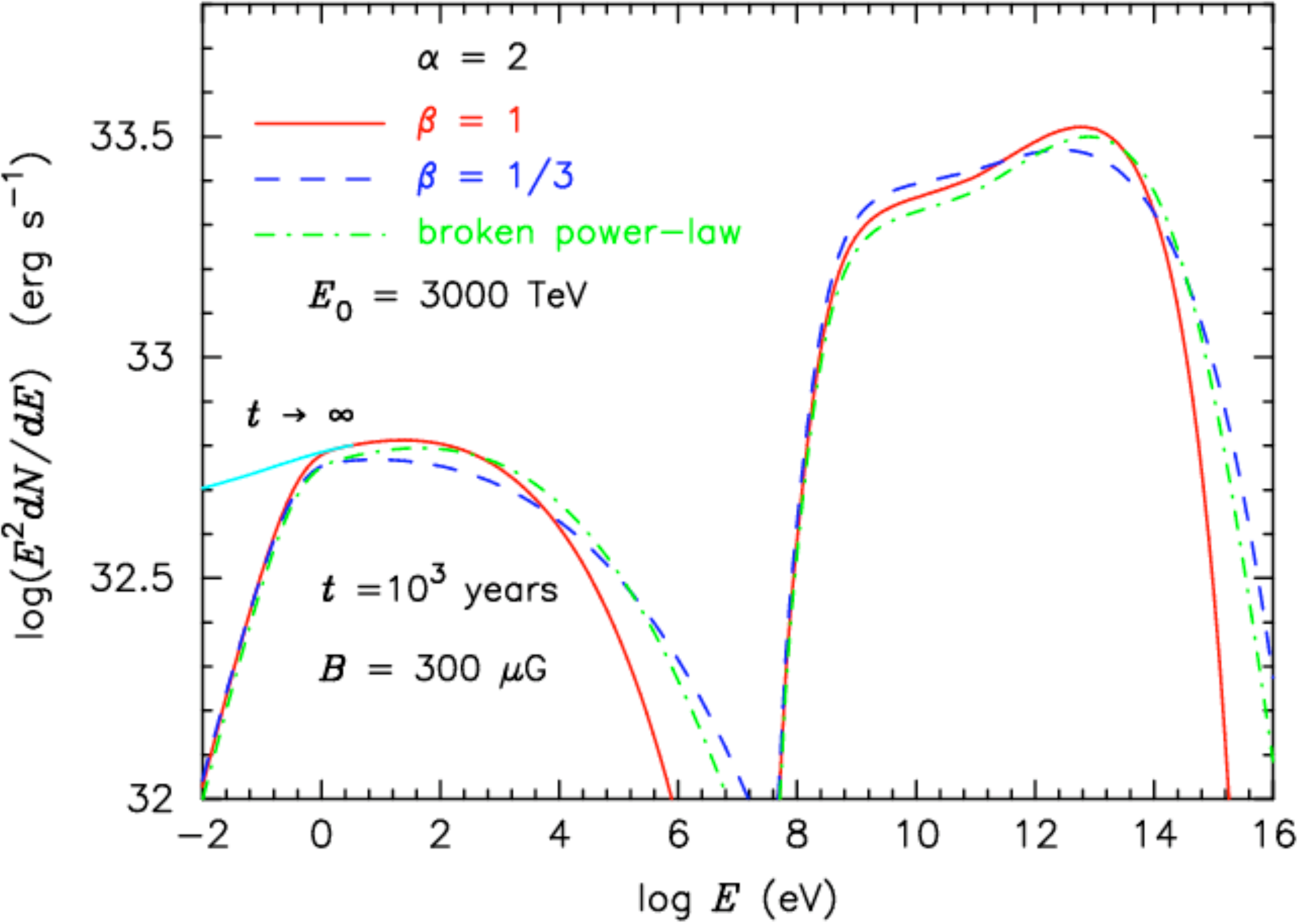, width=3.1in}\vspace{1mm}}
{\baselineskip 5.5pt\renewcommand{\baselinestretch}{0.05}\footnotesize \noindent
{\bf Fig.~21}\quad Broad-band radiation of a PeVatron initiated by interactions of protons with the 
ambient gas. The curves are based on three different proton distributions: a single power-law with 
an exponential cutoff ($E^{-\alpha} \exp({-E/E_0})^\beta$, with $\alpha=2$, $E_0=3$~PeV) but 
different indices $\beta=1$ (solid curve) and $\beta=1/3$ (dashed curve), respectively,  and a broken 
power-law where the spectral index changes at $E=1$~PeV from $\alpha=2$ to $\alpha=3$. From Ref.~[23].}
\vspace{2mm}

\noindent With respect to PeVatron considerations, the relevance of CR escape from the shell confinement has 
become more evident in the last years. Efficient acceleration and confinement of multi-TeV particles in SNRs can 
last less than several hundred years after the explosion, see e.g. [215], effectively constraining the number of 
SNRs emitting as PeVatrons. However, these escaping CRs could still be observable if certain conditions in the 
surrounding environment are fulfilled, such as the presence of massive molecular clouds which could provide a 
dense target to trap those running-away CRs for hadronic interactions, before they diffuse away and get integrated 
in the CR sea. Depending on the distance between the SNR and the molecular cloud (of typical mass of $10^{4}
M_\odot$), the time of particle injection into the interstellar medium and parameters such as the diffusion coefficient, 
different gamma-ray spectral shapes might be expected, reaching TeV flux levels detectable by current instruments 
[213,216].

\noindent Finally, the search for CRs accelerators also needs to be extended to other source classes. Different 
types of accelerators have been proposed to contribute to the Galactic CRs, such as superbubbles, see e.g. 
[217-219], or the remnants of gamma-ray bursts in our Galaxy [220]. Moreover, the CR flux measured at Earth 
might deviate from the Galactic one, if it would be dominated by a single nearby source (or a few local sources)
in the surrounding of the solar system. The recent CR "anomalies" discovered by the PAMELA experiment, 
such as the very high content of positrons in their leptonic component [221], or the significant differences in 
the energy spectra of protons and alpha-particles [222], might be due to an oversimplification of a CR
single-class-of-accelerator origin. In fact, a recent analysis [223] has reported variations of the CR spectrum in 
nearby molecular clouds embedded in the Gould Belt, which may indicate such deviation. Constraints on the 
CR diffusion coefficient in clouds nearby to SNRs also indicate a lower value than the one measured from the 
composition of the local CR spectrum (see, e.g. [224]), suggesting that the CR diffusion coefficient might not be 
isotropic, which could lead to differences in the CR spectrum in different regions of the Galaxy.\\
In this concern, the Galactic plane survey planned with the CTA observatory could result in the discovery of 
new and exciting classes of PeVatron gamma-ray sources of unknown origin [22].

\vspace*{5mm}\noindent{3.2\quad Relativistic Outflows and AGNs}\vspace{3.5mm}

\noindent Powerful relativistic outflows and jets, with speeds approaching that of light, have been 
seen in a number of astrophysical sources including Active Galactic Nuclei (AGNs), Pulsars and 
Microquasars (MQs). If these outflows point towards us, as thought to be the case in e.g. blazars, 
relativistic effects can dramatically change their appearance. The dynamics of and the non-thermal 
processes in these outflows are important fields of active research. Some examples of current 
frontiers relevant to TeV Astronomy are 
mentioned below: 

\vspace*{3mm} \noindent {\it 3.2.1\quad Relativistic pulsar winds:}\\
Outside the light cylinder, pulsars are believed to launch highly magnetised, electron-positron 
winds with terminal flow speeds possibly reaching Lorentz factors up to $\Gamma_w \sim 10^6$ 
(for recent review see e.g., [225-227]). At the location where the interaction with the ambient medium 
(SNR or ISM) causes the wind to slow down, a "termination" shock is formed. It is here where 
efficient particle acceleration is thought to energise the particles radiating in what is observed 
as pulsar wind nebula (PWN). The theory of pulsar winds is usually based on the MHD description of 
relativistic outflows, with its electromagnetic structure being strongly dependent on the current 
it carries. In the simplest (force-free) approximation the plasma flow in the relativistic case 
becomes asymptotically radial. The magnetic field is considered frozen into the flow, so that 
the field structure in the far wind zone becomes almost purely azimuthal/toroidal. The pulsar 
wind is commonly believed to be "cold"  (i.e., characterised by ordered motion only; in the 
co-moving frame all particle are stationary) and therefore mostly unobservable (e.g., there is 
no synchrotron radiation). Bulk inverse Compton scattering of ambient photons, however, 
could still reveal its presence ([228,229]; see also below). In the current MHD picture (cf. Fig.
22), the wind is launched as a highly magnetically (Poynting flux) dominated flow with $\sigma
\equiv B^2/4\pi n m_e c^2 \Gamma_w  \gg 1$ (ratio of Poynting to kinetic energy flux in the 
unshocked wind). Spectral modeling for the Crab Nebula suggests that close to its termination 
shock most of this energy should have been converted to the kinetic energy of the bulk motion 
(i.e., acceleration of the wind), resulting in $\sigma\ll 1$ [230]. How and where exactly this is 
achieved, however, is not yet completely understood ("$\sigma$-problem"). In an unconfined 
(radial) ideal MHD outflow the asymptotic Lorentz would eventually become comparable to 
that of the fast magnetosonic mode $\Gamma_{w,\infty} \simeq \sqrt{\sigma}$ only, implying 
$\sigma_{\infty}\simeq (\sigma_0\Gamma_0)^{2/3} \gg 1$ (e.g., typically $\sigma_0\sim 
10^4$, $\Gamma_0\sim 100$), i.e. the wind would stay Poynting-dominated. If this were to 
be true, efficient first-order Fermi acceleration at the termination shock would not be possible 
(due to a low compression ratio and a relativistic downstream speed).\\ 
The recent detection of pulsed gamma-ray emission between $\sim(100-400)$ GeV from the 
Crab pulsar [57,58] now provides for the first time strong observational support for the presence 
of a (cold) ultra-relativistic pulsar wind and indicates that wind acceleration to $\Gamma_w 
\simgt 10^5$ could already occur very close to the pulsar, i.e., at distances between (20-50) 
light-cylinder radii) [60], see also Fig.~4. Inverse Compton up-scattering (in the KN regime) of 
the pulsed magnetospheric X-ray emission by pulsar wind electrons seems to be responsible 
for VHE emission up to $\sim \Gamma_w m_e c^2 = 500~(\Gamma_w/5\times10^5)$ GeV.

\vspace{3mm}
\centerline{\psfig{figure=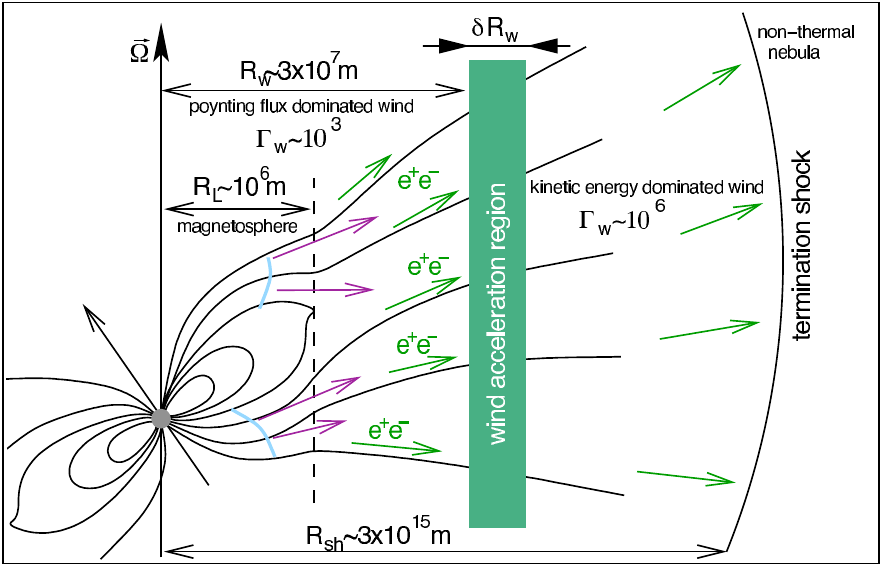, width=3in}\vspace{1mm}}
{\baselineskip 10.5pt\renewcommand{\baselinestretch}{1.05}\footnotesize \noindent
{\bf Fig.~22}\quad Anticipated pulsar wind structure in the case of the Crab.
From Ref.~[60].}
\vspace{4mm}

\noindent The apparent rapid conversion of Poynting to kinetic energy flux is certainly not 
easily accommodated in most of the current scenarios. It could be that this conversion attests
to a longitudinal current distribution that is smaller than the Goldreich-Julian one [231]. On 
the other hand, if the pulsar wind would still possess a non-negligible magnetisation ($\sigma
\simgt 0.1$) approaching the termination shock, cross-field transport would be suppressed, 
making efficient operation of first-order Fermi particle acceleration unlikely. In such a case, 
excitation of fast growing micro-turbulence (with $\delta B/B\gg 1$ on scales $\lambda< r_L$) 
would be required, but seems rarely to occur under the noted conditions [232]. Magnetic field 
annihilation by shock-driven reconnection in a striped wind [233-235] could perhaps offer 
an escape route, both by producing "non-thermal" (power-law type) particle distributions in 
the reconnection electric field, and by generating sufficiently strong turbulence for further 
Fermi acceleration. Alternatively, other non-ideal MHD effects (such as conversion of the 
wind from a sub-luminal MHD into a superluminal electromagnetic wave, expected to be 
accompanied by a substantial energy transfer from fields to particles [236]) may become 
important in outer wind regions. Similar challenges (perhaps even more severe given the 
fact the here the termination shock is located much closer to the pulsar) are encountered in 
the gamma-ray binary context (cf. 3.2.3). Further VHE studies like the ones noted above 
may soon help us to settle a fundamental issue in current pulsar wind theory.

\vspace*{3mm} \noindent {\it 3.2.2\quad Relativistic jets in AGNs:}\\   
AGNs are observed to exhibit collimated (often two-sided) relativistic outflows extending
from sub- up to hundreds of kilo-parsecs. Radio VLBI/VLBA observations of blazar jets 
have revealed significant apparent superluminal motion ($\beta_a=v_a/c >1$) of 
individual jet components on parsec-scale propagating away from the core. The velocity 
distribution of the fastest measured radio jet components peaks at $\beta_a \sim 10$, 
but seems to possess a tail extending up to $\sim50$ [237,238]. This indicates that 
characteristic flow speeds in AGN jets may reach bulk flow Lorentz factors up to $\Gamma
\sim \beta_a/\beta \sim 50$. Current evidence, however, suggests that the apparent 
(parsec-scale) radio jet speeds in the TeV-detected HBL sources are much slower than 
the speeds in the above noted, radio-selected blazars [239]. In fact, the apparent lack of 
superluminal speeds in the parsec-scale radio jets of TeV-HBLs seems to suggest that 
the bulk Lorentz factor in these object is only modest ($\Gamma\sim 2-3$), in contrast to 
estimates of the (sub-parsec-scale) bulk Lorentz factors based on simple SED modelling. 
This could attest to the presence of some velocity gradient in the flow, such as a 
(longitudinally) decelerating flow [241,242] or a (transversally) structured jet with a fast 
moving spine and a slower moving sheath [242]. If this were to be true, the radiative 
interplay between different zones and the possibility of further particle acceleration [243], 
could easily lead to complex emission patterns, thereby severely limiting classical 
inferences based on one-zone synchrotron-Compton modelling. Note that a spine-sheath 
(fast flow/slow flow)-type topology may be naturally expected within the context of current 
MHD models for the formation of relativistic jets. In principle, these models rely for power 
extraction on a source of rotation, i.e., either on the accretion disk (in a Blandford-Payne 
scenario) or the spin of the black hole itself (in a Blandford-Znajek scenario), see e.g. [244,245]
for a recent illustration. 
The extent to which one of these scenarios dominates in AGNs is not fully understood [246]. 
There are recent indications that the black hole spin is less important for powering AGN jets 
than previously believed [247], and that the radio power output in FR~I type sources, believed 
to be the parent population of BL Lacs, is consistent with a rather modest spin distribution 
[248]. Perhaps most likely a mixture occurs in which a pair-dominated, fast-moving ($\Gamma 
\simgt 10$) ergospheric-driven jet is encompassed by a slower-moving ($\Gamma\simlt 10$), 
disk-driven electron-proton jet. Clearly, proper modelling of the VHE source states in TeV 
blazars can shed light on these issues by constraining the required distribution of bulk flow 
Lorentz factors and the required power content in non-thermal particles.

\noindent The observed variability in VHE blazars with evidence for sub-hour variability (cf. 
Table~2) in several sources points to a compact VHE emitting region of size $R \leq D c \Delta 
t_{\rm VHE} (1+z) \simeq 10^{15} (D/10) (\Delta t_{\rm VHE}/1\mathrm{h})$ cm. Assuming a fiducial 
jet opening angle $i \sim 1/\Gamma$, as often done, this would (without further considerations) 
translate into distances of $d \sim \Gamma R \ll 1$ pc. While this often appears to be a good 
working assumption, such an inference seems severely challenged in the case of the distant 
VHE blazar PKS~1222+216 (4C 21.35, $z=0.432$, BH mass of a few times $10^8M_{\odot}$). 
PKS~1222+216 is a bright FSRQ where the strong disk UV continuum ($L_d \sim 10^{46}$ 
erg/s) is thought to photo-ionize what is called to be the "broad line region" (=BLR; often taken 
as a quasi-spherical ensemble of gas clouds moving with speeds of several 1000 km/s at 
distances of $d\sim 0.1$ pc). If the VHE gamma-rays were to originate inside the BLR, strong 
$\gamma\gamma$-absorption features (due to pair production in the dense BLR UV photon 
field) should become apparent in the observed VHE spectrum. Despite this expectation, 
however, the (EBL-deabsorbed) VHE spectrum as seen by MAGIC in June 2010 (coinciding 
with a high GeV source state) seems well compatible with a single power-law of photon index 
$2.7\pm0.3$ between 3 GeV and 400 GeV, without evidence for an intrinsic cut-off [249], see Fig.
23. The obvious inference that the VHE emission in this source may simply originate from beyond 
the BLR has spurred considerable theoretical efforts and led to a rapid proliferation of models, 
e.g. [251-254]. One possibility is that efficient gamma-ray production occurs when the 
parsec-scale jet experiences a strong re-collimation shock [250,251]. 

\vspace{3mm}
\centerline{\psfig{figure=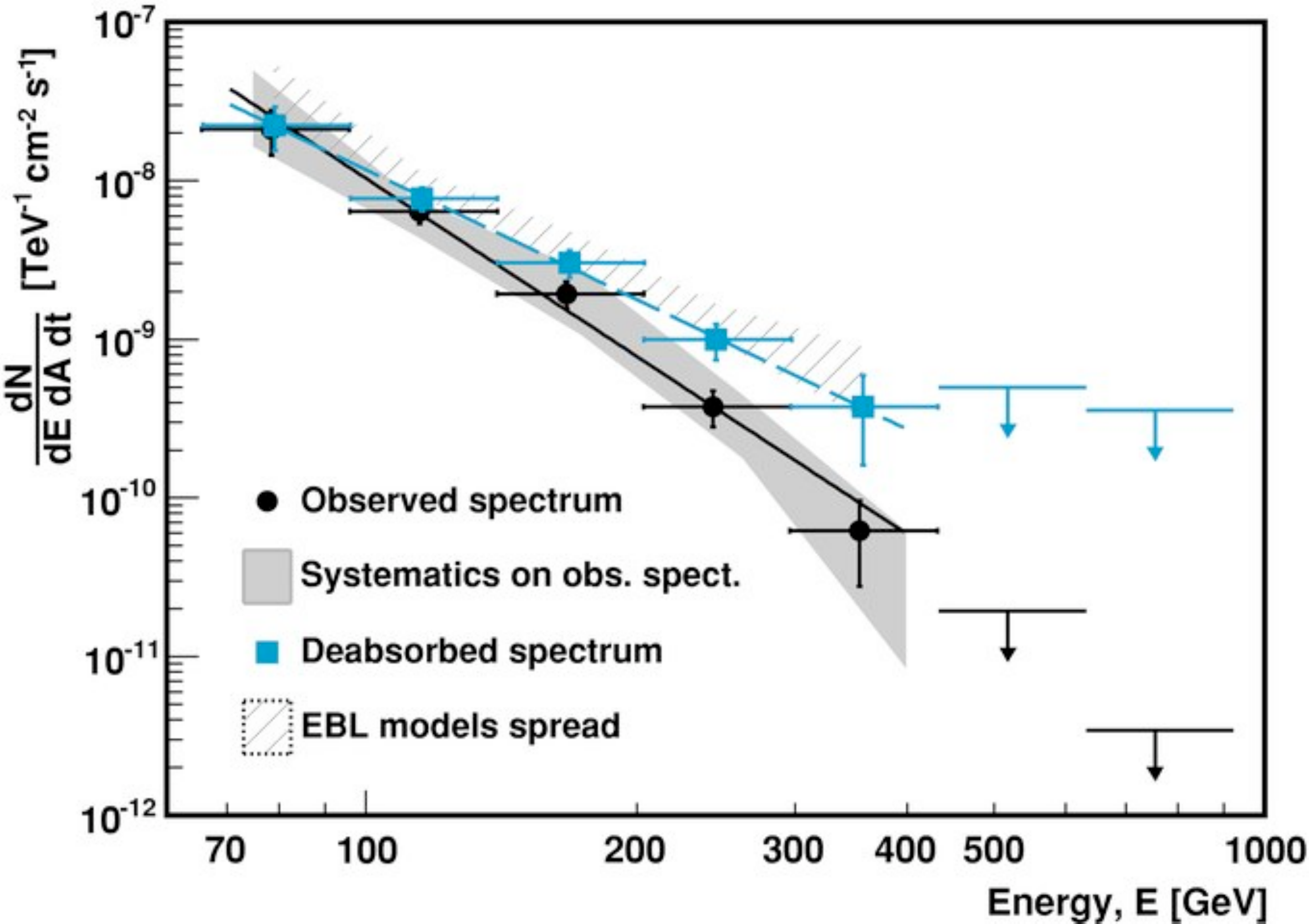, width=3in}\vspace{1mm}}
{\baselineskip 10.5pt\renewcommand{\baselinestretch}{1.05}\footnotesize \noindent
{\bf Fig.~23}\quad Differential energy spectrum of FSRQ source PKS 1222+216 as measured 
by MAGIC. Upper limits are shown as black arrows. The solid line is the best fit of the measured
spectrum to a power law. From Ref.~[249], reproduced by permission of the AAS.}
\vspace{4mm}

\noindent Alternatively, the VHE gamma-rays could also result from energetic, secondary leptons 
produced in photo-hadronic (n+$\gamma\rightarrow \pi$) and $\gamma\gamma$-interactions of 
neutrons and UHE $\gamma$-rays with infrared photons of the more distant dust torus, provided 
efficient UHE proton acceleration in the inner jet, generating an escaping beam of UHE neutrons 
via photo-hadronic processes, and Doppler factor $D\sim 100$ do occur [252]. Other models 
invoke oscillations of VHE source photons into axion-like particles [293] or magnetic reconnection 
events [254]. What seems to be in need of further study at this stage, however, is the extent to which 
more a realistic BLR modelling could perhaps allow to relax some of the opacity constraints (see 
[255] for a first step). As the BLR is not spatially resolved on direct images, reliable information on 
its detailed structure and kinematics is usually not straightforwardly available. It seems likely, for 
example, that in reality the BLR is more complicated, e.g., [256-257], being composed of multiple 
(geometrically different) regions (e.g., an inner, disk-ionized and an outer, jet-energized part; 
spherical or axial symmetric etc) and that the inferred BLR velocity is a superposition of different 
components (turbulence, in- and outflows, rotation etc). Interestingly, the H$_{\beta}$ emission line 
flux in PKS 1222+216 seemed to have varied rather little during the past years (within a factor 
$\simlt 1.7$), despite relatively large (factor $\sim 4.3$) optical continuum variations [259,260], 
which may indicate that the overall BLR is not strongly affected by the observed variable and 
supposedly highly beamed jet emission. Optical and VHE studies along this line could obviously 
help to shed more light on the BLR structure in powerful bazars.

\noindent For M87 (being more massive than PKS~1222+216 by an order of magnitude) the 
observed day-scale VHE activity is already comparable to the light travel time across 
Schwarzschild-radius distances. Under the condition of moderate Doppler boosting ($D
\simlt 3$, as expected for a misaligned jet-source) the observed variable and hard VHE 
emission should originate from near to the black hole. This is facilitated by the fact that 
M87 is highly under-luminous (compared to Eddington, and in contrast to PKS~1222+216), 
making it possible for VHE $\gamma$-rays to escape from its inner jet environment. A 
multitude of VHE emission scenarios, relying on either hadronic and/or leptonic radiation 
processes, have been developed and applied within recent years to get insight into the 
real nature of the source (see [170] for review and references). In the context of leptonic
models, one interesting variant assumes relativistic (Petschek-type) reconnection as 
acceleration mechanism of electrons to high energy and external Compton scattering 
to account for the VHE emission [261,262]. Efficient reconnection could possibly lead
to an additional relativistic velocity component of the ejected plasma (in the bulk jet frame),
making differential Doppler boosting possible, thereby circumventing the problem of a 
modest Doppler factor for the general bulk flow. It is, however, not clear at the moment
whether the required conditions (e.g., high magnetisation for an electron-proton jet, 
negligible guide field) are likely to prevail in AGN jet. Within hadronic models, on the 
other hand, one recent development utilises VHE gamma-ray production via inelastic 
proton-proton (pp) collisions, assuming that interactions of a red giant star or a massive, 
dense gas cloud with the base of the jet could occasionally introduce the required high 
amount of target matter [263]. Recent modelling suggests that such a scenario could 
possibly account for the observed temporal and spectral properties if the jet is powerful 
enough and a sufficiently large fraction of it could be channeled into VHE gamma-ray 
production. 
An alternative explanation has been pursued in the context of magnetospheric models 
(see [202] for review and references), motivated by the observed day-scale VHE variability 
(comparable to Schwarzschild scales) and the observed radio-VHE connection in 2008 
[175]. In a recent scenario [264], pair creation in a hot accretion flow (ADAF) is considered 
to lead to the injection of primary electrons and positrons into magnetospheric vacuum 
gaps, in which they are quickly accelerated to very high Lorentz factors. Curvature 
emission and, in particular, Compton up-scattering of (ADAF) disk photons then results 
in $\gamma$-rays with a spectrum extending up to $10^4$ TeV. For low accretion rates, 
photons with energies below several TeV could escape $\gamma\gamma$-absorption.
On the other hand, IC photons with energies above $\sim10$ TeV could interact with the 
ambient radiation field initiating pair cascades just above the gap. This would lead to a 
large pair multiplicity and ensures that a force-free outflow becomes established, 
accounting for the jet feature as seen in radio VLBA.\\
Obviously, due to its proximity and low luminosity, M87 seems to bear the rare potential 
to give us important insights into the near black hole environment of active galaxies. 
Future, more sensitive VHE observations will be important to study the spectral variability 
and further constrain the time scales of the VHE flux variations, while simultaneous radio 
observations will have the potential to pin down the location of the VHE emission on 
spatial scales $\simlt 10^2 r_s$.

\noindent Perhaps the most prominent example for extreme short-time VHE variability 
is the BL Lac object PKS 2155-304. In July 2006, clear evidence for minute-timescale 
($t_v \sim 200$ sec) VHE variability was found during a dramatic outburst (with flux 
level varying between 1 and 15 Crab units) [137]. The remarkable data set made a 
detailed temporal characterisation of the flaring period possible, showing that the 
variability power is distributed over (temporal) frequencies according to a red-noise-type 
PSD (power spectra density) $\propto \nu^{-\beta}$, with $\beta\simeq 2$ in the frequency 
range $(10^{-4}-10^{-2})$Hz, see Fig.~24 [137,265] (cf. also [266] for a related PSD GeV 
variability analysis below $\sim10^{-6}$ Hz). 

\vspace{3mm}
\centerline{\psfig{figure=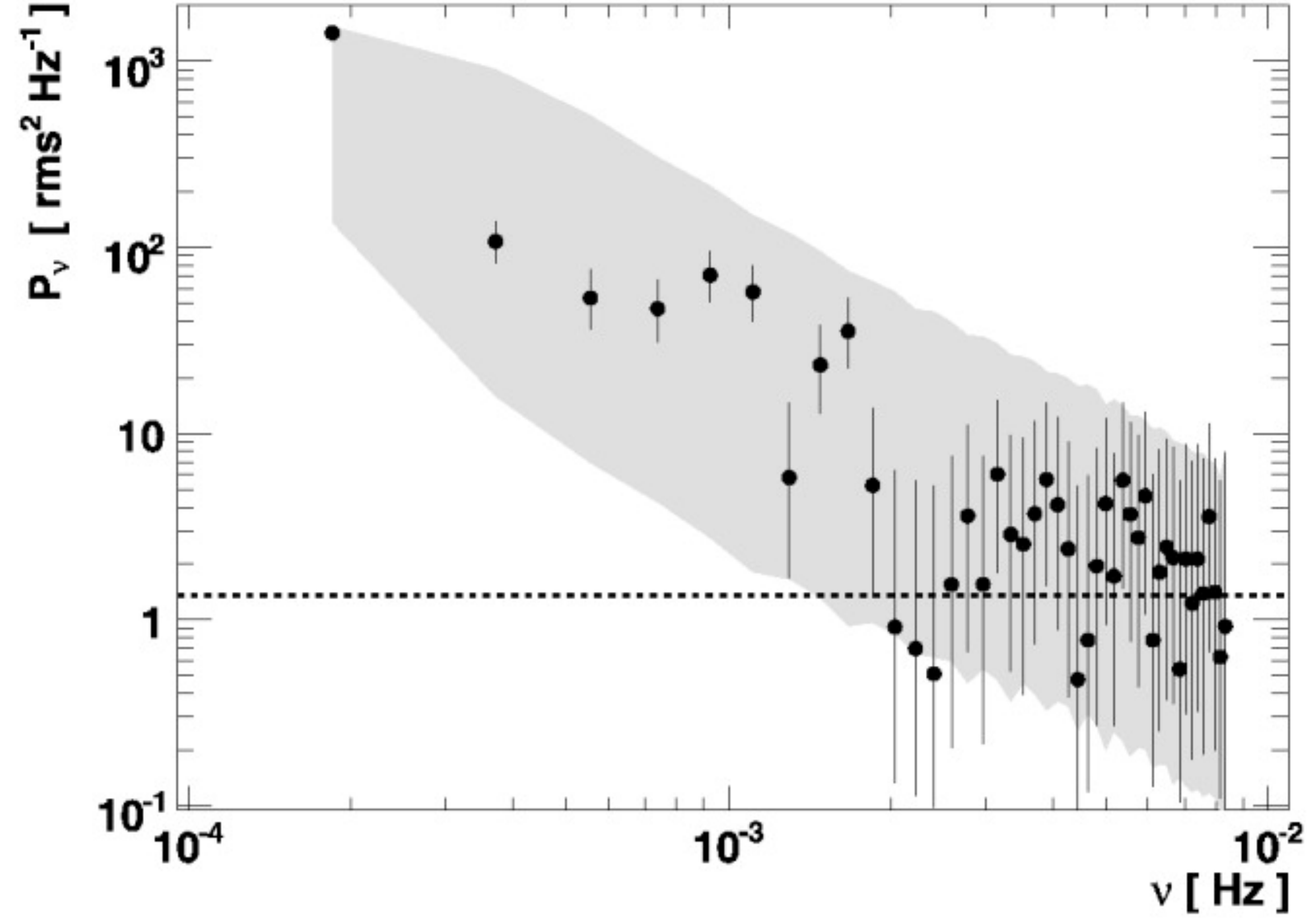, width=3.3in}\vspace{1mm}}
{\baselineskip 10.5pt\renewcommand{\baselinestretch}{1.05}\footnotesize \noindent
{\bf Fig.~24}\quad Power spectrum density (PSD) of the light curve of PKS 2155-304. 
The grey shaded area corresponds to the 90\% confidence interval for a light curve 
with a PSD $\propto \nu^{-\beta}$. The dotted horizontal line shows the average noise 
level. From Ref.~[137].}
\vspace{4mm}

\noindent Similarly to (previous) results in the X-ray domain, a correlation between 
(absolute) rms (root mean square) variability amplitude and mean VHE flux has been 
found. This is known to be characteristic of a non-linear, log-normal stochastic process 
and suggests that the underlying process driving the VHE variability is of multiplicative 
(not additive) nature (see [267], but cf. [268] for a possible caveat). 
The extreme short-term variability has attracted considerable attention, with a number
of interesting models developed to account for it (e.g., spine-sheath [269], reconnection/
mini-jets [270], jet-star interactions [271], non-ideal MHD effects [272] or turbulence [273]). 
Like in Mkn~501, the VHE variability in PKS~2155-304 seems to occur on timescales 
much shorter than the horizon light-crossing time for a typical black hole mass inferred 
from its host galaxy luminosity. This could suggest that the variability originates in a small 
region within a highly relativistic outflow having bulk Lorentz factors $\sim50$ [274] (but 
see [275] for an interesting alternative based on time-dependent modelling). The detailed 
temporal characteristics of the flare (PSD and log-normality) have not yet received as 
much attention. This situation needs to be remedied as it could offer the potential to 
distinguish between the different models proposed. Linear RMS-flux relations, for 
example, are known from black hole X-ray binaries and believed to be an essential
property of the accretion flow around black holes [276], likely produced by small, 
independent fluctuations in the accretion rate [277]. An application of such a scenario
to PKS 2155-204 is possible, but seems to require the presence of a supermassive 
binary black hole system [278]. The increased sensitivity of future instruments like CTA
will allow us to probe the temporal characteristics of flaring TeV blazars beyond the 
usual minimum variability considerations and thereby offer an important diagnostic tool 
to uncover the real nature of the source.

\vspace*{3mm} \noindent {\it 3.2.3\quad Relativistic outflows in galactic VHE binaries:}\\
Galactic VHE binaries are composed of a compact object (a stellar mass black hole in
accreting binary systems=Microquasar [MQ] scenario as in e.g., Cygnus X-1; or a 
ms-pulsar in pulsar wind binaries as in PSR B1259-63) and a bright companion star 
orbiting around each other. Complex interactions between the stellar wind and the outflow 
produced by the compact object (a mildly relativistic MQ jet, or an ultra-relativistic pulsar 
wind) are naturally expected to occur, potentially triggering efficient dissipation and 
gamma-ray production through a variety of processes (see [279] for review). Current 
research frontiers in this context include the following:\\
(1) The interactions of a relativistic pulsar wind with the wind from a stellar companion 
can result in the formation of an extended cometary-shaped radio structure rotating 
with the orbital period [280]. Electrons are thought to be efficiently energised in the 
termination shock region, producing HE and VHE gamma-rays by inverse Compton 
up-scattering of the stellar UV photons whilst being advected away [280,281]. In the 
case of the pulsar wind binary PSR B1259-63, observational evidence for variable and 
extended (AU-scale) radio emission has been found, demonstrating that extended (jet-like) 
structures can be produced in pulsar wind binaries [282]. Rotating jet-like radio features 
have also been detected in LS 5039 and LS~I +61 303 [283-285], whose natures 
are not yet established. In analogy to PSR B1259-63, these findings could suggest a 
pulsar wind binary interpretation. On the other hand, jet-stellar outflow interactions may 
be capable of producing similar features [286,288], and MQ jet precession (with 
variable Doppler boosting) has been recently favoured for LS 5039 based on the finding 
that its radio emission displays double-sided structures changing to one-sided ones [284]. 
The situation could be even more complicated in that the extended radio emission may 
partly also be related to synchrotron emission from secondary pairs generated in 
$\gamma\gamma$-interactions [287]. What seems to be an open issue for a pulsar wind 
interpretation when applied to LS~I +61 303 is that for a reasonable $\gamma$-ray 
production efficiency the required PW momentum flux significantly exceeds the stellar 
(Be) wind one, making the formation of elongated cometary-shaped radio structures 
unlikely [288].\\
(2) The location of the VHE emitting region is not yet understood and different (not 
mutually exclusive) emitting sites are conceivable (see also above). In principle, VHE 
gamma-rays could be produced inside the binary (pulsar wind zone or jet base), on 
binary scales (wind collision/interaction region) or on even larger scales (jet termination). 
In the case of LS~5039, the VHE data favour an origin of the TeV emission on binary 
($\sim 10^{12}$ cm) scales [90,289]. The possibility of efficient particle acceleration 
(at shocks) then becomes an open issue. If particle acceleration occurs inside a MQ jet, 
only mildly relativistic flow speeds may be expected. If, on the other hand, particle 
acceleration is believed to occur in the shock region formed by the encounter of the 
pulsar wind and the stellar wind (standoff distance much closer to the pulsar compared 
to the termination shock for isolated pulsars), efficient first-order Fermi acceleration may 
be difficult to achieve (cf. section 3.2.1.). A comparison of the HE and VHE emission in 
PSR B1259-63 indicates that the spectacular GeV flare seen by Fermi is not accompanied 
by a similar flare in the TeV regime, suggesting that the GeV and the TeV emission are 
produced by different mechanisms [290]. Similarly, in the case of LS~5039 and LS~I
+61 303, the exponential cut-off at a few GeV, combined with the hard spectrum at 
TeV and the anti-correlation between the two bands, points towards different origins 
for their GeV and TeV emissions [291]. Detailed numerical simulations of stellar/pulsar
wind or stellar wind/MQ jet interactions on different scales, see e.g. [292-294], 
are expected to improve our understanding of the dominant VHE emitting region.\\
(3) Finally, interactions of the jet with the ambient medium/ISM (termination) may also 
result in efficient particle acceleration facilitated by, e.g., re-collimation, forward or 
reverse shocks, and possibly lead to VHE gamma-ray production that may eventually 
become detectable [295]. If so, then this could allow to more tightly constrain the jet 
kinetic power and the magnetic field in the interaction region.\\ 
While further high-resolution radio observations bear the promise to reveal the 
true origin of the changing radio morphology in gamma-ray binaries, high(er) 
sensitivity VHE observations with dense temporal coverage are expected to shed 
new light on the non-thermal mechanisms responsible for TeV gamma-ray production,
and will eventually help to disclose the real nature of these sources.

\vspace*{6.0mm} \hrule\vspace{2mm} 
\noindent {\large \usefont{T1}{fradmcn}{m}{n}\xbt 4\quad Conclusion and Perspectives}\vspace{2.5mm}

The remarkable achievements of  observational gamma-ray astronomy over the last decade, and 
the recent theoretical and phenomenological studies of acceleration and radiation processes in 
astrophysical environments, fully justify the further exploration of the sky at high and very-high 
energies. Although generally the main motivations of gamma-ray astronomy remain unchanged, 
the recent observational results have revealed new features which in many cases require revisions 
of existing theoretical models and formulations of new concepts. It is expected that over the next 
decade the ongoing operation of {\it Fermi} will be accompanied by observations with the current 
(H.E.S.S., MAGIC, VERITAS) and planned (CTA, HAWC) ground-based detectors. The data obtained 
in the enormous energy range from 100~MeV to 1~PeV will provide very deep insight into a number 
of problems of high-energy astrophysics and fundamental physics.\\

\noindent {\baselineskip
10.5pt\renewcommand{\baselinestretch}{1.05}\footnotesize {\bf
Acknowledgements}~~~FMR acknowledges LEA support, and EOW support by a RYC 
fellowship.}\vspace{7mm}

\hrule\vspace{2mm}

\noindent {\large
\usefont{T1}{fradmcn}{m}{n}\bf References}\vspace{3.1mm}
\parskip=0mm \baselineskip 15pt\renewcommand{\baselinestretch}{1.25} \footnotesize
\parindent=9mm
\setlength{\baselineskip}{12.3pt}   

\newcommand{\aj}{AJ}
\newcommand{\apj}{ApJ}
\newcommand{\apjs}{ApJS}
\newcommand{\apjl}{ApJ}
\newcommand{\mnras}{MNRAS}
\newcommand{\pasj}{PASJ}
\newcommand{\aap}{A\&A}
\newcommand{\nat}{Nature}
\newcommand{\na}{New Astr.}
\newcommand{\araa}{ARA\&A}
\newcommand{\apss}{Ap\&SS}
\newcommand{\prd}{Phys. Rev. D}
\newcommand{\jcap}{JCAP}
\noindent 
{\bf [1] } F. Aharonian, J. Buckley, T. Kifune and G. Sinnis, Reports on Progress in Physics, 2008, 71: 096901
\\[1mm] {\bf [2] } J.A. Hinton and W. Hofmann, ARA\&A, 2009, 47: 523  
\\[1mm] {\bf [3] } A.U. Abeysekara et al. (HAWC Collaboration), APh, 2012, 35: 641 
\\[1mm] {\bf [4] } M. Actis et al. (CTA consortium), Experimental Astronomy, 2011, 32: 193 
\\[1mm] {\bf [5] } A.V. Plyasheshnikov, F.A. Aharonian and H.J. V{\"o}lk, JPhG, 2000, 26: 183 
\\[1mm] {\bf [6] } F.A. Aharonian et al., APh, 2001, 15: 335 
\\[1mm] {\bf [7] } L.O'C. Drury, RPPh, 1983, 46: 973 
\\[1mm] {\bf [8] } R. Blandford and D. Eichler, Physics Reports, 1987, 1154: 1
\\[1mm] {\bf [9] } J. Albert et al. (MAGIC Collaboration), \aap, 2007, 747: 937 
\\[1mm] {\bf [10] } V.A. Acciari et al. (VERITAS Collaboration), \apj, 2010, 714: 163 
\\[1mm] {\bf [11] } F.A. Aharonian et al. (HEGRA Collaboration), \aap, 2001, 370: 112 
\\[1mm] {\bf [12] } V.A. Acciari et al. (VERITAS Collaboration), \apjl, 2011, 730: L20 
\\[1mm] {\bf [13] } F. Acero et al. (H.E.S.S. Collaboration), \aap, 2010, 516: A62 
\\[1mm] {\bf [14] } R. Enomoto et al. (Cangaroo Collaboration), \nat, 2002, 416: 823 
\\[1mm] {\bf [15] } F.A. Aharonian et al. (H.E.S.S. Collaboration), \nat, 2004, 432: 75 
\\[1mm] {\bf [16] } F.A. Aharonian et al. (H.E.S.S. Collaboration), \aap, 2005, 437: L7 
\\[1mm] {\bf [17] } F.A. Aharonian et al. (H.E.S.S. Collaboration), \apj, 2009, 692: 1500 
\\[1mm] {\bf [18] } A. Abramowski et al. (H.E.S.S. Collaboration), \aap, 2011, 531: A81 
\\[1mm] {\bf [19] } W.W. Tian et al., \apj, 2010, 712: 790 
\\[1mm] {\bf [20] } W.W. Tian et al., \apj, 2008, 679: L85 
\\[1mm] {\bf [21] } F.A. Aharonian et al. (H.E.S.S. Collaboration), \aap, 2008, 484: 435 
\\[1mm] {\bf [22] } F. Acero et al. (for the CTA Collaboration), APh, 2012, to appear (arXiv:1209.0582) 
\\[1mm] {\bf [23] } F.A. Aharonian, Very high energy cosmic gamma radiation : a crucial window on the extreme Universe, World Scientific Publishing, 2004 
\\[1mm] {\bf [24] } M.A. Malkov and L. O'C Drury, Reports of Progress in Physics, 2001, 64: 429 
\\[1mm] {\bf [25] } Y. Uchiyama et al., \nat, 2007, 449: 576 
\\[1mm] {\bf [26] } J. Vink and M. Laming, \apj, 2003, 584: 758 
\\[1mm] {\bf [27] } V.N. Zirakashvili and F.A. Aharonian, \aap, 2007, 465: 695 
\\[1mm] {\bf [28] } V.N. Zirakashvili and F.A. Aharonian, \apj, 2010, 708: 965 
\\[1mm] {\bf [29] } T. Tanaka et al. (Fermi Collaboration), \apjl, 2011, 740: L51 
\\[1mm] {\bf [30] } F. Giordano et al. (Fermi Collaboration), \apjl, 2012, 744: L2 
\\[1mm] {\bf [31] } A.A. Abdo et al. (Fermi Collaboration), \apj, 2011, 734: 28. 
\\[1mm] {\bf [32] } A.A. Abdo et al. (Fermi Collaboration), \apjl, 2010, 710: L2. 
\\[1mm] {\bf [33] } T. Inoue et al., \apj, 2012, 744: 71 
\\[1mm] {\bf [34] } Y. Uchiyama et al.,\apjl, 2010, 723: L122 
\\[1mm] {\bf [35] } A.M. Atoyan et al.,\aap, 2000, 355: 211 
\\[1mm] {\bf [36] } A.M. Atoyan and C. Dermer, \apjl, 2012, 749: L26 
\\[1mm] {\bf [37] } H. V\"olk et al., \aap, 2008, 483: 529 
\\[1mm] {\bf [38] } T. Tanaka et al., \apjl, 2008, 685: 988 
\\[1mm] {\bf [39] } B. Katz and E. Waxman, \jcap, 2008, 1: 18  
\\[1mm] {\bf [40] } D.C. Ellison et al., \apj, 2010, 712: 287 
\\[1mm] {\bf [41] } J. Aleksi\'c et al. (MAGIC Collaboration), \aap, 2012, 541: 11   
\\[1mm] {\bf [42] } A. Fiasson et al. (H.E.S.S. Collaboration), Proceedings of the 31st ICRC, \'Lodz 2009. 
\\[1mm] {\bf [43] } F.A. Aharonian et al. (H.E.S.S. Collaboration), \aap, 2008, 481: 401 
\\[1mm] {\bf [44] } J. Albert et al. (MAGIC Collaboration), \aap, 2007, 664: L87 
\\[1mm] {\bf [45] } V.A. Acciari et al. (VERITAS Collaboration), \apjl, 2009, 698: L113 
\\[1mm] {\bf [46] } M. Tavani et al. (AGILE Collaboration), \apjl, 2010, 710: L151 
\\[1mm] {\bf [47] } A.A. Abdo et al. (Fermi Collaboration), \apj, 2010, 712: 459 
\\[1mm] {\bf [48] } A.A. Abdo et al. (Fermi Collaboration), \apjs, 2010, 187: 460 
\\[1mm] {\bf [49] } A.G. Muslimov and A.K. Harding, \apj, 2003, 588: 430 
\\[1mm] {\bf [50] } A.G. Muslimov and A.K. Harding, \apj, 2004, 606: 1143 
\\[1mm] {\bf [51] } K.S. Cheng et al. \apj, 2000, 537: 965 
\\[1mm] {\bf [52] } S.S. Komissarov, MNRAS, 2002, 336: 759 
\\[1mm] {\bf [53] } J.G. Kirk, Phys. Rev. Lett., 2004, 92:18  
\\[1mm] {\bf [54] } R. B\"uhler et al. \apj, 2012, 749: 26 
\\[1mm] {\bf [55] } A.A. Abdo et al. (Fermi Collaboration), \apj, 2010, 708: 1254 
\\[1mm] {\bf [56] } E. Aliu et al. (MAGIC Collaboration), Science, 2008, 322: 1221 
\\[1mm] {\bf [57] } J. Aleksic et al. (MAGIC Collaboration), A\&A, 2012: 540, 69 
\\[1mm] {\bf [58] } E. Aliu et al. (VERITAS Collaboration), Science, 2011, 334: 69 
\\[1mm] {\bf [59] } J. Takata et al., MNRAS, 2006, 366:1310 
\\[1mm] {\bf [60] } F.A. Aharonian, S. Bogovalov, D. Khangulyan, Nature, 2012: 482, 507 
\\[1mm] {\bf [61] } E. de O\~na Wilhelmi et al. (CTA Collaboration), APh, 2012, in press (arXiv:1209.0357) 
\\[1mm] {\bf [62] } B.M. Gaensler and P.O. Slane, \araa, 2006, 44: 17 
\\[1mm] {\bf [63] } A. Abramowski et al. (H.E.S.S. Collaboration), \aap, 2012, 545: L2 
\\[1mm] {\bf [64] } F.A. Aharonian et al. (HEGRA Collaboration), \apj, 2004, 641: 897 
\\[1mm] {\bf [65] } T.C. Weekes et al., \apj, 1989, 342: 379 
\\[1mm] {\bf [66] } C.F. Kennel and F.V. Coroniti, \apj, 1984, 283: 694 
\\[1mm] {\bf [67] } S.V. Bogovalov et al., \mnras, 2005, 358: 705 
\\[1mm] {\bf [68] } D. Volpi et al., \aap, 2008, 485: 337 
\\[1mm] {\bf [69] } J. Albert et al. (MAGIC Collaboration), \apj, 2008, 674: 1037 
\\[1mm] {\bf [70] } T. Tanimori et al. (CANGAROO Collaboration), \apjl, 1998, 497: L25 
\\[1mm] {\bf [71] } F.A. Aharonian et al. (H.E.S.S. Collaboration), \apj, 2006, 457: 899 
\\[1mm] {\bf [72] } M. Tavani et al. (AGILE Collaboration), Science, 2011, 331: 736 
\\[1mm] {\bf [73] } S.S. Komissarov and Y.E. Lyubarsky, \mnras, 2004, 349: 779 
\\[1mm] {\bf [74] } M. Lyutikov et al., \mnras, 2012, 2769: 
\\[1mm] {\bf [75] } W. Bednarek and W. Idec, \mnras, 2011, 414: 2229 
\\[1mm] {\bf [76] } A.P. Lobanov et al., \aap, 2011, 533: A10 
\\[1mm] {\bf [77] } Q. Yuan et al. \apjl, 2011, 730: L15 
\\[1mm] {\bf [78] } F.A. Aharonian et al., \mnras, 1997, 291: 162 
\\[1mm] {\bf [79] } J. Albert et al. (MAGIC Collaboration), \apj, 2008, 675: L28 
\\[1mm] {\bf [80] } F.A. Aharonian et al. (H.E.S.S. Collaboration), \aap, 2005, 435: L17 
\\[1mm] {\bf [81] } F.A. Aharonian et al. (H.E.S.S. Collaboration), \aap, 2012, in press 
\\[1mm] {\bf [82] } F.A. Aharonian et al. (H.E.S.S. Collaboration), \aap, 2006, 460: 365 
\\[1mm] {\bf [83] } F.A. Aharonian et al. (H.E.S.S. Collaboration), \aap, 2012, in press  
\\[1mm] {\bf [84] } J.M. Blondin et al., \apj, 2001, 563: 806 
\\[1mm] {\bf [85] } F.A. Aharonian et al. (H.E.S.S. Collaboration), \aap, 2005, 432: L25  
\\[1mm] {\bf [86] } A. Djannati-Atai et al. (H.E.S.S. Collaboration), Proc. 30th ICRC, Merida 2007 
\\[1mm] {\bf [87] } F. Acero et al. (H.E.S.S. Collaboration), Proc. 32nd ICRC, Beijing 2011 (arXiv:1201.0481) 
\\[1mm] {\bf [88] } P. Hofverberg et al. (H.E.S.S. Collaboration), Proc. 32nd ICRC, Beijing 2011 (arXiv:1112.2901) 
\\[1mm] {\bf [89] } W. Bednarek, MNRAS, 2006, 368: 579
\\[1mm] {\bf [90] } D. Khangulyan, F. Aharonian and V. Bosch-Ramon, MNRAS, 2008, 383: 467  
\\[1mm] {\bf [91] } A. Sierpowska-Bartosik and D.F. Torres, \apj, 2008, 30: 239
\\[1mm] {\bf [92] } G. Dubus, \aap, 2006, 451: 9
\\[1mm] {\bf [93] } D. Khangulyan et al., IJMPD, 2008, 17: 1909 
\\[1mm] {\bf [94] } M. Tavani and J. Arons, \apj, 1997, 477: 439
\\[1mm] {\bf [95] } J.G. Kirk et al., APh, 1999, 10: 31
\\[1mm] {\bf [96] } F.A. Aharonian et al. (H.E.S.S. Collaboration), \aap, 2005, 442: 1
\\[1mm] {\bf [97] } G. Maier et al. (VERITAS and H.E.S.S. Collaboration), Proc. 32nd ICRC, Beijing 2011 (arXiv:1111.2155)
\\[1mm] {\bf [98] }  J. Aleksic et al. (MAGIC Collaboration), \apj,  2012, 754L: 10A
\\[1mm] {\bf [99] }  F.A. Aharonian et al. (H.E.S.S. Collaboration), Science, 2005, 309: 746
\\[1mm] {\bf [100] } J. Albert et al. (MAGIC Collaboration), Science, 2006, 312: 1771
\\[1mm] {\bf [101] } J. Albert et al. (MAGIC Collaboration), \apj 2009, 693: 1
\\[1mm] {\bf [102] } V.A. Acciari et al. (VERITAS Collaboration), \apj, 2008, 679, 2: 1427
\\[1mm] {\bf [103] }  A. Abramowski et al. (H.E.S.S. Collaboration), \aap, 2012, 541: A5.
\\[1mm] {\bf [104] } J. Albert et al. (MAGIC Collaboration), \apj, 2007, 665: L51
\\[1mm] {\bf [105] }  A.A. Abdo et al. (Fermi Collaboration), \apj 2011, 736: L11
\\[1mm] {\bf [106] } P.H.T. Tam et al., \apj, 2011, 736:  L10
\\[1mm] {\bf [107] } D. Khangulyan et al. \apj, 2012, 752:  L17 
\\[1mm] {\bf [108] } S. D. Bongiorno et al., \apj, 2011, 736:  L11
\\[1mm] {\bf [109] } P.C. Gregory, \apj, 2002, 525: 427
\\[1mm] {\bf [110] } A.A. Zdziarski et al., MNRAS, 2010, 403: 1873
\\[1mm] {\bf [111] }  V.A. Acciari et al. (VERITAS Collaboration), \apj, 2011, 738: 3
\\[1mm] {\bf [112] } R. A. Ong (VERITAS Collaboration), AN, 2011, 3153
\\[1mm] {\bf [113] } D. Hadasch et al. (Fermi Collaboration), \apj, 2012, 749: 1
\\[1mm] {\bf [114] } M.J. Coe et al. (Fermi Collaboration) Science, 2012, 335: 6065 
\\[1mm] {\bf [115] } M. Chernyakova et al., ApJ, 2011, 726: 60 
\\[1mm] {\bf [116] } F. Aharonian et al. (H.E.S.S. Collaboration), A\&A, 2009, 503: 817 
\\[1mm] {\bf [117] } F. Aharonian and A. Neronov, ApJ, 2005, 619: 306 
\\[1mm] {\bf [118] } F. Aharonian and A. Neronov, Ap\&SS, 2005, 300: 255 
\\[1mm] {\bf [119] } T. Linden, E. Lovegrove and S. Profumo, ApJ, 2012, 753: 41 
\\[1mm] {\bf [120] } F. Aharonian et al. (H.E.S.S. Collaboration), A\&A, 2004, 425: L13 
\\[1mm] {\bf [121] } F. Acero et al. (H.E.S.S. Collaboration), MNRAS, 2010, 402: 1877  
\\[1mm] {\bf [122] } F. Aharonian et al. (H.E.S.S. Collaboration), Nature, 2006, 439: 365  
\\[1mm] {\bf [123] } M. Su, T.R. Slatyer and D.P. Finkbeiner, ApJ, 2010, 724: 1044 
\\[1mm] {\bf [124] } R.M. Crocker and F. Aharonian, Phys. Rev. Lett., 2011, 106: 101102 
\\[1mm] {\bf [125] } P. Mertsch and S. Sarkar, Phys. Rev. Lett., 2011, 107: 091101 
\\[1mm] {\bf [126] } K.S. Cheng et al., ApJ, 2011, 731: L17  
\\[1mm] {\bf [127] } C. Lunardini and S. Razzaque, Phys. Rev. Lett., 2012, 108: 221102  
\\[1mm] {\bf [128] } K.S. Cheng et al., ApJ, 2012m 746: 116  
\\[1mm] {\bf [129] } M. Ackermann et al. (Fermi-LAT Collaboration), ApJ, 2011, 743: 171 
\\[1mm] {\bf [130] } F. Aharonian et al., A\&A, 1999, 349: 11 
\\[1mm] {\bf [131] } M. Tluczykont et al. (for the H.E.S.S. Collaboration), POS(Texas2010) 2010, 197 
\\[1mm] {\bf [132] } V. Acciari et al., ApJ, 2010, 715: L49 
\\[1mm] {\bf [133] } G. Fossati et al., MNRAS, 1998: 299, 433  
\\[1mm] {\bf [134] } D. Donato et al., A\&A, 2001: 375, 739  
\\[1mm] {\bf [135] } P. Giommi et al., MNRAS, 2012: 420, 2899  
\\[1mm] {\bf [136] } M. B\"ottcher, in: Fermi meets Jansky, 2010, arXiv: 1006.5048  
\\[1mm] {\bf [137] } F. Aharonian et al. (H.E.S.S. Collaboration), ApJ, 2007, 664: L71 
\\[1mm] {\bf [138] } J. Albert et al. (MAGIC Collaboration), ApJ, 2007, 664: 71 
\\[1mm] {\bf [139] } E. Aliu et al. (MAGIC Collaboration), ApJL, 2009, 692: L29 
\\[1mm] {\bf [140] } J. Aleksic et al. (MAGIC Collaboration), ApJ, 2010, 710: 634 
\\[1mm] {\bf [141] } H. Krawczynski et al., ApJ, 2004, 601: 151 
\\[1mm] {\bf [142] } F.A. Aharonian, NewA, 2000, 5: 377 
\\[1mm] {\bf [143] } M. Barkov et al., ApJ, 2012, 749: 119 
\\[1mm] {\bf [144] } E. Aliu et al. (VERITAS Collaboration), ApJ, 2012, 750: 94 
\\[1mm] {\bf [145] } P. Fortin et al. (for the H.E.S.S. Collaboration), Proc. of the 25th Texas Symposium on Relativistic Astrophysics, PoS(Texas2010), 199 
\\[1mm] {\bf [146] } A. Abramowski et al. (H.E.S.S. Collaboration), APh, 2011, 34: 738 
\\[1mm] {\bf [147] } F. Aharonian et al. (H.E.S.S. Collaboration), Nature, 2006, 440: 1018 
\\[1mm] {\bf [148] }  J. Albert et al. (MAGIC Collaboration), Science, 2008, 320: 1752 
\\[1mm] {\bf [149] } A.M. Taylor, I. Vovk and A. Neronov, A\&A, 2011, 529: 144 
\\[1mm] {\bf [150] } F. Aharonian et al. (H.E.S.S. Collaboration), A\&A, 2007, 475: L9 %
\\[1mm] {\bf [151] } M. Meyer, M. Raue and D. Horns, A\&A, 2012, 542: 59 
\\[1mm] {\bf [152] } M. Ackermann et al. (Fermi-LAT Collaboration), Science, 2012, DOI: 10.1126/science1227160 
\\[1mm] {\bf [153] } A. Neronov, D. Semikoz and A.M. Taylor, A\&A, 2012, 514: 31 
\\[1mm] {\bf [154] } K. Katarzynski et al., MNRAS, 2007, 368: L52 
\\[1mm] {\bf [155] } E. Lefa, F.M. Rieger and F.A. Aharonian, ApJ, 2011, 740: 64  
\\[1mm] {\bf [156] } O. Zacharopoulou et al., MNRAS, 2011, 738: 157  
\\[1mm] {\bf [157] } F. Aharonian et al., Preprint 2012, arXiv:1206.6715 
\\[1mm] {\bf [158] } J. Albert et al. (MAGIC Collaboration), Phys. Lett. B, 2008, 668: 253 
\\[1mm] {\bf [159] } J. Aleksic et al. (MAGIC Collaboration), ApJ, 2012, 748: 46  
\\[1mm] {\bf [160] } E. Massaro et al., A\&A, 2009, 495: 691 
\\[1mm] {\bf [161] } P.L. Nolan et al. (Fermi-LAT Collaboration), ApJS, 2012, 199: 31 
\\[1mm] {\bf [162] } A. Abdo et al. (Fermi-LAT Collaboration), ApJ, 2010, 720: 912 
\\[1mm] {\bf [163] } J. Albert et al. (MAGIC Collaboration), Science, 2008, 5884: 1752 
\\[1mm] {\bf [164] } Y. Becherini et al. (H.E.S.S. Collaboration), AIP Conf. Proc., 2012, 1505: 490 
\\[1mm] {\bf [165] } F. Aharonian et al. (H.E.S.S. Collaboration), ApJL, 2009, 695: L40 
\\[1mm] {\bf [166] } A.A. Abdo et al. (Fermi-LAT Collaboration) ApJ, 2010, 719: 1433 
\\[1mm] {\bf [167] } F.M. Rieger, Mem. S.A.It. 2012, 83:127 
\\[1mm] {\bf [168] } R-Z. Yang et al., A\&A, 2012, 542: A19 
\\[1mm] {\bf [169] } F. Aharonian et al. (HEGRA Collaboration), A\&A, 2003, 403: L1 
\\[1mm] {\bf [170] } F.M. Rieger and F. Aharonian, MPLA, 2012, 27: 12300301 
\\[1mm] {\bf [171] } F. Aharonian et al. (H.E.S.S. Collaboration), Science, 2006, 314: 1424 
\\[1mm] {\bf [172] } J. Albert et al. (MAGIC Collaboration), ApJ, 2008, 685: L23 
\\[1mm] {\bf [173] } E. Aliu et al. (VERITAS Collaboration), ApJ, 2012, 746: 141 
\\[1mm] {\bf [174] } A. Abramowksi et al., ApJ, 2012, 746: 151 
\\[1mm] {\bf [175] } V.A. Acciari et al. (VERITAS, MAGIC, VLBA M87 and H.E.S.S. Collab.), Science, 2009, 325: 444 
\\[1mm] {\bf [176] } J. Aleksic et al. (MAGIC Collaboration), ApJL, 2010, 723: 207  
\\[1mm] {\bf [177] } T.A. Rector, J.T. Stocke and E.S. Perlman, ApJ, 1999, 516: 145 
\\[1mm] {\bf [178] } J. Aleksic et al. (MAGIC Collaboration), A\&A, 2012, 539: 2  
\\[1mm] {\bf [179] } A.M. Brown and J. Adams, MNRAS, 2011, 413: 2785  
\\[1mm] {\bf [180] } M. Kadler et al., A\&A, 2012, 538: L1  
\\[1mm] {\bf [181] } F. Acero et al. (H.E.S.S. Collaboration), Science, 2009, 326: 1080 
\\[1mm] {\bf [182] } A. Abramowksi et al. (H.E.S.S. Collaboration), ApJ, 2012, 757: 158 
\\[1mm] {\bf [183] } V.A. Acciari et al. (VERITAS Collaboration), Nature, 2009, 462: 770 
\\[1mm] {\bf [184] } A.R. Bell, APh, 2012, in press (doi: 10.1016/j.astropartphys.2012.05.022)  
\\[1mm] {\bf [185] } P. Meszaros, AIP Conf. Proc., 2009, 1085: 38 
\\[1mm] {\bf [186] } F. Aharonian et al. (H.E.S.S. Collaboration), A\&A, 2009, 495: 505 
\\[1mm] {\bf [187] } V. Acciari et al. (VERITAS Collaboration), ApJ, 2011, 743: 62. 
\\[1mm] {\bf [188] } J. Albert et al. (MAGIC Collaboration), ApJ, 2007, 667: 358 
\\[1mm] {\bf [189] } A. Ackermann et al. (Fermi-LAT Collaboration), ApJ, 2011, 729: 114 
\\[1mm] {\bf [190] } P. Blasi, S. Gabici and G. Brunetti, IJMPA 2007, 22: 681 
\\[1mm] {\bf [191] } A.M. Atoyan and H.J. Voelk, ApJ, 2000, 535: 45 
\\[1mm] {\bf [192] } V. Petrosian, A. Bykov and Y. Raphaeli, Space Science Reviews, 2008, 134: 191 
\\[1mm] {\bf [193] } G. Vannoni et al., A\&A, 2011, 536: A56  
\\[1mm] {\bf [194] } S. Inoue, F.A. Aharonian and N. Sugiyama, ApJ, 2005, 628: L9 
\\[1mm] {\bf [195] } S.R. Kelner and F.A. Aharonian, Phys. Rev. D, 2008, 78: 034013 
\\[1mm] {\bf [196] } A. Pinzke, C. Pfrommer and L. Bergstr\"om, Phys. Rev. D, 2011, 84: 123509  
\\[1mm] {\bf [197] } J. Aleksic et al. (MAGIC Collaboration), A\&A, 2012, 541: A99 
\\[1mm] {\bf [198] } T. Arlen et al. (VERITAS Collaboration), ApJ, 2012, 757: 123 
\\[1mm] {\bf [199] } F. Aharonian et al. (H.E.S.S. Collaboration), A\&A, 2009, 502: 437 
\\[1mm] {\bf [200] } A. Abramowski et al. (H.E.S.S. Collaboration), A\&A, 2012, in press (arXiv:1208.1370) 
\\[1mm] {\bf [201] } A. Abramowski et al. (H.E.S.S. Collaboration), ApJ, 2012, 750: 123 
\\[1mm] {\bf [202] } F.M. Rieger, IJMPD, 2011, 20: 1547 
\\[1mm] {\bf [203] } G. Pedaletti, S.J. Wagner and F.M. Rieger, ApJ, 2011, 738: 142 
\\[1mm] {\bf [204] } G.V. Kulikov and G.~B. Kristiansen, J Exp Theor. Phys., 1958, 35: 441
\\[1mm] {\bf [205] } V. Hess, Physikalische Zeitschrift, 1912, 13:1084
\\[1mm] {\bf [206] } V.L. Ginzburg and S.I. Syrovatskii, The Origin of Cosmic Rays, New York: Macmillan, 1964
\\[1mm] {\bf [207] } T.K. Gaisser, Cosmic rays and particle physics, Cambridge University Press, 1990
\\[1mm] {\bf [208] } V.S. Berezinskii et al., Astrophysics of Cosmic Rays, Amsterdam 1990, edited by V.L. Ginzburg
\\[1mm] {\bf [209] } L.O' Drury, F.A. Aharonian and H.J. V\"olk, \aap, 1994, 287: 959
\\[1mm] {\bf [210] } F.A. Aharonian, L.O. Drury and H.J. V\"olk, \aap, 1994, 285: 654
\\[1mm] {\bf [211] } E.G. Berezhko and H. V\"olk, \aap, 2008, 492: 695
\\[1mm] {\bf [212] } G. Morlino et al., \mnras, 2009, 392: 240
\\[1mm] {\bf [213] } S. Gabici, F.A. Aharonian and S. Casanova, \mnras, 2009, 396: 1629
\\[1mm] {\bf [214] } D.C. Ellison and A.M. Bykov, ApJ, 2011, 731: 87 
\\[1mm] {\bf [215] } S. Casanova et al., \pasj, 2010, 62: 765
\\[1mm] {\bf [216] } F.A. Aharonian and A.M. Atoyan, \aap, 1996, 309: 917
\\[1mm] {\bf [217] } E. Parizot et al., \aap, 2004, 424: 747
\\[1mm] {\bf [218] } J.C. Higdon and R.E. Lingenfelter, \apj, 2005, 628: 738
\\[1mm] {\bf [219] } A.A. Bykov et al., AIP Conf. Proc., 2012, 1505: 46
\\[1mm] {\bf [220] } A. Atoyan et al., \apjl, 2006, 642: L153
\\[1mm] {\bf [221] } O. Adriani et al., \nat, 2009, 458: 607 
\\[1mm] {\bf [222] } O. Adriani et al., Science, 2011, 332: 69 
\\[1mm] {\bf [223] } A. Neronov, D.V. Semikoz and A.M. Taylor, Phys. Rev. Lett., 2012, 108: 051105 
\\[1mm] {\bf [224] } S. Gabici et al., SF2A, 2010, (arXiv:1009.5291)
\\[1mm] {\bf [225] } J.G. Kirk, Y. Lyubarsky and J. Petri, ASSL 2009, 357: 421  
\\[1mm] {\bf [226] } J. Arons, ASSL 2009, 357: 373 
\\[1mm] {\bf [227] } B. Bucciantini, Int. J. Mod. Phys. Conf. Ser. 2012, 8: 120 
\\[1mm] {\bf [228] } S.V. Bogovalov and F.A. Aharonian, MNRAS, 2000: 313, 504 
\\[1mm] {\bf [229] } L. Ball and J.G. Kirk, APh, 2000: 12, 335 
\\[1mm] {\bf [230] } C.F. Kennel and F.V. Coroniti, ApJ, 1984: 283, 710 
\\[1mm] {\bf [231] } V.S. Beskin and R.R. Rafikov, MNRAS, 2000: 313, 433 
\\[1mm] {\bf [232] } M. Lemoine and G. Pelletier, AIP 2012: 1439, 194  
\\[1mm] {\bf [233] } J. Petri and Y. Lyubarsky, A\&A 2007: 473, 683 
\\[1mm] {\bf [234] } Y. Lyubarsky and M. Liverts, ApJ, 2008: 682, 1436  
\\[1mm] {\bf [235] } L. Sironi and A. Spitkovsky, ApJ 2011: 741, 39 
\\[1mm] {\bf [236] } I. Arka and J.G. Kirk, ApJ, 2012: 745, 108  
\\[1mm] {\bf [237] } M.L. Lister et al., AJ, 2009: 138, 1874  
\\[1mm] {\bf [238] } B.G. Piner et al., ApJ, 2012: 758, 84 
\\[1mm] {\bf [239] } B.G. Piner et al., ApJ, 2010: 723, 1150 
\\[1mm] {\bf [240] } M. Georganopoulos and D. Kazanas,  ApJ, 2003, 599: L5  
\\[1mm] {\bf [241] } A. Levinson, ApJ, 2007, 671: L29 
\\[1mm] {\bf [242] } G. Ghisellini, F. Tavecchio and M. Chiaberge, A\&A, 2005, 432: 401 
\\[1mm] {\bf [243] } F.M. Rieger and P. Duffy, ApJ, 2004, 617: 155 
\\[1mm] {\bf [244] } J.C. McKinney, MNRAS, 2006, 368: 1561 
\\[1mm] {\bf [245] } O. Porth and C. Fendt, ApJ, 2010, 709: 1100  
\\[1mm] {\bf [246] } M. Livio, G.I. Ogilvie and J.E. Pringle, ApJ, 1999, 512: 100  
\\[1mm] {\bf [247] } J.W. Broderick and F.P Fender, MNRAS, 2011, 417: 184  
\\[1mm] {\bf [248] } R. Daly, MNRAS, 2011, 414: 1253  
\\[1mm] {\bf [249] } J. Aleksic et al. (MAGIC Collaboration), ApJL, 2011, 730: L8  
\\[1mm] {\bf [250] } O. Bromberg and A. Levinson, ApJ, 2009, 699: 1274  
\\[1mm] {\bf [251] } F. Tavecchio et al., A\&A, 2011, 534: 86  
\\[1mm] {\bf [252] } C.D. Dermer, K. Murase and H. Takami, ApJ, 2012, 755: 14 
\\[1mm] {\bf [253] } F. Taveccio et al. , Phys. Rev. D, 2012, 86: 085036 
\\[1mm] {\bf [254] } K. Nalewajko et al., MNRAS, 2012, 425: 2519 
\\[1mm] {\bf [255] } F.~Tavecchio and G.~Ghisellini, preprint 2012, arXiv: 1209.2291  
\\[1mm] {\bf [256] } C.M. Gaskell, New Astronomy Review, 2009, 53: 140  
\\[1mm] {\bf [257] } B. Czerny and K. Hryniewicz, A\&A, 2011, 525: L8 
\\[1mm] {\bf [258] } T.G. Arshakian et al., MNRAS, 2010, 401: 1231  
\\[1mm] {\bf [259] } P.S. Smith, G.D. Schmidt and B.T. Jannuzi, Fermi Symposium 2011, arXiv:1110.6040  
\\[1mm] {\bf [260] } E.P. Farina et al., MNRAS, 2012, 424: 393  
\\[1mm] {\bf [261] } D. Giannios, D.A. Uzdenksy and M.C. Begelman, MNRAS, 2010, 402: 1649 
\\[1mm] {\bf [262] } Y-D. Cui et al., ApJ, 2012, 746: 177 
\\[1mm] {\bf [263] }  M.V. Barkov, V. Bosch-Ramon and F.A. Aharonian, ApJ, 2012, 755: 170 
\\[1mm] {\bf [264] } A. Levinson and F. Rieger, ApJ, 2011, 730: 123  
\\[1mm] {\bf [265] } A. Abramowski et al. (H.E.S.S. Collaboration), A\&A, 2010, 520: A83 
\\[1mm] {\bf [266] } A.A. Abdo et al. (Fermi-LAT Collaboration), ApJ, 2010, 722: 520  
\\[1mm] {\bf [267] } P. Uttley, I.M. McHardy and S. Vaughan, MNRAS, 2005, 359: 345 
\\[1mm] {\bf [268] } J. Biteau and B. Giebels, A\&A, 2012, in press (arXiv:1210.2045)
\\[1mm] {\bf [269] } G. Ghisellini and F. Tavecchio, MNRAS, 2008, 386: L28   
\\[1mm] {\bf [270] } D. Giannios, D.A. Uzdenksy and M.C. Begelman, MNRAS, 2009, 395: L29  
\\[1mm] {\bf [271] } M. Barkov et al., ApJ, 2012, 749: 119  
\\[1mm] {\bf [272] } J.G. Kirk and I. Mochol, ApJ, 2011, 729: 104  
\\[1mm] {\bf [273] } R. Narayan and T. Piran, MNRAS, 2012, 420: 604 
\\[1mm] {\bf [274] } M.C. Begelman, A.C. Fabian and M.J. Rees , MNRAS, 2008, 384: L19 
\\[1mm] {\bf [275] } T. Boutelier, G. Henri and P-O. Petrucci, MNRAS, 2008, 390: 73  
\\[1mm] {\bf [276] } L.M. Heil, S. Vaughan and P. Uttley, MNRAS, 2012, 422: 2620 
\\[1mm] {\bf [277] } Y.E. Lyubarskii, MNRAS 1997, 292: 679  
\\[1mm] {\bf [278] } F.M. Rieger and F. Volpe, A\&A, 2010, 520: 23 
\\[1mm] {\bf [279] } V. Bosch-Ramon and D. Khangulyan, IJMPD, 2009, 18: 347  
\\[1mm] {\bf [280] } G. Dubus, A\&A, 2006, 456: 801  
\\[1mm] {\bf [281] } G. Dubus, B. Cerrutti and G. Henri, A\&A, 2010, 516: 18 
\\[1mm] {\bf [282] } J. Moldon et al., ApJ, 2011, 732: L10 
\\[1mm] {\bf [283] } M. Ribo et al., A\&A, 2008, 481: 17 
\\[1mm] {\bf [284] } M. Massi, E. Ros, and L. Zimmermann, A\&A, 2012, 540: A142 
\\[1mm] {\bf [285] } J. Moldon, M. Ribo and J. Paredes, AIP Conf Proc., 2012, 1505: 386 (arXiv: 1210.7702) 
\\[1mm] {\bf [286] } M. Perucho, V. Bosch-Ramon and D. Khangulyan, A\&A, 2010, 512: L4   
\\[1mm] {\bf [287] } V. Bosch-Ramon and D. Khangulyan, PASJ, 2011, 63: 1023  
\\[1mm] {\bf [288] } G.E. Romero et al., A\&A, 2007, 474, 15  
\\[1mm] {\bf [289] } T. Takahashi et al., ApJ, 2009, 697: 592  
\\[1mm] {\bf [290] } A. Abramowski et al. (H.E.S.S. Collaboration), A\&A, 2013, in press (arXiv:1301.3930) 
\\[1mm] {\bf [291] } V. Zabalza et al., A\&A, 2013 to appear (arXiv:1212.3222) 
\\[1mm] {\bf [292] } J. Takata et al., ApJ, 2012, 750: 70 
\\[1mm] {\bf [293] } S.V. Bogovalov et al., MNRAS, 2012, 419: 3426  
\\[1mm] {\bf [294] } M. Perucho and V. Bosch-Ramon, A\&A, 2012, 539: 57 
\\[1mm] {\bf [295] } P. Bordas et al., A\&A, 2009, 497: 325 
\\[1mm] {\bf [296] } H. Gast et al. (H.E.S.S. Collaboration), Proc. 32nd ICRC, Beijing 2011 (arXiv:1204.5860) 
\\[1mm] {\bf [297] } http://www.mpi-hd.mpg.de/hfm/HESS/pages/home/som/ 2011/01/  
\\[1mm] {\bf [298] } A.A. Abdo et al. (Fermi-LAT Collaboration), ApJ, 2009, 701: L123 
\\[1mm] {\bf [299] } A.A. Abdo et al. (Fermi-LAT Collaboration), ApJ, 2009, 706: L56 
\\[1mm] {\bf [300] } A. Abramowski et al. (H.E.S.S. Collaboration), PRL, 2013, in press (arXiv:1301.1173) 
\\[1mm] {\bf [301] } D.I. Jones et al., ApJL, 2012, 747: L12 
\\[1mm] {\bf [302] } Q. Yuan, S. Liu and X. Bi, ApJ, 2012, 761: 133 
\\[1mm] {\bf [303] } S. Sahu, B. Zhang and N. Fraija, Phys Rev D, 2012, 85: 043012  
\\[1mm] {\bf [304] } K. Mannheim, D. Els\"asser and O. Tibolla, 2012, APh, 35: 797 
\\[1mm] {\bf [305] } S. Ohm and J. Hinton, 2013, MNRAS, 429: L70 

\end{multicols}
\end{document}